\documentclass{imsart}

\RequirePackage{amsthm,amsmath,amsfonts,amssymb}
\RequirePackage[numbers]{natbib}
\RequirePackage[colorlinks,citecolor=blue,urlcolor=blue]{hyperref}
\RequirePackage{graphicx}

\arxiv{2212.08446}
\startlocaldefs
\theoremstyle{plain}

\newtheorem{theorem}{Theorem}[section]
\newtheorem{lemma}[theorem]{Lemma}
\theoremstyle{definition}
\newtheorem{definition}[theorem]{Definition}
\newtheorem*{example}{Example}

\newtheorem{assumption}{Assumption}
\theoremstyle{remark}

\endlocaldefs
\usepackage{sectsty, secdot, bm, mathrsfs}
\usepackage{booktabs, threeparttable, enumitem}

\newcommand{\bX}{{\bf{X}}}
\newcommand{\bZ}{\mathbf{Z}}
\newcommand{\bH}{\mathbb{H}}

\newcommand{\bx}{\bm{x}}
\newcommand{\bV}{\bm{V}}

\newcommand{\bSigma}{\bm{\Sigma}}

\newcommand{\bbeta}{\bm{\beta}}

\newcommand{\btheta}{\bm{\theta}}
\newcommand{\bgamma}{\bm{\gamma}}

\newcommand{\bW}{\bm{W}}

\newcommand{\mbE}{\mathbb{E}}
\newcommand{\mbR}{\mathbb{R}}

\newcommand{\bA}{\mathbf{A}}

\newcommand{\Var}{\mathrm{Var}}

\newcommand{\rmPr}{\mathrm{Pr}}
\newcommand{\calG}{\mathcal{G}}

\newcommand{\bS}{\bm{S}}
\newcommand{\bzero}{\bm{0}}
\newcommand{\bQ}{\bm{Q}}
\newcommand{\calL}{\mathcal{L}}
\newcommand{\bLambda}{\bm{\Lambda}}

\newcommand{\bet}{{\bm{\eta}}}

\newcommand{\mA}{\mathcal{A}}

\newcommand{\brebW}{\breve{\bW}}
\newcommand{\bnu}{\bm{\nu}}
\newcommand{\bC}{\mathbf{C}}
\newcommand{\bD}{\mathbf{D}}
\newcommand{\bxi}{\bm{\xi}}

\newdimen\biblioindent    \biblioindent=30pt

 at 7truept

\def\bU{\mathbf{U}}

\def\bV{\bm{V}}

\def\diag{\mathrm{diag}}

\def\tr{\mathrm{tr}}
\def\argmin{\mbox{argmin}}

\def\beq{\begin{equation}}
	\def\eeq{\end{equation}}
\def\beqn{\begin{eqnarray}}
	\def\eeqn{\end{eqnarray}}
\def\beqnn{\begin{eqnarray*}}
	\def\eeqnn{\end{eqnarray*}}

\def\be{\bm{e}}

\def\mA{\mathcal{A}}

\def\bbeta{\bm{\beta}}

\def\btheta{\bm{\theta}}
\def\bgamam{\bm{\gamma}}
\def\bW{\bm{W}}
\def\bGamma{\bm{\Gamma}}

\begin{document}
\begin{frontmatter}
\title{Score function-based tests for ultrahigh-dimensional linear models}

\begin{aug}
\author[A]{\fnms{Weichao}~\snm{Yang}\ead[label=e1]{yangweichao@mail.bnu.edu.cn}},
\author[B]{\fnms{Xu}~\snm{Guo}\ead[label=e2]{xustat12@bnu.edu.cn}}
\and
\author[C]{\fnms{Lixing}~\snm{Zhu}\ead[label=e3]{lzhu@bnu.edu.cn}}
\address[A]{School of Statistics,
Beijing Normal University\printead[presep={,\ }]{e1}}
\address[B]{School of Statistics,
	Beijing Normal University\printead[presep={,\ }]{e2}}

\address[C]{Center for Statistics and Data Science,
Beijing Normal University at Zhuhai\printead[presep={,\ }]{e3}}
\end{aug}

\begin{abstract}
In this paper, we investigate score function-based tests to check the significance of an ultrahigh-dimensional sub-vector of the model coefficients when the nuisance parameter vector is also ultrahigh-dimensional in linear models. We first reanalyze and extend a recently proposed score function-based test to derive, under weaker conditions, its limiting distributions under the null and local alternative hypotheses. As it may fail to work when the correlation between testing covariates and nuisance covariates is high, we propose an orthogonalized score function-based test with two merits: debiasing to make the non-degenerate error term degenerate and reducing the asymptotic variance to enhance power performance. Simulations evaluate the finite-sample performances of the proposed tests, and a real data analysis illustrates its application.
\end{abstract}



\end{frontmatter}

\section{Introduction}

It is important to check whether the covariates of interest contribute to the response, given the other covariates. In linear regression models, this is formulated as testing whether the parameter vector of interest is equal to zero. This paper studies inference of an ultrahigh-dimensional parameter vector of interest with an ultrahigh-dimensional nuisance parameter vector. This problem is of great importance in practice. For instance, researchers may aim to test whether a gene pathway, consisting of ultrahigh-dimensional genes for the same biological functions, is important for a certain clinical outcome, given the other ultrahigh-dimensional genes.

For this challenging problem, there are several proposals available in the literature. The coordinate-based maximum tests have been proposed recently. See for instance \cite{ning2017general,zhang2017simultaneous,dezeure2017high,ma2021global,wu2021model}. These methods are computationally expensive because many penalized optimization implementations with ultrahigh-dimensional parameter vector are involved.  
A Wald-type test was suggested by \cite{guo2021group}, which is computationally low cost. They imposed the boundedness of the eigenvalues of the covariance matrix.
To tackle the problem in the study of the asymptotic properties, brought by too small variance (\cite{guo2021group} pointed out), a positive tuning parameter over the sample size is added to the estimated variance. As the limiting null distribution remains unknown, their test with the critical values determined by the standard normal distribution  is conservative in theory (see the discussion on page 11 of \cite{guo2021group}). 

Score function-based testing procedures are also popular. 
When the dimension of the nuisance parameter vector is low or diverging at a relatively slow rate and the parameter vector of interest is high-dimensional, the references include \cite{goeman2006testing}, \cite{zhong2011tests}, \cite{guojrssb2016} (for generalized linear models), \cite{cui2018test}, and \cite{guo2022conditional}. The recent development of score function-based testing procedure is made by  \cite{chen2022testing} who  extended the score function-based test of \cite{guojrssb2016} to handle ultrahigh-dimensional nuisance parameter vector. This test is suitable under dense alternative hypotheses and the numerical studies supported some merits of their test. However, their results require that the eigenvalues of the covariance matrix are all bounded and the dimension of the parameter vector of interest grows polynomially with the sample size to guarantee nontrivial power. Further the limiting distributions under the local alternatives are not established.

The above observations motivate us to further study the score function-based test for linear models, extend the results in the literature and propose new test to handle high correlation between nuisance and testing covariates.  To be specific,  we will do the following. First, we need to reanalyze, under weaker conditions, the properties of the test statistic and extend the results to the case where the testing parameter and nuisance parameter vectors are ultrahigh-dimensional simultaneously at the rates up to the exponential of the sample size. To this end, we derive the limiting distributions under the null and local alternative hypotheses. Second, when the correlation between the covariates of interest and the nuisance covariates is strong, a non-negligible bias causes the tests in \cite{chen2022testing} fail to work. Therefore, we propose an orthogonalization procedure to reduce the possible bias. Although this technique has been adopted in the recent high-dimensional inference literature \citep{zhang2014confidence,van2014asymptotically,javanmard2014confidence,belloni2015uniform,chernozhukov2018double}, to the best of our knowledge, it has not been applied to constructing test statistics based on the quadratic norm of the score function for ultrahigh-dimensional testing parameter vector. Two merits shown in our investigation are as follows. The orthogonalization can debiase the error terms and convert the non-degenerate error terms to degenerate, thus relaxing the correlation assumption between the covariates of interest and nuisance covariates; it can also reduce the variance of the test statistic and thus enhance the power performance, which was not observed in the literature.

Technically,  we establish the asymptotic normality of the two proposed test statistics in a different way from those used by \cite{zhong2011tests, guojrssb2016, cui2018test} for the quadratic norm-based test statistics. Instead of calculating the relatively complex {spectral} norm of the ultrahigh-dimensional sample matrix, we derive the order of element-max norm of the ultrahigh-dimensional $U$-statistics with the help of maximal inequalities established in \cite{chernozhukov2015comparison} and \cite{chen2018gaussian}. The technique developed in this paper can be useful for other high-dimensional inference problems.

The rest of the paper is organized as follows. Section \ref{sec2} re-analyzes the test statistic in \cite{chen2022testing} to handle the case with  higher dimensional parameter vector of interest and presents the  limiting distributions under both null and local alternative hypotheses. The failure of this test statistic in the high correlation case is also discussed. Section \ref{sec3} introduces the orthogonalization procedure. Section \ref{sec4} contains an oracle inference procedure to illustrate the merits of the orthogonalization approach. Further, Section \ref{sec5} develops an orthogonalization-based test in the general case and derives the relevant asymptotic analysis.  Section \ref{sec6} presents simulation studies and a real data analysis. 
Section \ref{sec7} offers some conclusions. 
The Appendix includes detailed proofs of the theoretical properties, technical lemmas, and additional simulation results.

Before closing this section, we introduce some necessary notations. For a $d$-dimension vector $\bU$,  write $\lVert \bU\rVert_r = (\sum_{k=1}^{d}|U_{k}|^{r})^{1/r}$ and $\lVert \bU \rVert_{\infty} = \max_{1\leq k\leq d}\lvert U_{k}\rvert$ to denote $L_r$ and $L_{\infty}$ norms of $\bU$, where $U_{k}$ is the $k$-th element of $\bU$. Further define $\|\bU\|_{0}=\#\left\{k: U_{k} \neq 0\right\}$. A random variable $X$ is $\operatorname{\mathit{sub-Gaussian}}$ if the moment generating function (MGF) of $X^2$ is bounded at some point, namely $\mathbb{E}\exp(X^2/K^2)\leq2$, where $K$ is a positive constant. A random vector $\bX$ in $\mbR^p$ is called $\operatorname{\mathit{sub-Gaussian}}$ if $x^\top\bX$ are sub-gaussian random variables for all $x\in\mbR^p$. For $a,b\in\mathbb{R}$, write $a\vee b=\max\{a,b\}$. {For $p_1\times p_2$ dimensional matrix $\bA$, write $\lambda_{\rm{max}}(\bA)$ to denote the spectral norm of $\bA$.} Further define $\lVert\bA\rVert_{0} = \#\{(i,j): A_{ij} \neq 0 \}$ and $\lVert\bA\rVert_{F} = \{\tr(\bA\bA^\top)\}^{1/2}$, where $A_{ij}$ is the $(i,j)$-th element of $\bA$.

\section{The score function-based test and  new results}
\label{sec2}

Let $Y\in\mbR$ be the response variable along with the covariates $\bX = (X_{1},\ldots,X_{p_{\bbeta}})^\top\in\mbR^{p_{\bbeta}}$ and $\bZ = (Z_{1},\ldots,Z_{p_{\bgamma}})^\top\in\mbR^{p_{\bgamma}}$. Consider the following linear model:
\begin{align}
	\label{linearmodel1}
	Y = \bbeta^\top\bX + \bgamma^\top\bZ + \epsilon,
\end{align}
where $\epsilon$ is the random error satisfying $\mbE(\epsilon) = 0$ and $\mbE(\epsilon^{2}) = \sigma^{2}$. Let $\bV = (\bX^\top,\bZ^\top)^\top$ and $\bSigma$ be the covariance matrix of  $\bV$. Without loss of generality, assume that $\mbE(\bV) = \bzero$, $\bSigma$ is positive definite and $\epsilon$ is uncorrelated with $\bV$. Our primary interest is to detect whether $\bX$ contributes to the response $Y$ or not given the other covariates, which is testing the following inference problem:
\begin{align}
	\label{question1}
	\bH_{0}:\bbeta = \bm{0},\qquad \mathrm{versus}\qquad \bH_{1}:\bbeta \neq \bm{0}.
\end{align}

To test the above hypothesis, we can construct test statistics based on score functions. An advantage of score function-based tests is that we do not need to estimate the parametric vector of interest.
To be precise, consider the following $L_2$ loss function:
\begin{align*}
	\mathcal{L}(\bbeta,\bgamma) = \mbE (Y-\bbeta^\top\bX - \bgamma^\top\bZ)^2/2,
\end{align*}
and the corresponding score function of $\bbeta$:
\begin{align*}
	\partial\mathcal{L}(\bbeta,\bgamma)/\partial\bbeta=\nabla_{\bbeta}\mathcal{L}(\bbeta,\bgamma) = -\mbE\{(Y-\bbeta^\top\bX - \bgamma^\top\bZ)\bX\}.
\end{align*}
Then $\nabla_{\bbeta}\mathcal{L}(\bzero,\bgamma) = \bzero$ corresponds to  $\bH_0$; otherwise, to $\bH_{1}$. 

A test statistic can be based on the quadratic norm $\nabla_{\bbeta}\mathcal{L}(\bzero,\bgamma)^\top\nabla_{\bbeta}\mathcal{L}(\bzero,\bgamma)$. As the nuisance parameter $\bgamma$ is unknown, we can replace $\bgamma$ with {an estimator} $\hat{\bgamma}$.
Now suppose $\{\bX_i,\bZ_i,Y_i\}_{i=1}^{n}$ is a random sample from the population $(\bX,\bZ,Y)$. \cite{guojrssb2016} proposed the following test statistic based on the quadratic norm of the score function:
\begin{align}
	\label{teststatisticguo}
	T_{n} = \frac{1}{n}\sum_{i\neq j}(Y_i - \hat\bgamma^\top\bZ_i)(Y_j - \hat\bgamma^\top\bZ_j)\bX_i^\top\bX_j,
\end{align}
where $\hat\bgamma$ is the least squares estimator. Clearly $T_n$ can be seen as a $U$-statistic type estimator of $(n-1)\nabla_{\bbeta}\mathcal{L}(\bzero,\hat\bgamma)^\top\nabla_{\bbeta}\mathcal{L}(\bzero,\hat\bgamma)$. In the asymptotic analysis, the growth rate of the dimension of $\bgamma$ is required to be slower than $n^{1/4}$. Recently, \cite{chen2022testing} extended it to handle the ultrahigh-dimensional nuisance parameter situation. They obtained the estimator $\hat\bgamma$ by solving the following penalized problem,
\begin{align}
	\label{bgammaestimator}
	\hat\bgamma = \mathop{\argmin}\limits_{\bgamma\in\mbR^{p_{\bgamma}}}\frac{1}{2n}\sum_{i=1}^{n}(Y_i - \bgamma^\top\bZ_i)^2 + \lambda_{Y}\lVert\bgamma\rVert_{1},
\end{align}
where $\lambda_Y$ is the tuning parameter and $\bgamma$ has a sparse structure. Other penalties such as SCAD and MCP, are also applicable. To deal with the case with ultrahigh-dimensional parameter vector of interest and relax some conditions, 
we conduct a further investigation for their test firstly.

Compared with existing high-dimensional literature in which the dimension of nuisance parameter is much less than sample size, there are some technical difficulties listed as following.
\begin{itemize}
    \item There is no explicit formula for the penalized estimator. When the dimension of the nuisance parameter vector is low or diverging at relatively slow rate, the nuisance parameter was estimated by least square estimation or maximum likelihood estimation in \cite{guojrssb2016,guo2022conditional}. These estimators aforementioned have explicit formulas. The theoretical results can be proved by pluging the explicit forms of the estimators in relevant test statistics. However, when the nuisance parameter vector is ultrahigh-dimensional, these estimators are no longer feasible. Instead we use penalized estimation procedures to deal with the ultrahigh-dimensional unknown nuisance parameter vector. This brings great challenge now since the penalized estimators do not have explicit formulas.

    \item It is difficult to calculate the spectral norm of an ultrahigh-dimensional sample matrix. To establish the asymptotic normality, we need to derive the order of the following term 
    $$(\widehat{\bgamma}-\bgamma)^\top\frac{1}{n}\sum_{i\neq j}\bZ_i\bZ_j^\top\bX_i^\top\bX_j(\widehat{\bgamma}-\bgamma).$$
    When the dimension of nuisance covariates $\bZ$ is relatively low, previous literature \citep{zhong2011tests, guojrssb2016, cui2018test} calculated the spectral norm of $n^{-1}\sum_{i\neq j}\bZ_i\bZ_j^\top\bX_i^\top\bX_j$. However, when the dimension of nuisance covariates $\bZ$ is ultrahigh, it is very difficult to obtain the order of the spectral norm of the above ultrahigh-dimensional matrix. Alternatively we turn to derive the order of element-max norm of the related ultrahigh-dimensional sample matrix with the help of maximal inequalities established in \cite{chernozhukov2015comparison} and \cite{chen2018gaussian}. This is a new technical approach for analyzing score function-based tests. 
\end{itemize}

\subsection{Limiting null distribution}
\label{limitnulldisofsec2}

Let $\bSigma_{\bX}$ and $\bSigma_{\bZ}$ be the covariance matrices of the covariates $\bX$ and $\bZ$ respectively.
{Denote $p_{\bbeta}$, $p_{\bgamma}$ as the dimension of $\bbeta$ and $\bgamma$, and let $p = p_{\bbeta} + p_{\bgamma}$.
	Denote $\bgamma_{\phi} = \bSigma_{\bZ}^{-1}\mbE(\bZ Y)$, and $\bgamma_{\phi} = \bgamma$ under $\bH_{0}$.
Let $s$ be a positive integer and represents the sparsity level of $\bgamma_{\phi}$. Let $\varrho^2 = \max_{1\leq k\leq p_{\bgamma}}\lVert\mbE(Z_{k}\bX)\rVert_{2}^2$ and $\varpi^2 = s^2\log p_{\bgamma}\varrho^2$. Here $\varrho^2$ describes the dependence of covariates of interest and nuisance covariates. Next, under some technical assumptions, we study the asymptotic null distribution of the test statistic $T_n$ with $\hat\bgamma$ in \eqref{bgammaestimator}.

\begin{assumption}
	\label{assumptionbmatrix2}
	$\tr(\bSigma^4_{\bX})=o(\tr^2(\bSigma_{\bX}^2))$ and $\tr(\bSigma_{\bX}^2)\rightarrow\infty$ as $(n,p_{\bbeta})\rightarrow\infty$.
\end{assumption}

\begin{assumption}
	\label{assumptionb5}
	$\bV$ can be expressed as
	\begin{align*}
		\bV = \bGamma \bnu,
	\end{align*}
	where $\bGamma$ is a $p \times m$ dimensional matrix with $p\leq m$.
	The $L_2$ norms of row vectors in $\bGamma$ are uniformly bounded.
	$\bnu$ is an $m$-dimensional sub-Gaussian random vector with mean zero and identity covariance matrix.
\end{assumption}

\begin{assumption}
	\label{assumptionb6}
	$\lVert \hat{\bgamma} - \bgamma_{\phi} \rVert_{1} = O_p(s\sqrt{\log p_{\bgamma}/n})$.
\end{assumption}

\begin{assumption}
	\label{assumptionb8}
	$\log p_{\bgamma} = O(n^b)$ for some constant $0<b<1/3$.
\end{assumption}

\begin{assumption}
	\label{assumptionb7}
	$\epsilon$ is $\operatorname{\mathit{sub-Gaussian}}$ with bounded $\operatorname{\mathit{sub-Gaussian}}$ norm.
\end{assumption}
Assumption \ref{assumptionbmatrix2} frequently appeared in the literature \citep{chen2010two, guojrssb2016, cui2018test} and is required in applying the martingale central limit theorem. If all the eigenvalues of $\bSigma_{\bX}$ are bounded, then Assumption \ref{assumptionbmatrix2} holds. In the previous works such
as \cite{ma2021global}, \cite{guo2021group}, \cite{chen2022testing}, they required that the eigenvalues of $\bSigma$ are all bounded, which is stronger than our assumption \ref{assumptionbmatrix2}. Assumption \ref{assumptionb5} says that $\bV$ can be expressed as a linear transformation of an $m$-dimensional sub-Gaussian vector $\bnu$ with zero mean and unit variance.
This assumption is similar to the pseudo-independence assumption, which is widely used in the literature such as \cite{bai1996effect, chen2009effects, chen2010two, zhong2011tests, cui2018test}. The boundedness assumption of $L_2$ norms of row vectors in $\bGamma$ is imposed to ensure that sub-Gaussian norms of the components of $\bV$ are uniformly bounded. Assumption \ref{assumptionb6} requires the $L_1$ error bound of $\hat\bgamam$ at the order of $s(\log p_{\bgamma}/n)^{1/2}$. Many estimators, such as Lasso, SCAD, and MCP, can achieve such a rate of convergence.
{See for instance \cite{loh2015regularized}.} Assumption \ref{assumptionb8} allows the dimension of the nuisance parameter in an exponential order of the sample size. Assumption \ref{assumptionb7} is standard in the analysis for high-dimensional linear models.

\begin{theorem}
	\label{limitnulldisofguosum}
	Under $\bH_{0}$ in \eqref{question1} and Assumptions~\ref{assumptionbmatrix2} -- \ref{assumptionb7}, and the following two conditions:
	\begin{align}
		\label{condition1inguosum}
		\varpi^2\vee (\log p_{\bgamma})^{1/2}\varpi\lambda_{\rm{max}}^{1/2}(\bSigma_{\bX}) = o(\sqrt{\bLambda_{\bX}}),
	\end{align}
	and
	\begin{align}
		\label{condition2inguosum}
		s(\log p_{\bgamma})^{3/2}/\sqrt{n} = o(1),
	\end{align}
	we have
	\[
	\frac{T_n}{\sqrt{2\bLambda_{\bX}}}\rightarrow N(0,1)\quad \text{in distribution}
	\]
	as $(n,p_{\bbeta},p_{\bgamma})\rightarrow\infty$, where  $\bLambda_{\bX} = \sigma^4\mathrm{tr}(\bSigma^2_{\bX})$.
\end{theorem}

Note that \cite{zhang2017simultaneous} assumed $s^*(\log p)^{3/2}/\sqrt{n} = o(1)$. Here $s^*$ represents the sparsity level of the entire parameter vector $(\bbeta^\top, \bgamma^\top)^\top$. Under null hypothesis, $s^*(\log p)^{3/2}/\sqrt{n} = o(1)$ becomes $s(\log p)^{3/2}/\sqrt{n} = o(1)$. Since $p$ can be much larger than $p_{\bgamma}$, the sparsity requirement in \eqref{condition2inguosum} is still weaker than that in \cite{zhang2017simultaneous}. 
Condition \eqref{condition1inguosum} restricts the sparsity, the dimensions of the nuisance and testing parameter, and the correlations among the covariates. Generally speaking, if the correlations are weak and the dimension of the testing parameter is high, the sparsity level and the dimension of the nuisance parameter can be very high.
Therefore,  when the correlation between $\bX$ and $\bZ$ is weak, $T_{n}$ still has a tractable limiting null distribution.
Particularly, when $\bSigma$ has bounded eigenvalues, $\varrho^2$ and $\lambda_{\rm{max}}(\bSigma_{\bX})$ can be bounded by a constant. Thus condition \eqref{condition1inguosum} can be simplified as
$s^2\log p_{\bgamma} = o(\sqrt{p_{\bbeta}}).$ Furthermore, if $p_{\bbeta}= cn^2$ with $c>0$, condition \eqref{condition1inguosum} can be further simplified as $s^2\log p_{\bgamma} = o(n)$, which is weaker than the conditions on sparsity and dimension in previous works (for example $s = o(\sqrt{n}/(\log p)^{3/2})$  in \cite{zhang2017simultaneous}; $s = o(\sqrt{n}/(\log p)^{3})$ in \cite{ma2021global}). With larger $p_{\bbeta}$, the condition $s^2\log p_{\bgamma} = o(\sqrt{p_{\bbeta}})$ is milder. This implies that $T_n$ has tractable null distribution even when both $p_{\bbeta}$ and $p_{\bgamma}$ are of exponential order of $n$.

To formulate the testing procedure based on Theorem \ref{limitnulldisofguosum}, we use
\[R_{1n}= \hat{\sigma}^{4}\frac{1}{2\tbinom{n}{4}}\sum_{i< j<k<l}^{n}(\bX_{i} - \bX_{j})^\top(\bX_{k} - \bX_{l})(\bX_{j} - \bX_{k})^\top(\bX_{l} - \bX_{i})\]
to estimate $\bLambda_{\bX}$, where $\hat{\sigma}^{2}$
is a consistent estimator of the error variance $\sigma^2$, such as the one in \cite{sun2012scaled}.
Under the null hypothesis, $R_{1n}$ is a ratio consistent estimator of $\bLambda_{\bX}$. The consistency of $R_{1n}$ has been discussed by many authors   such as \cite{zhong2011tests,cui2018test,guo2022conditional}.
Combining Theorem \ref{limitnulldisofguosum} and Slutsky Theorem, we reject $\bH_{0}$ at a significance level $\alpha$ if
\[
T_n\geq z_{\alpha}\sqrt{2R_{1n}}.
\]
Here $z_{\alpha}$ is the upper-$\alpha$ quantile of standard normal distribution.

\subsection{Power analysis}
\label{poweranalysisofsec2}

Next, we study the limiting distribution of $T_n$ under a class of alternative hypotheses. Let $\mbE(\bX\bZ^\top)=:\bSigma_{\bX\bZ}=\bSigma_{\bZ\bX}^\top$ be the covariance matrix between $\bX$ and $\bZ$ and $\bet = \bX - \bW^\top\bZ$, where
\begin{align}
	\label{bW}
	\bW = \bSigma_{\bZ}^{-1}\bSigma_{\bZ\bX}.
\end{align}
Let $\bSigma_{\bet}$ be the covariance matrix of $\bet$.

Define the following family of local alternatives:
\[
\mathscr{L}_1(\bbeta) = \biggl\{\bbeta\in\mbR^{p_{\bbeta}}\bigg\vert \bbeta^\top\bSigma_{\bet}\bbeta = o(1),\, \bbeta^\top\bSigma_{\bet}^{2}\bbeta = o\biggl(\frac{\bLambda_{\bX}}{n\varpi^2}\biggr) \,\,\text{and}\,\,\bbeta^\top\bSigma_{\bet}\bSigma_{\bX}\bSigma_{\bet}\bbeta = o\biggl(\frac{\bLambda_{\bX}}{n}\biggr)\biggr\}.
\]
Similar definitions of local alternative sets have been also considered in \cite{zhong2011tests,cui2018test,guo2022conditional}. The class $\mathscr{L}_{1}(\bbeta)$ prescribes a small difference between $\bbeta$ and $\bzero$. For illustration, suppose the eigenvalues of $\bSigma$ are bounded by a constant, then $\mathscr{L}_{1}(\bbeta)$ can be simplified as
\[
    \mathscr{L}_{1}(\bbeta) = \biggl\{
    \bbeta\in\mbR^{p_{\bbeta}}\bigg\vert \lVert\bbeta\rVert_{2} = o\biggl(\min\biggl(1,\sqrt{\frac{p_{\bbeta}}{ns^{2}\log p_{\bgamma}}}\biggr)\biggr)
    \biggr\}.
\]
We have the following theorem.
\begin{theorem}
	\label{limitpowerofguosum}
	Under the conditions in Theorem \ref{limitnulldisofguosum},  for $\bbeta\in\mathscr{L}_1(\bbeta)$, we  have
	\[
	\frac{T_n - n\bbeta^\top\bSigma_{\bet}^{2}\bbeta}{\sqrt{2\bLambda_{\bX}}}\rightarrow N(0,1)\quad \text{in distribution}
	\]
	as $(n,p_{\bbeta},p_{\bgamma})\rightarrow\infty$.
\end{theorem}

Theorem \ref{limitpowerofguosum} indicates that the asymptotic power of the test statistic $T_{n}$ under the local alternatives $\mathscr{L}_1(\bbeta)$ is given by
\begin{align}
    \label{powerfunctionofguosum}
    \phi_{1n} = \Phi\biggl(-z_{\alpha} + \frac{n\bbeta^\top\bSigma_{\bet}^{2}\bbeta}{\sqrt{2\bLambda_{\bX}}}\biggr),
\end{align}
where $\Phi(\cdot)$ denotes the standard normal cumulative distribution function.
Equation \eqref{powerfunctionofguosum} implies that the proposed test has nontrivial power as long as the signal-to-noise ratio $n\bbeta^\top\bSigma_{\bet}^{2}\bbeta/\sqrt{2\bLambda_{\bX}}$ does not vanish to 0 as $(n,p_{\bbeta})\rightarrow\infty$.

It is noteworthy that the power performance of $T_{n}$ relies not only on the difference between $\bbeta$ and $\bzero$ but also on $\bSigma_{\bet}$. $T_{n}$ gains more power when the difference between $\bbeta$ and $\bzero$ is larger. Additionally, $T_{n}$ gains more power when the eigenvalues of $\bSigma_{\bet}$ are larger. Further recall that $\bLambda_{\bX} = \sigma^4\mathrm{tr}(\bSigma^2_{\bX})$. Thus when the variance of the error term is small, the power of $T_n$ can also be large. 

Compared with \cite{chen2022testing}, we now establish the asymptotic distribution of $T_n$ under local alternative hypotheses. Both the  parameter vector of interest and nuisance parameter vector are allowed to be ultrahigh-dimensional.

\subsection{The failure of $T_n$}
\label{sec2.3}

As discussed, the asymptotic distribution of $T_n$ can be established when the correlation between $\bX$ and $\bZ$ is weak and $p_{\bbeta}$ is relatively large compared with $p_{\bgamma}$. However, in highly correlated cases, condition \eqref{condition1inguosum} fails as $\varrho^2$ can be divergent in this situation.
This may cause the failure of the asymptotic normality.
Practically, the situation that $\bX$ and $\bZ$ are correlated can be motivated by the study where $\bZ$ can represent some confounding variables, which can affect many  variables in $\bX$.
For instance, in genetic studies, the effect of certain segments of the deoxyribonucleic acid on the gene expression may be confounded by population structure and microarray expression artifacts \citep{listgarten2010correction}. The confounding biases can yield an inflated distribution of test statistics in genome-wide association studies \citep{bulik2015ld}. 
To illustrate this problem intuitively, consider the following toy example.
\begin{example}
	\label{example2}
	Generate the covariates according to the following model:
	\begin{align}
		\label{eqofexample2}
		\bX = \bW^\top\bZ + \bet,
	\end{align}
	where $\bet$ is a $p_{\bbeta}$-dimensional random vector and $\bet$ is independent of $\bZ$.  $\bW$ is a $p_{\bgamma}\times p_{\bbeta}$ dimensional matrix with non-zero corners
	\begin{align}
		\label{bWexample}
		\bW^\top = \begin{pmatrix}
			\bm 0 & \bm 0 & \bm 0\\
			\bQ^\top & \bm 0 & \bQ^\top
		\end{pmatrix},
	\end{align}
	where $\bQ$ is a $d_{\bZ}\times 2d_{\bX}$ dimensional matrix and all the elements of $\bQ$ equal $0.5$. In this example, we set $\bbeta = \bzero$ and $\bgamma\neq\bzero$. The null hypothesis $\bH_{0}$ then holds. We derive $\varrho^2$ increases with the increase of $d_{\bX}$
	by some calculation.
	Let $d_{\bZ} = 3$ and vary $d_{\bX}$ from $0$ to $15$.
\end{example}
Left panel of Figure \ref{figureofexample2} plots the empirical sizes of $T_n$. As $d_{\bX}$ increases, the empirical size rises rapidly, and the significance level cannot be maintained. Right panel of Figure \ref{figureofexample2} reports the empirical probability density function of $T_n$, which can be well approximated by the standard normal distribution when $d_{\bX}=0$. However, the deviation gradually increases with the increase of $d_{\bX}$. This toy example illustrates that when $\bX$ and $\bZ$ are not weakly dependent, $T_n$ is not applicable.
\begin{figure}[ht]
	\centering
	\includegraphics[width=\textwidth]{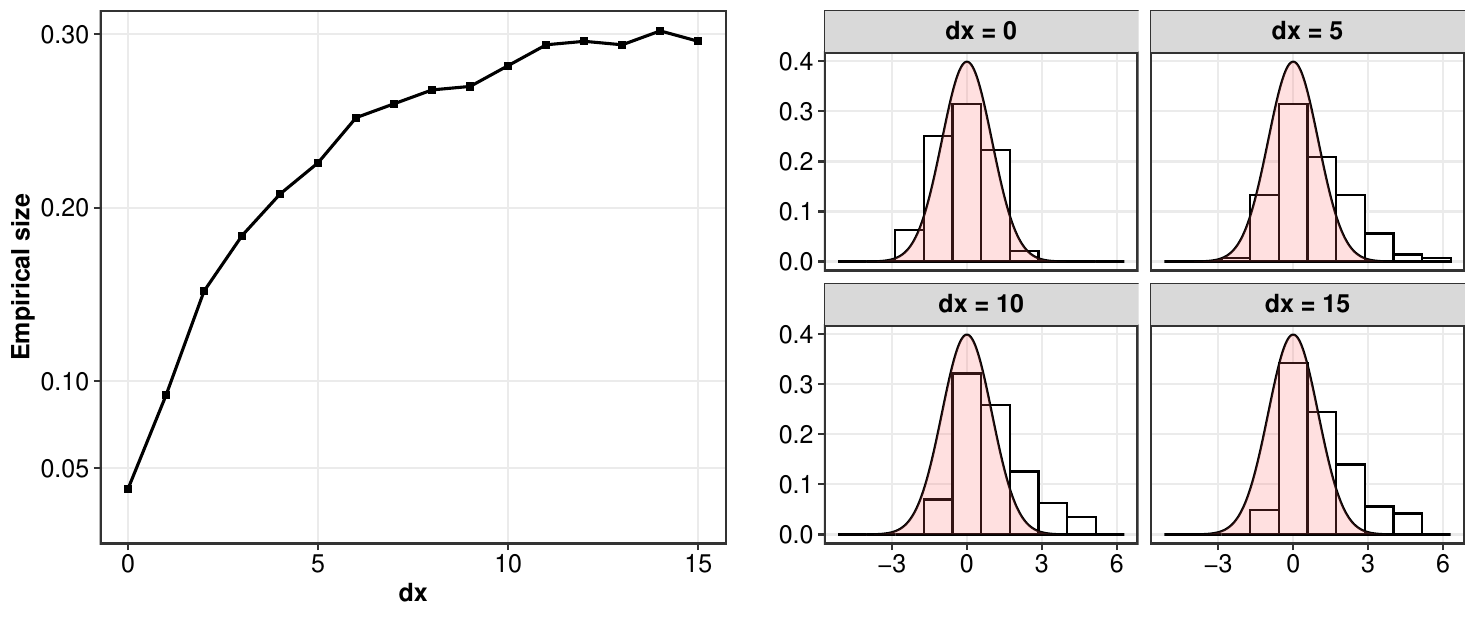}
	\caption{Left panel presents the empirical sizes of $T_n$ with different $d_{\bX}$.
		Right panel presents the empirical probability density function of $T_n$ with different $d_{\bX}$ ($d_{\bX}$ = 0,5,10,15). The pink shade represents the probability density function of the standard normal distribution. Set $n=100$, $p = 600$ and $p_{\bbeta} = p_{\bgamma}$. We generate 500 replications and reject the null hypothesis at the significance level $\alpha = 0.05$. More details can be seen in Scenario 3 in Section \ref{sec6}.}
	\label{figureofexample2}
\end{figure}

In the proof of Theorem \ref{limitnulldisofguosum}, the following error term depends on the correlation between $\bX$ and $\bZ$:
\begin{align}
	\label{error1}
	\operatorname{ERROR}_1= (\bgamma - \hat\bgamma)^\top\frac{1}{n}\sum_{i\neq j}\bZ_i\bX_i^\top\bX_j\bZ_j^\top(\bgamma - \hat\bgamma),
\end{align}
where $n^{-1}\sum_{i\neq j}\bZ_i\bX_i^\top\bX_j\bZ_j^\top$ is a $p_{\bgamma}\times p_{\bgamma}$ dimensional random matrix, and the $(k,l)$-th element is a non-degenerate U-statistic with non-zero mean $\mbE(Z_{k}\bX)^\top\mbE(Z_{l}\bX)$.
The order of $\operatorname{ERROR}_1$ is $O_p(s^2\log p_{\bgamma}\varrho^2)$. Thus condition \eqref{condition1inguosum} is required to reduce the impact of bias term $\operatorname{ERROR}_1$ on the asymptotic behavior of $T_n$.
Clearly the bias is no longer negligible if $s^2\log p_{\bgamma}\varrho^2\geq C\sqrt{\bLambda_{\bX}}$ for some constant $C$.
In the following, we suggest an orthogonalization-based test to reduce the bias term.

\section{An orthogonal score function-based test}
\label{sec3}

As discussed in subsection \ref{sec2.3}, the estimation error of $\hat\bgamma$ may make the test statistic $T_n$ fail to work when the correlation of $\bX$ and $\bZ$ is high. To make the score function immune to the estimation error of $\hat\bgamma$, we consider orthogonalizing the score function of $\bbeta$,
\begin{align*}
	\mathcal{S}(\bbeta,\bgamma) = \nabla_{\bbeta}\mathcal{L} - \bW^\top\nabla_{\bgamma}\mathcal{L}\quad \mathrm{with}\quad \bW^\top = \nabla_{\bbeta\bgamma}\calL(\nabla_{\bgamma\bgamma}\calL)^{-1}.
\end{align*}
Here $\mathcal{L}=\mathcal{L}(\bbeta,\bgamma) = \mbE (Y-\bbeta^\top\bX - \bgamma^\top\bZ)^2/2,
\nabla_{\bbeta}\mathcal{L}=\partial\mathcal{L}/\partial\bbeta, \nabla_{\bbeta\bgamma}\calL=\partial\mathcal{L}^2/\partial\bbeta\partial\bgamma$. $\nabla_{\bgamma}\mathcal{L}$ and $\nabla_{\bgamma\bgamma}\calL$ are similarly defined.

The main idea of orthogonalization is to construct a statistic for the target parameter, which is locally insensitive to the nuisance parameter. The orthogonalization plays an important role in high-dimensional inference problems and has been successfully applied in the recent literature. See for example, \cite{belloni2015uniform}, \cite{ning2017general} and \cite{belloni2018uniformly}. However, the current adoption of orthogonalization only focuses on low-dimensional parameters. Actually the coordinate-based maximum tests firstly consider orthogonalization for each element of testing parameter vector and then take the maximum of all individual test statistics for each elements. To the best of our knowledge, orthogonalization has not been investigated for test statistics based on the quadratic norm of the score function for ultrahigh-dimensional testing parameter vector.

In our model setting, $\mathcal{S}(\bbeta,\bgamma)$ is equal to
\begin{align*}
	\mathcal{S}(\bbeta,\bgamma) = -\mbE\{(Y - \bbeta^\top\bX - \bgamma^\top\bZ)(\bX - \bW^\top\bZ)\}.
\end{align*}
Again $\bH_0$ corresponds to $\mathcal{S}(\bzero,\bgamma) = \bzero$. Compared with $\mathcal{L}(\bbeta,\bgamma)$, we replace $\bX$ with $\bX - \bW^\top\bZ$ now. We can then construct a test statistic based on $\mathcal{S}(\bzero,\bgamma)^\top\mathcal{S}(\bzero,\bgamma)$.
For the sample $\{\bX_{i},\bZ_{i},Y_{i}\}_{i=1}^{n}$,
define:
\begin{align}
	\label{teststatisticstar1}
	M_{n}^{*} = \frac{1}{n}\sum_{i\neq j}(Y_i - \bgamma^\top\bZ_i)(Y_j - \bgamma^\top\bZ_j)(\bX_i - \bW^\top\bZ_i)^\top(\bX_j - \bW^\top\bZ_j).
\end{align}
We construct a final test statistic in Sections \ref{sec4} and \ref{sec5}.

\section{The oracle inference}
\label{sec4}

To illustrate the merits of the orthogonalization technique, we first consider the case with given $\bW$. Recall $\bW$ is defined in \eqref{bW} and $\bW = \bSigma_{\bZ}^{-1}\bSigma_{\bZ\bX}$. A sufficient condition for known $\bW$ is that the joint distribution of $(\bX^\top,\bZ^\top)^\top$ is known in advance. This assumption is given in the recent literature on high-dimensional statistics, such as the model-X knockoff procedure in \cite{candes2018panning}.
Based on the term in \eqref{teststatisticstar1},  consider the following test statistic:
\begin{align}
	\label{teststatisticoracle1}
	M_{n}^{\text{o}} = \frac{1}{n}\sum_{i\neq j}(Y_i - \hat\bgamma^\top\bZ_i)(Y_j - \hat\bgamma^\top\bZ_j)(\bX_i - \bW^\top\bZ_i)^\top(\bX_j - \bW^\top\bZ_j).
\end{align}
We obtain the estimator $\hat\bgamma$ by solving a penalized least squares problem in \eqref{bgammaestimator}.

\subsection{Limiting null distribution}

Recall $\bet$ is defined in subsection \ref{poweranalysisofsec2} and $\bet = \bX - \bW^\top\bZ$.
$\bSigma_{\bet}$ is the covariance matrix of $\bet$.
Give the following assumption.

\begin{assumption}
	\label{assumptionbmatrix}
	$\tr(\bSigma^4_{\bet})=o(\tr^2(\bSigma_{\bet}^2))$ and $\tr(\bSigma_{\bet}^2)\rightarrow\infty$ as $(n,p_{\bbeta})\rightarrow\infty$.
\end{assumption}

Assumption \ref{assumptionbmatrix} is a counterpart of Assumption \ref{assumptionbmatrix2} in the case of the orthogonal score function.
Theorem \ref{limitnulldisoforaclesum} states the limiting null distribution of the oracle test statistic $M_{n}^{\text{o}}$ in \eqref{teststatisticoracle1}.

\begin{theorem}
	\label{limitnulldisoforaclesum}
	Under $\bH_0$, and Assumptions~\ref{assumptionb5}-\ref{assumptionb7}, \ref{assumptionbmatrix} and condition \eqref{condition2inguosum},
	we have
	\[
	\frac{M_{n}^{o}}{\sqrt{2\bLambda_{\bet} }}\rightarrow N(0,1)\quad \text{in distribution}
	\]
	as $(n,p_{\bbeta},p_{\bgamma})\rightarrow\infty$, where $\bLambda_{\bet} =\sigma^4 \mathrm{tr}(\bSigma_{\bet}^2)$.
\end{theorem}
Notably, compared with Theorem \ref{limitnulldisofguosum}, condition \eqref{condition1inguosum} is removed in Theorem \ref{limitnulldisoforaclesum}. Except for  Assumption \ref{assumptionbmatrix}, there are no additional restrictions on the relationship between the covariates in Theorem \ref{limitnulldisoforaclesum}.  The dependence requirement is greatly relaxed. The proof for this theorem is similar to that for Theorem \ref{limitnulldisofguosum}, but the error term becomes
\[
\operatorname{ERROR}^{\text{o}} = (\bgamma - \hat{\bgamma})^\top\frac{1}{n}\sum_{i\neq j}\bZ_i\bet_{i}^\top\bet_{j}\bZ_j^\top(\bgamma - \hat\bgamma),
\]
where $\bet_{i} = \bX_i - \bW^\top\bZ_i$ and $n^{-1}\sum_{i\neq j}\bZ_i\bet_{i}^\top\bet_{j}\bZ_j^\top$ is a $p_{\bgamma}\times p_{\bgamma}$ dimensional matrix with zero-mean degenerate $U$-statistic components. Benefitting from the zero-mean property of $n^{-1}\sum_{i\neq j}\bZ_i\bet_{i}^\top\bet_{j}\bZ_j^\top$, the order of the bias term $\operatorname{ERROR}^{\text{o}}$ is much smaller than that of $\operatorname{ERROR}_1$. From the theory of $U$-statistics (see, e.g., \cite{serfling1980approximation}), the order of a typical degenerate $U$ statistic is $O_p(n^{-1})$. Thus,   by the property of degenerate $U$ statistic, $n^{-1}\sum_{i\neq j}\bZ_i\bet_{i}^\top\bet_{j}\bZ_j^\top$ is much easier to handle than $n^{-1}\sum_{i\neq j}\bZ_i\bX_{i}^\top\bX_{j}\bZ_j^\top$. In the proof, we show that $\operatorname{ERROR}^{\text{o}} = o_p(\sqrt{2\bLambda_{\bet}})$. In summary, the orthogonalization technique has two merits: debiasing and converting a non-degenerate $U$-statistic to a degenerate one. The first is frequently observed in the literature, such as \cite{zhu2006empirical}, \cite{zhang2014confidence}, and \cite{van2014asymptotically}.
We have not seen in the literature any study to show the second merit of the orthogonalization technique.

\subsection{Power analysis}
\label{poweranalysisofsec4}

To study the power performance of $M_{n}^{\text{o}}$, consider the following local alternatives:
\begin{align*}
	\mathscr{L}^{\text{o}}(\bbeta) = \biggl\{\bbeta\in\mbR^{p_{\bbeta}}\bigg\vert \bbeta^\top\bSigma_{\bet}\bbeta = o(1)\,\,\text{and}\,\,\bbeta^\top\bSigma_{\bet}^{3}\bbeta = o\biggl(\frac{\Lambda_{\bet}}{n}\biggr)\biggr\}.
\end{align*}

Compared with previous literature \citep{zhong2011tests,cui2018test}, we replace $\bSigma_{\bX},\,\bLambda_{\bX}$ with $\bSigma_{\bet},\,\bLambda_{\bet}$ in $\mathscr{L}^{o}(\bbeta)$. This modification is due to the construction of $M_{n}$. While previous test statistics are constructed based on $\bX_{i}^\top\bX_{j}$, the construction of $M_{n}$ is based on $\bet_{i}^\top\bet_{j}$. Similar to $\mathscr{L}_{1}(\bbeta)$, the class $\mathscr{L}^{o}(\bbeta)$ also describes a small discrepancy between $\bbeta$ and $\bzero$. Under the bounded eigenvalue assumption of $\bSigma_{\bet}$, $\mathscr{L}^{o}(\bbeta)$ can be simplified as 
\begin{align*}
    \mathscr{L}^{o}(\bbeta) = \biggl\{\bbeta\in\mbR^{p_{\bbeta}}\bigg\vert \lVert\bbeta\rVert_{2} = o\biggl(\min\biggl(1,\sqrt{\frac{p_{\bbeta}}{n}}\biggr)\biggr)\biggr\}.
\end{align*}
We have the following theorem.
\begin{theorem}
	\label{limitpoweroforaclesum}
	Under conditions in Theorem \ref{limitnulldisoforaclesum}, and for $\bbeta\in\mathscr{L}^{\text{o}}(\bbeta)$, we derive
	\[
	\frac{M_{n}^{\text{o}} - n\bbeta^\top\bSigma_{\bet}^{2}\bbeta}{\sqrt{2\bLambda_{\bet} }}\rightarrow N(0,1)\quad \text{in distribution}
	\]
	as $(n,p_{\bbeta},p_{\bgamma})\rightarrow\infty$.
\end{theorem}

Similarly,  the asymptotic power function  of $M_{n}^{\text{o}}$  under the local alternatives is
\begin{align}
	\label{powerfunctionoforaclesum}
	\phi^{\text{o}}_{n} = \Phi\biggl(-z_{\alpha} + \frac{n\bbeta^\top\bSigma_{\bet}^{2}\bbeta}{\sqrt{2\bLambda_{\bet}}}\biggr).
\end{align}
Equation \eqref{powerfunctionoforaclesum} implies that the proposed test has nontrivial power as long as the signal-to-noise ratio $n\bbeta^\top\bSigma_{\bet}^{2}\bbeta/\sqrt{2\bLambda_{\bet}}$ does not vanish to 0 as $(n,p_{\bbeta})\rightarrow\infty$.

Comparing with Theorem \ref{limitpowerofguosum},  the Pitman asymptotic relative efficiency (ARE) of $M_{n}^{\text{o}}$ concerning the $T_n$ is
\[
\operatorname{ARE}(M_{n}^{\text{o}},T_n) =
\biggl\{\frac{\tr(\bSigma_{\bX}^2)}{\tr(\bSigma_{\bet}^2)}\biggr\}^{1/2}.
\]
By the definition of $\bet$, $\tr(\bSigma_{\bX}^{2})\geq \tr(\bSigma_{\bet}^{2})$. Thus the power of $M_{n}^{o}$ is higher than $T_n$.
This result is important as the orthogonalization technique can simultaneously reduce bias and variance, thus improving power performance.

\section{The test with unknown \texorpdfstring{$\bW$}{\textbf{W}}}
\label{sec5}

In this case, the test statistic is defined by using a plug-in estimator of $\bW$:
\begin{align}
	\label{teststatistic1}
	M_{n} = \frac{1}{n}\sum_{i\neq j}(Y_i - \hat\bgamma^\top\bZ_i)(Y_j - \hat\bgamma^\top\bZ_j)(\bX_i - \hat\bW^\top\bZ_i)^\top(\bX_j - \hat\bW^\top\bZ_j).
\end{align}
We obtain the estimator $\hat\bgamma$ by solving a penalized least squares problem in \eqref{bgammaestimator}.
Suppose we use data $\{\bX_{i},\bZ_{i}\}_{i=1}^{n^\prime}$ to estimate $\bW$. The $k$-th column of $\bW$ can be estimated by
\begin{align}
    \label{bWestiamtor}\tag{16}
    \hat\bW_k = \mathop{\argmin}\limits_{\bW_k\in\mbR^{p_{\bgamma}}}\frac{1}{2n^\prime}\sum_{i=1}^{n^\prime}(X_{ik} - \bW_k^\top\bZ_i)^2 + \lambda_{X_k}\lVert\bW_k\rVert_{1},
\end{align}
where $X_{ik}$ is the $k$-th component of $\bX_i$, and $\lambda_{X_k}$ is the tuning parameter.

\subsection{Limiting null distribution}

We let $\lVert\bW\rVert_{0}\leq s^{\prime}$, where $s^{\prime}$ is a positive integer and represents the sparsity level of $\bW$.
Let $\varphi^2 = \max_{1\leq k\leq p_{\bgamma}}\lVert\mbE(Z_{k}\bZ)\rVert_{2}^{2}$ and $\vartheta^2 = s^2s^\prime(\log p_{\bgamma})^{2}\varphi^2/n^{\prime}$.
Next, under the following assumption, we study the asymptotic null distribution of the test statistic $M_{n}$.

\begin{assumption}
	\label{assumption8}
	The estimator $\hat\bW$ is independent of the data $\{\bX_i,\bZ_i,Y_i\}_{i=1}^{n}$, and
	$\lVert\hat\bW - \bW\rVert_{F} = O_p(\sqrt{s^{\prime}\log p_{\bgamma}/n^\prime})$
	for some positive integer $n^\prime$.
\end{assumption}
Assumption \ref{assumption8} requires the Frobenius norm bound of $\hat{\bW}$ in the order of $(s^\prime\log p_{\bgamma}/n^\prime)^{1/2}$, which can be satisfied by most of existing high-dimensional estimators. See section 9.7 in \cite{highprobability2019} for instance. Besides, $n^\prime$ represents the sample size used to estimate $\bW$. We can estimate $\bW$ using additional data of $\bX$ and $\bZ$ if we have, otherwise, applying the sample-splitting approach. See, for instance, \cite{belloni2012sparse} and \cite{chernozhukov2018double}.
We also note that the independence  between $\hat\bW$ and $\{\bX_i,\bZ_i,Y_i\}_{i=1}^{n}$  can make the asymptotic analysis more easily. However,  { the simulation study shows that the proposed test statistic $M_n$ still works well
	numerically even when we estimate $\bW$ based on the same data set. Therefore, we guess this independence might not be necessary, although we have not yet got rid of it in the technical deduction.}
\begin{theorem}
	\label{limitnulldisofsum}
	Under $\bH_0$,  Assumptions
	\ref{assumptionb5}-\ref{assumptionb7},
	\ref{assumptionbmatrix}, and \ref{assumption8}, condition \eqref{condition2inguosum}, and
	\begin{align}
		\label{condition1insum}
		\vartheta^2 \vee (\log p_{\bgamma})^{1/2}\vartheta\biggl(\lambda_{\rm{max}}(\bSigma_{\bet}) + \lambda_{\rm{max}}(\bSigma_{\bZ})\frac{s^\prime\log p_{\bgamma}}{n^\prime}\biggr)^{1/2}\vee \lambda_{\rm{max}}(\bSigma_{\bZ})\frac{s^\prime\log p_{\bgamma}}{n^\prime} = o(\sqrt{\bLambda_{\bet}}),
	\end{align}
	we have
	\[
	\frac{M_{n}}{\sqrt{2\bLambda_{\bet} }}\rightarrow N(0,1)\quad \text{in distribution}
	\]
	as $(n,p_{\bbeta},p_{\bgamma})\rightarrow\infty$, where $\bLambda_{\bet}$ is defined in Theorem \ref{limitnulldisoforaclesum}.
\end{theorem}
Compared with Theorem \ref{limitnulldisofguosum}, condition \eqref{condition1inguosum} is replaced by condition \eqref{condition1insum} that handles the impact caused by the estimation error of $\hat\bW$.
As discussed, the error term caused by the correlation between $\bX$ and $\bZ$ is reduced by the orthogonalization technique, but the estimation error is brought by  $\hat\bW$.
Now we compare condition \eqref{condition1insum} with \eqref{condition1inguosum}. By the formula of these conditions, it suffices to compare $\vartheta^2$ with $\varpi^2$. Note that
\begin{align}
	\label{ratiobetweenvarthetaandvarrho}
	\frac{\vartheta^2}{\varpi^2} = \frac{\varphi^2}{\varrho^2}\frac{s^\prime\log p_{\bgamma}}{n^{\prime}}.
\end{align}
If ratio \eqref{ratiobetweenvarthetaandvarrho} is small, condition \eqref{condition1insum} can be weaker than \eqref{condition1inguosum}.
When the relationship between nuisance covariates is weak while the relationship between covariates of interest and nuisance covariates is high, the ratio \eqref{ratiobetweenvarthetaandvarrho} can be small. When the matrix $\bW$ is sparse, or $n^\prime$ is large, the ratio \eqref{ratiobetweenvarthetaandvarrho} can also be small.

Similar to the test construction in Section \ref{limitnulldisofsec2}, we estimate $\bLambda_{\bet}$ by
\[
R_{n} = \hat{\sigma}^{4}\frac{1}{2\tbinom{n}{4}}\sum_{i< j<k<l}^{n}(\hat{\bet}_{i} - \hat{\bet}_{j})^\top(\hat{\bet}_{k} - \hat{\bet}_{l})(\hat{\bet}_{j} - \hat{\bet}_{k})^\top(\hat{\bet}_{l} - \hat{\bet}_{i}),
\]
where $\hat\bet_{i} = \bX_{i} - \hat{\bW}^\top\bZ_{i}$. Under the null hypothesis and conditions in Theorem \ref{limitnulldisofsum}, $R_{n}$ is a ratio consistent estimator of $\Lambda_{\bet}$; see the details in Supplementary Material. We reject $\bH_{0}$ at the significance level $\alpha$ if
\[
M_{n}\geq z_{\alpha}\sqrt{2R_{n}}.
\]
Here $z_{\alpha}$ is the upper-$\alpha$ quantile of standard normal distribution.

\subsection{Power analysis}

Consider the class of local alternatives as follows:
\[
\mathscr{L}(\bbeta) = \biggl\{\bbeta\in\mbR^{p_{\bbeta}}\bigg\vert\bbeta\in\mathscr{L}^{\text{o}}(\bbeta)\quad\text{and}\quad \bbeta^\top\bSigma_{\bet}^{2}\bbeta = o\biggl(\frac{\Lambda_{\bet}}{n(\vartheta^2\vee \lambda_{\max}(\bSigma_{\bZ})s^\prime\log p_{\bgamma}/n^\prime)}\biggr)\biggr\}.
\]
Compared to $\mathscr{L}^{o}(\bbeta)$, class $\mathscr{L}(\bbeta)$ has more restrictions on $\bbeta$. This is because we have to handle the extra error caused by the estimation of $\bW$. Similar to $\mathscr{L}^{o}(\bbeta)$, the class $\mathscr{L}(\bbeta)$ also illustrates a small dissimilarity between $\bbeta$ and $\bzero$. Under the bounded eigenvalue assumptions of $\bSigma_{\bet}$ and $\bSigma_{\bZ}$, $\mathscr{L}(\bbeta)$ can be simplified as 
\begin{align*}
    \mathscr{L}(\bbeta) = \biggl\{\bbeta\in\mbR^{p_{\bbeta}}\bigg\vert \lVert\bbeta\rVert_{2} = o\biggl(\min\biggl(1,\sqrt{\frac{p_{\bbeta}}{n(1\vee s^2s^\prime (\log p_{\bgamma})^{2}/n^\prime)}}\biggr)\biggr)\biggr\}.
\end{align*}
Then the power performance is stated in the following theorem.
\begin{theorem}
	\label{limitpoweroftruesum}
	Assume conditions in Theorem \ref{limitnulldisofsum}, and for $\bbeta\in\mathscr{L}(\bbeta)$, we derive
	\[
	\frac{M_{n} - n\bbeta^\top\bSigma_{\bet}^{2}\bbeta}{\sqrt{2\bLambda_{\bet} }}\rightarrow N(0,1)\quad \text{in distribution}
	\]
	as $(n,p_{\bbeta},p_{\bgamma})\rightarrow\infty$.
\end{theorem}

According to Theorem \ref{limitpoweroftruesum}, $M_{n}$ have the same asymptotic power function with $M_{n}^{o}$ when $\bbeta\in\mathscr{L}(\bbeta)$. Thus under designed conditions, the test $M_{n}$ has the same power performance as the oracle test $M_{n}^o$ with a known $\bW$ asymptotically and is more powerful than $T_n$.

\section{Numerical studies}
\label{sec6}

\subsection{Simulations}
\label{sec6.1}

We first compare the performances among four tests: (1). the score function-based test  $T_n$
in Section \ref{sec2}; (2). the orthogonalied score function-based  test $M_n$  in Section \ref{sec5}; (3). the studentized bootstrap-assisted test $ST$ in \cite{zhang2017simultaneous}; and
(4). the Wald-type test $WALD$ 
in \cite{guo2021group}.

Generate data from the following ultrahigh-dimensional linear model:
\begin{align*}
	Y_{i} &= \bbeta^\top\bX_{i} + \bgamma^\top\bZ_{i} + \epsilon_{i},\quad i=1,\ldots,n,
\end{align*}
where the covariates $\bV_i=(\bX_i^\top,\bZ_i^\top)^\top$ are generated from the multivariate normal distribution. The details are given later, and the regression error $\epsilon_i \sim N(0,1)$ independent of $\bV_i$. Denote $s_{\bbeta}$ and $s_{\bgamma}$ as the sparsity levels of $\bbeta$ and $\bgamma$ respectively. The regression coefficients $\bbeta$ are set as: $\beta_{j}=b_0$ for $1 \leq j \leq s_{\bbeta}$ and $\beta_{j}=0$ otherwise. Similarly let $\gamma_{j}=g_0$ for $1 \leq j \leq s_{\bgamma}$ and $\gamma_{j}=0$ otherwise. Throughout the simulation study,  let $s_{\bgamma} = \lfloor 5\% p_{\bgamma} \rfloor$ and $g_0 = 0.5$.
There are three settings for the values of $\bbeta$:
\begin{enumerate}[itemsep=0pt, topsep=0pt, leftmargin=2.2cm, label={\bf Setting \arabic*}:]
	\item Consider $b_0 = 0$ to assess the empirical Type-I error.
	\item Let $s_{\bbeta} = \lfloor 5\% p_{\bbeta} \rfloor$ and $b_0\neq 0$ to assess the empirical power with sparse alternative.
	\item Let $s_{\bbeta} = \lfloor 50\%p_{\bbeta} \rfloor$  and $b_0\neq 0$ to assess the empirical power with dense alternative.
\end{enumerate}
The experiment is repeated $500$ times for each simulation setting to assess the empirical type-I error and power at the significance level $\alpha = 0.05$. 
The tuning parameter $\lambda_Y$ in \eqref{bgammaestimator} and $\lambda_{X_k}$ in \eqref{bWestiamtor} are selected by 10-fold cross-validations using the R-package \texttt{glmnet} \cite{glmnet2010}. Based on these settings, we consider the following two scenarios.

{\bf{Scenario 1.}} We aim to compare our tests with other testing methods in this scenario. The covariates are generated from the multivariate normal distribution $N_p(\bm{0}_p, \bSigma)$. Here $\bSigma = (\sigma_{ij})_{p\times p}$ follows the Toeplitz design, that is, $\sigma_{ij} = 0.5^{|i-j|}, i,j = 1, \ldots, p$. The sample size $n = 100$, the covariate dimension $p = 600$ and $p_{\bbeta} = p_{\bgamma}=300$. In the sparse alternative (setting 2) and the dense alternative (setting 3), we vary $b_0$ from $0$ to $\sqrt{\lVert\bgamma\rVert_{2}^{2}/s_{\bbeta}}$.


{\bf{Scenario 2.}} This scenario investigates the performance of our tests when $\bX$ and $\bZ$ are highly correlated.
The covariates are generated according to the following model:
\begin{align}
	\label{eq1ofexample2}
	\bX &= \bW^\top\bZ + \bet,
\end{align}
where $\bet$ is a $p_{\bbeta}$-dimensional random vector and $\bet$ is independent of $\bZ$. $\bet\sim N_{p_{\bbeta}}(\bzero_{p_{\bbeta}},\bSigma_{\bet})$ and $\bZ\sim N_{p_{\bgamma}}(\bzero_{p_{\bgamma}},\bSigma_{\bZ})$, where $\bSigma_{\bet}$ and $\bSigma_{\bZ}$ follow the Toeplitz design with $\rho=0.5$ respectively. $\bW$ is defined in \eqref{bWexample} in subsection \ref{sec2.3}.
Throughout the scenario, $d_{\bZ} = 3$ and $d_{\bX} = 10$.
The sample size $n = 100$, the predictor dimension $p = 600$ and $p_{\bbeta} = p_{\bgamma} = 300$. In the sparse alternative (setting 2) and the dense alternative (setting 3), we vary $b_0$ from $0$ to $\sqrt{\lVert\bgamma\rVert_{2}^{2}/s_{\bbeta}}$.

Figure \ref{figure2} displays the empirical size-power curves of the four tests in scenario 1.
It can be observed that $T_{n}$, $M_{n}$ and $ST$ tests control the size well.  $T_n$ and $M_n$ are generally more powerful than $ST$ under  the sparse and dense alternative hypotheses. Under the dense alternative, the empirical powers of $ST$ can be as low as the significance level. The empirical powers of $T_n$ and $M_n$ increase quickly as the signal strength $b_0$ becomes stronger.
{On the other hand, the $WALD$ test is very liberal to have very large empirical size when we use the tuning parameter $\tau =1$ recommended by \cite{guo2021group}. While the numerical studies in \cite{guo2021group} suggest that the $WALD$ test with $\tau=1$ can be very conservative in their setting. We have  also conducted different settings with different dimensions and sample sizes, and found that with different values $\tau= 0, 0.5, 1.0, 1.5, 2.0, 2.5, 3.0, 3.5$ the test can be either very liberal or very conservative. Thus selecting a proper tuning parameter is difficult in general.}
It is worth noticing that $T_n$ and $M_n$ have similar performances in this scenario. $T_n$ performs well enough when the correlation between $\bX$ and $\bZ$ is relatively weak.

\begin{figure}[ht]
	\centering
	\includegraphics[width=\textwidth]{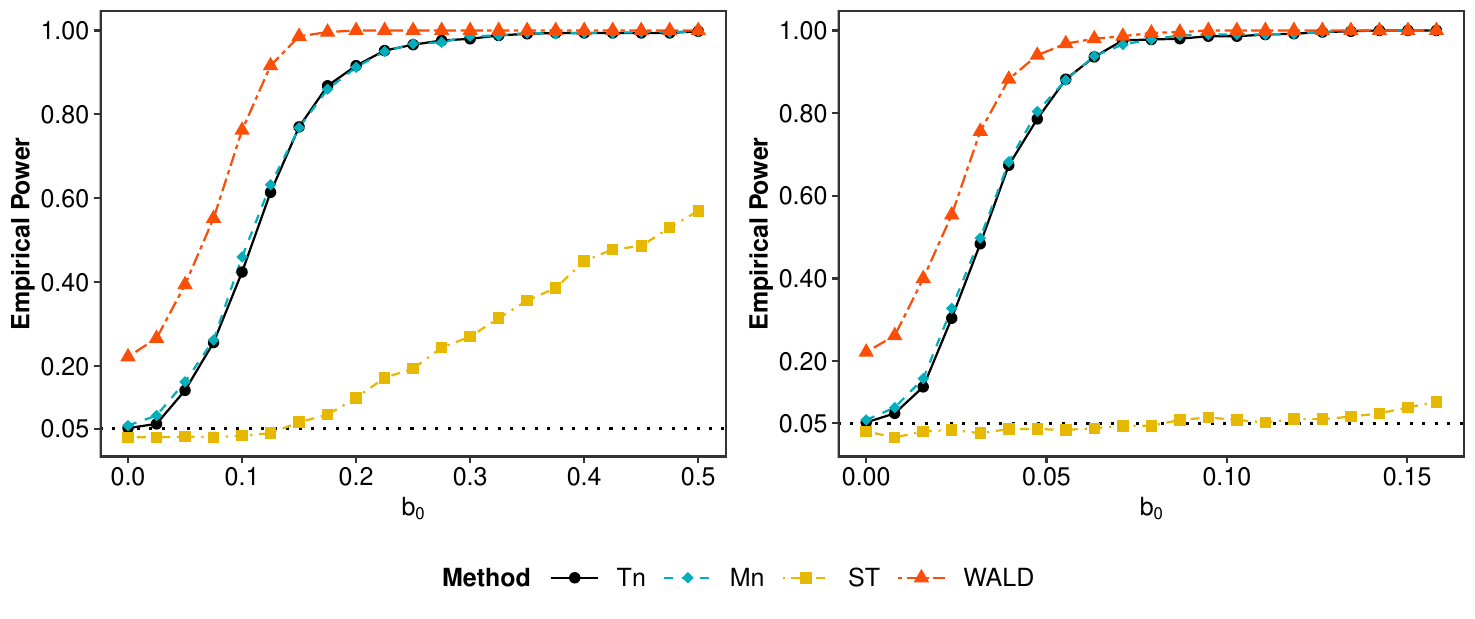}
	\caption{The left panel represents empirical sizes and powers of the $T_n, M_n$, $ST$ and $WALD$ in the sparse alternative (setting 2). The right panel corresponds to dense alternative (setting 3). The solid line with circle points, dash line with diamond points, dot-dash line with square points and two-dash line with triangle points represent the empirical sizes and powers of $T_n$, $M_n$, $ST$ and $WALD$, respectively.}
	\label{figure2}
\end{figure}

Figure \ref{figure3} displays the empirical size-power curves of $T_n$ and $M_n$ in scenario 2. We find that  $T_n$ is too liberal  to maintain the significance level.  On the contrary,  $M_n$ maintains the level well. As $b_0$ increases, although the empirical powers of  $T_n$ and $M_n$ increase rapidly, the empirical power of $T_n$ does not go to $1$ as the increase of $b_0$. In contrast, the empirical power of $M_n$ can increase to 1 quickly. The results show that  $M_n$ can also improve the power compared with $T_n$. This confirms the theory.
\begin{figure}[ht]
	\centering
	\includegraphics[width=\textwidth]{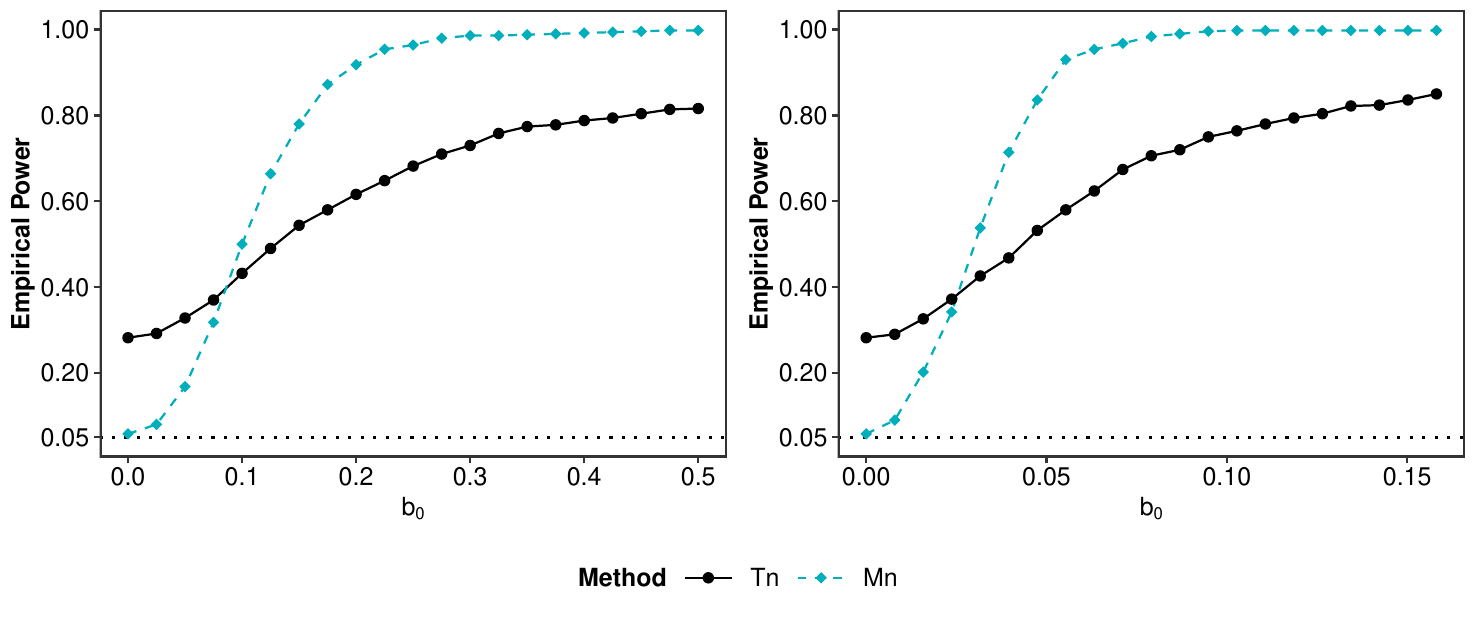}
	\caption{The left panel represents empirical sizes and powers of the $T_n$ and $M_n$ in the sparse alternatives (setting 2). The right panel corresponds to dense alternatives (setting 3). The solid line with circle points and dash line with diamond points represent the empirical sizes and powers of $T_n$ and $M_n$, respectively.}
	\label{figure3}
\end{figure}

The above simulation studies conclude that our proposed tests perform well even when the testing and nuisance parameters are both ultrahigh-dimensional.  When a relatively high correlation exists between the covariates of interest and the nuisance covariates, $M_n$ can enhance the power performance over $T_n$. These confirm the merits of the orthogonalization technique.

\subsection{Real data analysis}

We have employed our tests on the identical dataset as utilized in \cite{houtepen2016genome}, \cite{van2019exploratory} and \cite{guo2022high}, for investigating the role of DNA methylation in the regulation of human stress reactivity. This dataset is accessible for download at https://www.ebi.ac.uk/arrayexpress/experiments/E-GEOD-77445 and encompasses 385882 DNA methylation loci along with various variables pertaining to 85 individuals. 
\cite{houtepen2016genome} conducted a study to explore the role of DNA methylation in cortisol stress reactivity and its relationship with childhood trauma. 
Besides, it is of substantial interests to study the interactions between environmental factors and DNA methylation in genome-wide DNA methylation analysis.
Motivated by these observations, this article primarily focuses on the identification of interactions between DNA methylation and childhood trauma.

In this context, the variable $X_{e}$ represents the childhood trauma as a one-dimensional score from a childhood trauma questionnaire, and the response variable $Y$ denotes the increased area under the curve (iAUC) in cortisol after a stress test. Eight confounding variables are considered, which include age ($Z_1$), sex ($Z_2$), B cell proportion ($Z_3$), CD4 T
cell proportion ($Z_4$), CD8 T cell proportion ($Z_5$), Monocytes cell proportion ($Z_6$), Granulocytes
cell proportion ($Z_7$) and Natural Killer cell proportion ($Z_8$). Let $\bZ = (Z_{1},\ldots,Z_{8})^\top$ be the random vector comprising these eight confounding variables. The 385882 methylation sites are divided into 1930 sets, where 1929 sets contain 200 methylation sites each, and one set contains 82 methylation sites. Let $\bX_{g_k}$ be the random vector for the $k$th DNA methylation set. To illustrate the interactions between DNA methylation and childhood trauma, we use $\bX_{g_k}\times X_{e}$ to denote the interaction between the $k$th DNA methylation set and childhood trauma. For the $k$th DNA methylation set, the aim is to test whether it interacts with childhood trauma. This can be stated as:
\begin{align*}
	H_{0k}: \text{The $k$th DNA methylation set does not interact with childhood trauma.}
\end{align*} 
Specifically, the model for the $k$th methylation set is represented as follows: 
\begin{align*}
	Y = \bbeta_{k}^\top\bX_{g_k}\times X_{e} + \gamma_{e_k}X_{e} + \bgamma_{g_k}^\top\bX_{g_k} + \bgamma_{c_k}^\top\bZ + \epsilon_{k},
\end{align*}
where $\epsilon_{k}$ is the error term, and $(\bbeta_{k}^\top,\gamma_{e_k},\bgamma_{g_k}^\top,\bgamma_{c_k}^\top)^\top$ is the coefficient vector with $k\in\{1,\ldots,1930\}$. In the given model setting, the null hypothesis of interest is transformed as $H_{0k}:\bbeta_{k} = \bzero$.  The testing parameter is $\bbeta_{k}$, and the nuisance parameter is $\bgamma_{k} = (\gamma_{e_k},\bgamma_{g_k}^\top,\bgamma_{c_k}^\top)^\top$.

We employ $T_{n}$ and $M_{n}$ tests. The data is standardized, and a total of 1930 tests are conducted. Figure \ref{figurerealdataanalysis} illustrates the empirical distributions of the negative base-10 logarithm of the $p$-values for $T_{n}$ and $M_{n}$, respectively. After Bonferroni correction, with a significant level of $p$-value $<$ $\alpha_{\text{bon}} = 2.591\times 10^{-5}$, out of the 1930 tests, we identify 126 significant DNA methylation sets using the $T_{n}$ test and 149 significant sets using the $M_{n}$ test. Notably, $M_{n}$ identifies 18\% more significant DNA methylation sets than the $T_{n}$ test. The specific numbers of significant DNA methylation sets are provided in Section \ref{selectedgenesinrealdataanalysis} in the Appendix. It is evident that $M_{n}$ can detect the vast majority of sets that are significant according to the $T_{n}$ test. Moreover, there are many sets that are exclusively identified by the $M_{n}$ test and not by the $T_{n}$ test. These results affirm the excellent performance of $M_{n}$.

\begin{figure}[ht]
	\centering
	\includegraphics[width=\textwidth]{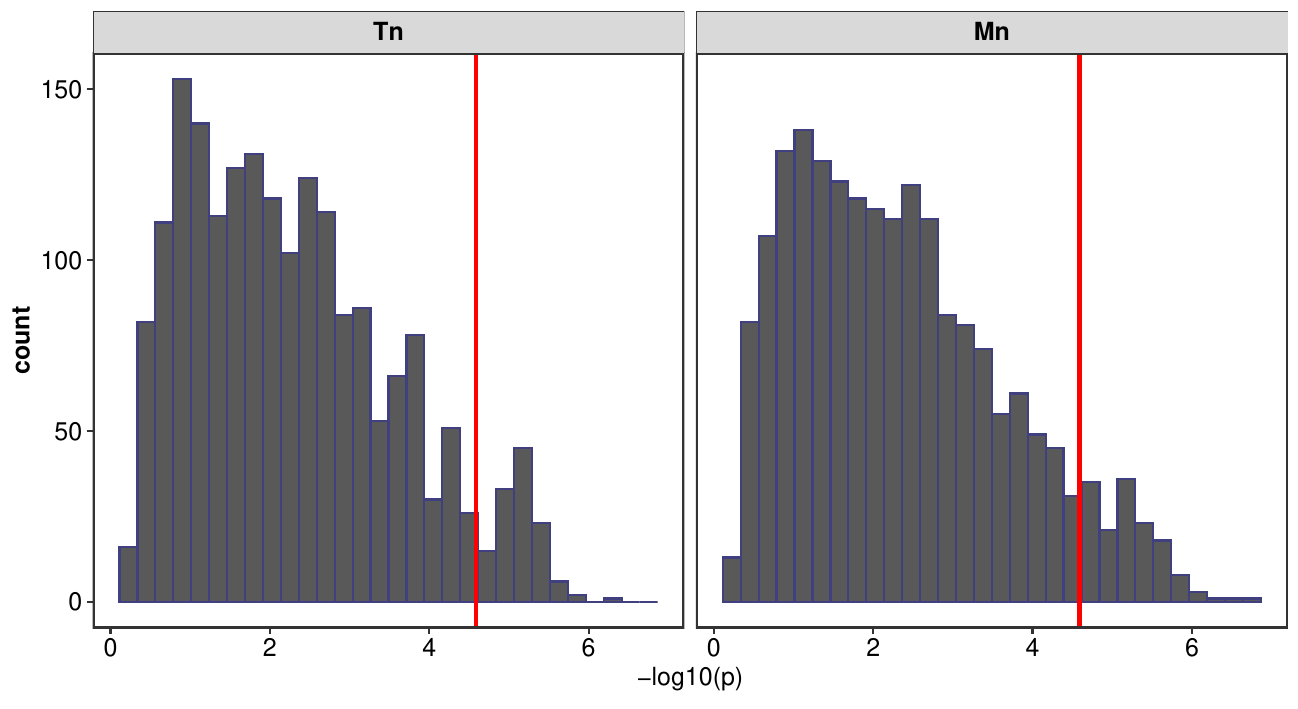}
	\caption{The left panel illustrates the empirical distribution of the negative base-10 logarithm of the p-values for the $T_{n}$ test, while the right panel corresponds to the $M_{n}$ test. In both panels, the red solid line represents $-\log_{10} (\alpha_{\text{bon}})$, where $\alpha_{\text{bon}} = 2.591\times 10^{-5}$ signifies the Bonferroni-adjusted significance level. }
	\label{figurerealdataanalysis}
\end{figure}

\section{Conclusions}
\label{sec7}

This paper considers testing the significance of ultrahigh-dimensional parameter vector of interest with ultrahigh-dimensional nuisance parameter vector. We first reanalyze the score function-based test under weaker conditions to show the limiting distributions under the null and local alternative hypotheses. We construct an orthogonalized score function-based test to handle the correlation between the covariates of interest and nuisance covariates. Our investigation shows that the orthogonalization technique can debiase the error term, convert the non-degenerate error terms to degenerate, and reduce the variance to achieve higher power than the non-orthogonalized score function-based test.

{Our procedure is very generic. Extensions to other regression models such as generalized linear regression models and partially linear single-index regression models \citep{cui2024estimation} are possible. We would investigate these extensions in near future.}


\section*{Appendix}
\label{appendix}

\begin{appendix}
	
Section \ref{notation} presents some notations.
Section \ref{proofoftheoremsinthemaintext} provides the proof of theorems in the main text.
Section \ref{someusefullemmas} consists of some technical lemmas which are used in the proof of theorems in the main text.
Section \ref{additionalsimulationresults} presents additional simulation results. Section \ref{selectedgenesinrealdataanalysis} gives numbers of significant DNA methylation sets in real data analysis in the main text. Section \ref{additionalrealdataanalysis} contains an additional real data analysis. 

\section{Notation}\label{notation}
For functions $f(n)$ and $g(n)$, we write $f(n)\lesssim g(n)$ to mean that $f(n)\leq cg(n)$ for some universal constant $c\in(0,\infty)$, and similarly, $f(n)\gtrsim g(n)$ when $f(n)\geq c'g(n)$ for some universal constant $c'\in(0,\infty)$. We write $f(n)\asymp g(n)$ when $f(n)\lesssim g(n)$ and $f(n)\gtrsim g(n)$ hold simultaneously.
The $\operatorname{\mathit{sub-Gaussian\ norm}}$ of a $\operatorname{\mathit{sub-Gaussian}}$ random variable $X$ is defined as $\lVert X\rVert_{\psi_2} {:=} \inf\{t>0:\mathbb{E}\exp(X^2/t^2)\leq2\}$.
Similarly, the $\operatorname{\mathit{sub-Exponential\ norm}}$ of a $\operatorname{\mathit{sub-Exponential}}$ random variable $Y$ is defined as $\lVert Y\rVert_{\psi_1} { :=} \inf\{t>0:\, \mathbb{E}\exp(\lvert Y\rvert/t)\leq2\}$.
The $\operatorname{\mathit{sub-Gaussian\ norm}}$ of a $\operatorname{\mathit{sub-Gaussian}}$ random vector $\bX$ in $\mbR^p$ is defined as $\lVert\bX\rVert_{\psi_2} := \sup_{x\in\mathbb{S}^{p-1}}\lVert x^\top\bX\rVert_{\psi_2}$. The $L_p$-$\mathit{norm}$ of a random variable $X$ in $\mbR$ is defined as $\lVert X\rVert_{p} := (\mbE\lvert X\rvert^{p})^{1/p}$. For $x\in\mbR$, $\lceil x\rceil$ represents the least integer greater than or equal to $x$. {For $p_1\times p_2$ dimensional matrix $\bA$, write $\lVert\bA\rVert_{\infty} = \max_{1\leq i\leq p_{1},1\leq j\leq p_{2}}\lvert A_{ij}\rvert$ to denote the element-max norm of $\bA$, where $A_{ij}$ is the $(i,j)$-th element of $\bA$.}

\section{Proof of theorems in the main text}
\label{proofoftheoremsinthemaintext}

To simplify the representations of the proofs, we give nine lemmas in Section \ref{someusefullemmas}. Therefore, we will cite them in the proofs. For brevity,  assume $\lVert\bnu\rVert_{\psi_2} = 1$. To simplify the notation, let $\epsilon_i = Y_i - \bbeta^\top\bX_{i} - \bgamma^\top\bZ_i$, $\hat{\epsilon}_i = Y_i - \hat{\bgamma}^\top\bZ_i$ and $\breve{\bgamma} = \bgamma_{\phi} - \hat{\bgamma}$. Under $\bH_{0}$, $\epsilon_{i} = Y_{i} - \bgamma^\top\bZ_i$. Let $\bet_i = \bX_i - \bW^\top\bZ_i$, $\hat{\bet}_i = \bX_i - \hat\bW^\top\bZ_i$, $\bGamma_{\bet} = \bGamma_{\bX} - \bW^\top\bGamma_{\bZ}$, $\bGamma_{\hat\bet} = \bGamma_{\bX} - \hat\bW^\top\bGamma_{\bZ}$ and $\breve{\bW} = \bW - \hat{\bW}$, where $\bGamma = (\bGamma_{\bX}^\top,\bGamma_{\bZ}^\top)^\top$ and $\bGamma$ is defined in Assumption \ref{assumptionb5}. Write $\bX_{i} = \bGamma_{\bX}\bnu_{i},\,\bZ_{i} = \bGamma_{\bZ}\bnu_{i},\,\bet_{i} = \bGamma_{\bet}\bnu_{i}$ and $\hat{\bet}_{i} = \bGamma_{\hat\bet}\bnu_{i}$. Further, write $\bSigma_{\bX} = \bGamma_{\bX}\bGamma_{\bX}^\top$, $\bSigma_{\bZ} = \bGamma_{\bZ}\bGamma_{\bZ}^\top$, $\bSigma_{\bet} = \bGamma_{\bet}\bGamma_{\bet}^\top$ and $\bSigma_{\hat\bet} = \bGamma_{\hat\bet}\bGamma_{\hat\bet}^\top$.
Write $q = \lceil 4/(1-3b) \rceil$.

\subsection{Proofs of Theorem \ref{limitnulldisofguosum}}


As $\bgamma_{\phi} = \bgamma$ and $\epsilon_{i} = Y_i - \bgamma^\top\bZ_i$ under $\bH_{0}$, we  decompose $T_{n}$ it as
\begin{align*}
	T_{n} &= \frac{1}{n}\sum_{i\neq j}\hat{\epsilon}_{i}\hat{\epsilon}_j\bX_i^\top\bX_j
	= \frac{1}{n}\sum_{i\neq j}(\epsilon_{i} + \breve{\bgamma}^\top\bZ_i)(\epsilon_{j} + \breve{\bgamma}^\top\bZ_j)\bX_i^\top\bX_j\\
	&= \frac{1}{n}\sum_{i\neq j}\epsilon_{i}\epsilon_{j}\bX_i^\top\bX_j + \frac{1}{n}\sum_{i\neq j}\breve{\bgamma}^\top\bZ_i\breve{\bgamma}^\top\bZ_j\bX_i^\top\bX_j\\
	&\quad\,\,+\frac{1}{n}\sum_{i\neq j}(\epsilon_{i}\breve{\bgamma}^\top\bZ_j + \breve{\bgamma}^\top\bZ_i\epsilon_{j})\bX_i^\top\bX_j\\
	&=: I_{1}^{\text{c}} + I_{2}^{\text{c}} + I_{3}^{\text{c}}.
\end{align*}
Similar to the proof of Theorem 3 in \cite{guojrssb2016}, we have
\[
\frac{I_{1}^{\text{c}}}{\sqrt{2\bLambda_{\bX}}}\rightarrow N(0,1)\quad \text{in distribution}
\]
as $(n,p_{\bbeta})\rightarrow\infty$. Lemma \ref{lemma14} in the Supplementary Material gives the detailed proof about $I_{1}^{\text{c}}$.
We now prove that $I_{2}^{\text{c}}$ and $I_{3}^{\text{c}}$ are $o_p(\sqrt{2\bLambda_{\bX}})$.

Following Lemma \ref{lemma37} in the Supplementary Material, we have
\begin{align}
	\label{ineq1ofproofofguosum}
	I_{2}^{\text{c}}\leq \biggl\lVert\frac{1}{n}\sum_{i\neq j}\bZ_i\bZ_j^\top\bX_i^\top\bX_j\biggr\rVert_{\infty}\lVert\breve{\bgamma}\rVert_{1}^{2},
\end{align}
Let $\bU_1^{\text{c}}$ be a $p_{\bgamma}\times p_{\bgamma}$ dimension matrix with $(k_1,k_2)$-th element
\[
U_{1,(k_1,k_2)}^{\text{c}} =  \frac{1}{n}\sum_{i\neq j}\frac{1}{2}(Z_{i,k_1}Z_{j,k_2} + Z_{i,k_2}Z_{j,k_1})\bX_i^\top\bX_j.
\]
Note that $\bU_{1}^{\text{c}} = n^{-1}\sum_{i\neq j}\bZ_i\bZ_j^\top\bX_i^\top\bX_j$ and all of its elements are $U$-statistics. By Hoeffding decomposition, we derive
\[
\frac{1}{n-1} U_{1,(k_1,k_2)}^{\text{c}} = \mbE(Z_{k_1}\bX)^\top\mbE(Z_{k_2}\bX) + 2S_{1,1,(k_1,k_2)}^{\text{c}} + S_{1,2,(k_1,k_2)}^{\text{c}},
\]
where
\[
S_{1,1,(k_1,k_2)}^{\text{c}} = \frac{1}{n}\sum_{i=1}^{n}g_{1,1,(k_1,k_2)}^{\text{c}}(\bX_i,\bZ_i)
\]
with $g_{1,1,(k_1,k_2)}^{\text{c}}(\bX_1,\bZ_1) = \{Z_{1,k_1}\bX_1 - \mbE(Z_{k_1}\bX)\}^\top\mbE(Z_{k_2}\bX)/2 + \{Z_{1,k_2}\bX_1 - \mbE(Z_{k_2}\bX)\}^\top\mbE(Z_{k_1}\bX)/2$,
\[
S_{1,2,(k_1,k_2)}^{\text{c}} = \frac{1}{n(n-1)}\sum_{i\neq j}^{n}g_{1,2,(k_1,k_2)}^{\text{c}}(\bX_i,\bZ_i;\bX_j,\bZ_j)
\]
with $g_{1,2,(k_1,k_2)}^{\text{c}}(\bX_1,\bZ_1;\bX_2,\bZ_2) = \{Z_{1,k_1}\bX_1 - \mbE(Z_{k_1}\bX)\}^\top\{Z_{2,k_2}\bX_2 - \mbE(Z_{k_2}\bX)\}/2 + \{Z_{1,k_2}\bX_1 - \mbE(Z_{k_2}\bX)\}^\top\{Z_{2,k_1}\bX_2 - \mbE(Z_{k_1}\bX)\}/2$.

For any $1\leq k_1,k_2\leq p_{\bgamma}$, $Z_{1,k_1}$ is a sub-Gaussian random variable and $\mbE(Z_{k_2}\bX)^\top\bX$ is a sub-Gaussian random variable with norm $\lVert\mbE(Z_{k_2}\bX)^\top\bX\rVert_{\psi_2}\lesssim \{\mbE(Z_{k_2}\bX)^\top\bSigma_{\bX}\mbE(Z_{k_2}\bX)\}^{1/2}$. Thus $\{Z_{1,k_1}\bX_1 - \mbE(Z_{k_1}\bX)\}^\top\mbE(Z_{k_2}\bX)$ is sub-Exponential with norm
\begin{align}
	\label{ineq1ofproofoflimitnulldisofguosum}
	\lVert \{Z_{1,k_1}\bX_1 - \mbE(Z_{k_1}\bX)\}^\top\mbE(Z_{k_2}\bX)\rVert_{\psi_1}&\leq \lVert Z_{1,k_1}\rVert_{\psi_2}\lVert\mbE(Z_{k_2}\bX)^\top\bX\rVert_{\psi_2}\notag\\
	&\lesssim \{\mbE(Z_{k_2}\bX)^\top\bSigma_{\bX}\mbE(Z_{k_2}\bX)\}^{1/2}.
\end{align}
Therefore we derive
\begin{align}
	\label{ineq2ofproofoflimitnulldisofguosum}
	\lVert\max_{1\leq i\leq n}\max_{1\leq k_1,k_2\leq p_{\bgamma}}g_{1,1,(k_1,k_2)}^{\text{c}}\rVert_{2}&\lesssim\log (np_{\bgamma})\max_{1\leq i\leq n}\max_{1\leq k_1,k_2\leq p_{\bgamma}}\lVert g_{1,1,(k_1,k_2)}^{\text{c}}\rVert_{\psi_1}\notag\\
	&\lesssim \log (np_{\bgamma})\max_{1\leq k\leq p_{\bgamma}}\{\mbE(Z_{k}\bX)^\top\bSigma_{\bX}\mbE(Z_{k}\bX)\}^{1/2}\notag\\
	&\lesssim \log (np_{\bgamma})\lambda_{\rm{max}}^{1/2}(\bSigma_{\bX})\varrho.
\end{align}
The first inequality follows from Lemma \ref{lemma36} in the Supplementary Material. The second inequality holds by inequality \eqref{ineq1ofproofoflimitnulldisofguosum}.

Denote $q = \lceil 4/(1-3b) \rceil$. We have
\begin{align}
	\label{ineq3ofproofoflimitnulldisofguosum}
	\lVert\max_{1\leq i\neq j\leq n}\max_{1\leq k_1,k_2\leq p_{\bgamma}}g_{1,2,(k_1,k_2)}^{\text{c}}\rVert_{4}&\leq\lVert\max_{1\leq i\neq j\leq n}\max_{1\leq k_1,k_2\leq p_{\bgamma}}g_{1,2,(k_1,k_2)}^{\text{c}}\rVert_{q}\notag\\
	&\lesssim \lVert\max_{1\leq i\neq j\leq n}\max_{1\leq k_1,k_2\leq p_{\bgamma}}Z_{i,k_1}Z_{j,k_2}\bX_{i}^\top\bX_{j}\rVert_{q}\notag\\
	&\lesssim \lVert\max_{1\leq i\neq j\leq n}\max_{1\leq k_1,k_2\leq p_{\bgamma}}Z_{i,k_1}Z_{j,k_2}\rVert_{2q}\lVert\max_{1\leq i\neq j\leq n}\bX_{i}^\top\bX_{j}\rVert_{2q}\notag\\
	&\lesssim \log (np_{\bgamma})\max_{1\leq k_1,k_2\leq p_{\bgamma}}\lVert Z_{1,k_1}Z_{2,k_2}\rVert_{\psi_1}n^{1/q}\tr^{1/2}(\bSigma_{\bX}^{2})\notag\\
	&\lesssim \log (np_{\bgamma})n^{1/q}\tr^{1/2}(\bSigma_{\bX}^{2}).
\end{align}
The first inequality holds by Liapounov inequality. The third inequality holds by Cauchy-Schwartz inequality, the fourth inequality holds by the property of Orlicz norm (Page 96 in \cite{vanderVaart.1996}), Lemmas \ref{lemma36} and \ref{lemmaofassumptionb5} in the Supplementary Material. The last inequality holds by Assumption \ref{assumptionb5}.
Similarly, we derive
\begin{align}
	\label{ineq4ofproofoflimitnulldisofguosum}
	\max_{1\leq k_1,k_2\leq p_{\bgamma}}\lVert g_{1,1,(k_1,k_2)}^{\text{c}}\rVert_{2}\lesssim \lambda_{\rm{max}}^{1/2}(\bSigma_{\bX})\varrho
\end{align}
and
\begin{align}
	\label{ineq5ofproofoflimitnulldisofguosum}
	\max_{1\leq k_1,k_2\leq p_{\bgamma}}\lVert g_{1,2,(k_1,k_2)}^{\text{c}}\rVert_{2}\leq \max_{1\leq k_1,k_2\leq p_{\bgamma}}\lVert g_{1,2,(k_1,k_2)}^{\text{c}}\rVert_{4}\lesssim \tr^{1/2}(\bSigma_{\bX}^2).
\end{align}
Applying Lemma \ref{lemma26} in the Supplementary Material, we derive
\begin{align}
	\label{ineq2ofproofofguosum}
	\mbE(\lVert\bU_{1}^{\text{c}}\rVert_{\infty}) \lesssim& n\varrho^2 + \{n^{1/2}(\log p_{\bgamma})^{1/2} + \log p_{\bgamma}\log (np_{\bgamma})\}\lambda_{\rm{max}}^{1/2}(\bSigma_{\bX})\varrho\notag\\
	&+ \{\log p_{\bgamma} + n^{-1/2+1/q}(\log p_{\bgamma})^{3/2}\log (np_{\bgamma})\}\tr^{1/2}(\bSigma_{\bX}^{2})\notag\\
	\lesssim& n\varrho^2 + n^{1/2}(\log p_{\bgamma})^{1/2}\lambda_{\rm{max}}^{1/2}(\bSigma_{\bX})\varrho + \log p_{\bgamma}\tr^{1/2}(\bSigma_{\bX}^{2}),
\end{align}
where the first inequality holds by Lemma \ref{lemma26} in the Supplementary Material and the inequalities \eqref{ineq2ofproofoflimitnulldisofguosum}-\eqref{ineq5ofproofoflimitnulldisofguosum}. The last inequality holds by the fact $\log p_{\bgamma} = O(n^b)$, where $0<b<1/3$.
Combining equations \eqref{ineq1ofproofofguosum} and \eqref{ineq2ofproofofguosum}, Assumption \ref{assumptionb6}, we have
\begin{align}
	\label{eq1ofproofofguosum}
	I_2^{\text{c}} = O_p\bigl\{s^2\log p_{\bgamma}\varrho^2 + n^{-1/2}s^{2}(\log p_{\bgamma})^{3/2}\lambda_{\rm{max}}^{1/2}(\bSigma_{\bX})\varrho + n^{-1}s^2(\log p_{\bgamma})^{2}\tr^{1/2}(\bSigma_{\bX}^2)\bigr\}.
\end{align}
Thus $I_2^{\text{c}} = o_p(\sqrt{\bLambda_{\bX}})$ when $s^2\log p_{\bgamma}\varrho^2 = o(\sqrt{\bLambda_{\bX}})$, $n^{-1/2}s^{2}(\log p_{\bgamma})^{3/2}\lambda_{\rm{max}}^{1/2}(\bSigma_{\bX})\varrho = o(\sqrt{\bLambda_{\bX}})$ and $n^{-1}s^2(\log p_{\bgamma})^{2} = o(1)$ hold simultaneously.

Similar to the proof of $I_{2}^{\text{c}}$, we derive
\begin{align}
	\label{ineq3ofproofofguosum}
	I_3^{\text{c}}\leq \biggl\lVert\frac{1}{n}\sum_{i\neq j}(\epsilon_{i}\bZ_j + \bZ_i\epsilon_{j})\bX_i^\top\bX_j\biggr\rVert_{\infty}\lVert\breve{\bgamma}\rVert_{1}.
\end{align}
Let $\bU_2^{\text{c}}$ be a $p_{\bgamma}$ dimension vector with $k$-th element
\[
U_{2,k}^{\text{c}} =  \frac{1}{n}\sum_{i\neq j}(\epsilon_iZ_{j,k} + Z_{i,k}\epsilon_{j})\bX_i^\top\bX_j.
\]
Note that $\bU_{2}^{\text{c}} = n^{-1}\sum_{i\neq j}(\epsilon_{i}\bZ_j + \bZ_i\epsilon_{j})\bX_i^\top\bX_j$ and all of its elements are $U$-statistics. By Hoeffding decomposition again, we derive
\[
\frac{1}{n-1} U_{2,k}^{\text{c}} = 2S_{2,1,k}^{\text{c}} + S_{2,2,k}^{\text{c}},
\]
where
\[
S_{2,1,k}^{\text{c}} = \frac{1}{n}\sum_{i=1}^{n}g_{2,1,k}^{\text{c}}(\bX_i,\bZ_i,\epsilon_{i})
\]
with $g_{2,1,k}^{\text{c}}(\bX_1,\bZ_1,\epsilon_{1}) = \epsilon_{1}\bX_1^\top\mbE(Z_{k}\bX)$,
\[
S_{2,2,k}^{\text{c}} = \frac{1}{n(n-1)}\sum_{i\neq j}^{n}g_{2,2,k}^{\text{c}}(\bX_i,\bZ_i,\epsilon_{i};\bX_j,\bZ_j,\epsilon_{j})
\]
with $g_{2,2,k}^{\text{c}}(\bX_1,\bZ_1,\epsilon_{1};\bX_2,\bZ_2,\epsilon_{2}) = \epsilon_{1}\bX_1^\top\{Z_{2,k}\bX_2 - \mbE(Z_{k}\bX)\} + \epsilon_{2}\bX_2^\top\{Z_{1,k}\bX_1 - \mbE(Z_{k}\bX)\}$.

Similar to the proof of inequality \eqref{ineq2ofproofoflimitnulldisofguosum}, we derive
\begin{align}
	\label{ineq6ofproofoflimitnulldisofguosum}
	\lVert\max_{1\leq i\leq n}\max_{1\leq k\leq p_{\bgamma}}g_{2,1,k}^{\text{c}}\rVert_{2} &\lesssim \lVert\max_{1\leq i\leq n}\max_{1\leq k\leq p_{\bgamma}}\epsilon_{i}\bX_{i}^\top\mbE(Z_{k}\bX)\rVert_{2}\notag\\
	&\lesssim \log (np_{\bgamma})\max_{1\leq k\leq p_{\bgamma}}\lambda_{\rm{max}}^{1/2}(\bSigma_{\bX})\varrho.
\end{align}
The last inequality is derived by Assumption \ref{assumptionb7}, Lemma \ref{lemma36} in the Supplementary Material, and the technique used in the last inequality in the proof for the inequality \eqref{ineq2ofproofoflimitnulldisofguosum}. Similar to the proof of \eqref{ineq3ofproofoflimitnulldisofguosum}, we derive
\begin{align}
	\label{ineq7ofproofoflimitnulldisofguosum}
	\lVert\max_{1\leq i\neq j\leq n}\max_{1\leq k\leq p_{\bgamma}}g_{2,2,k}^{\text{c}}\rVert_{4} &\leq \lVert\max_{1\leq i\neq j\leq n}\max_{1\leq k\leq p_{\bgamma}}g_{2,2,k}^{\text{c}}\rVert_{q}\notag\\
	&\lesssim \lVert\max_{1\leq i\neq j\leq n}\max_{1\leq k\leq p_{\bgamma}}\epsilon_{1}Z_{2,k}\bX_1^\top\bX_2\rVert_{q}\notag\\
	&\lesssim \log (np_{\bgamma})n^{1/q}\tr^{1/2}(\bSigma_{\bX}^2).
\end{align}
Similarly,
\begin{align}
	\label{ineq8ofproofoflimitnulldisofguosum}
	\max_{1\leq k\leq p_{\bgamma}}\lVert g_{2,1,k}^{\text{c}}\rVert_{2}\lesssim \lambda_{\rm{max}}(\bSigma_{\bX})\varrho^2
\end{align}
and
\begin{align}
	\label{ineq9ofproofoflimitnulldisofguosum}
	\max_{1\leq k\leq p_{\bgamma}}\lVert g_{2,2,k}^{\text{c}}\rVert_{2}\leq \max_{1\leq k\leq p_{\bgamma}}\lVert g_{2,2,k}^{\text{c}}\rVert_{4}\lesssim \tr^{1/2}(\bSigma_{\bX}^2).
\end{align}
The argument for proving the inequality \eqref{ineq2ofproofofguosum} yields
\begin{align}
	\label{ineq4ofproofofguosum}
	\mbE(\lVert\bU_{2}^{\text{c}}\rVert_{\infty}) \lesssim n^{1/2}(\log p_{\bgamma})^{1/2}\lambda_{\rm{max}}^{1/2}(\bSigma_{\bX})\varrho + \log p_{\bgamma}\tr^{1/2}(\bSigma_{\bX}^2).
\end{align}
Combining equations \eqref{ineq3ofproofofguosum} and \eqref{ineq4ofproofofguosum}, Assumption \ref{assumptionb6}, we have
\begin{align}
	\label{eq2ofproofofguosum}
	I_3^{\text{c}} = O_p\bigl\{s\log p_{\bgamma}\lambda_{\rm{max}}^{1/2}(\bSigma_{\bX})\varrho + n^{-1/2}s(\log p_{\bgamma})^{3/2}\tr^{1/2}(\bSigma_{\bX}^2)\bigr\}.
\end{align}
Thus $I_3^{\text{c}} = o_p(\sqrt{\bLambda_{\bX}})$ when $s\log p_{\bgamma}\lambda_{\rm{max}}^{1/2}(\bSigma_{\bX})\varrho = o(\sqrt{\bLambda_{\bX}})$ and $n^{-1/2}s(\log p_{\bgamma})^{3/2} = o(1)$ hold simultaneously.
The proof is concluded.

\subsection{The proof of Theorem \ref{limitpowerofguosum}}
Under the local alternatives, we decompose $T_{n}$ as:
\begin{align*}
	T_{n} =& \frac{1}{n}\sum_{i\neq j}\hat{\epsilon}_{i}\hat{\epsilon}_{j}\bX_{i}^\top\bX_{j}\\
	=& \frac{1}{n}\sum_{i\neq j}(\bbeta^\top\bet_{i} + \epsilon_{i} + \breve{\bgamma}^\top\bZ_{i})(\bbeta^\top\bet_{j} + \epsilon_{j} + \breve{\bgamma}^\top\bZ_{j})\bX_{i}^\top\bX_{j}\\
	=& \frac{1}{n}\sum_{i\neq j}(\bbeta^\top\bet_{i} + \epsilon_{i})(\bbeta^\top\bet_{j} + \epsilon_{j})\bX_{i}^\top\bX_{j}\\
	&+ \frac{1}{n}\sum_{i\neq j}\breve{\bgamma}^\top\bZ_{i}\breve{\bgamma}^\top\bZ_{j}\bX_{i}^\top\bX_{j}\\
	&+ \frac{1}{n}\sum_{i\neq j}(\epsilon_{i}\breve{\bgamma}^\top\bZ_{j} + \breve{\bgamma}^\top\bZ_{i}\epsilon_{j})\bX_{i}^\top\bX_{j}\\
	&+ \frac{1}{n}\sum_{i\neq j}(\bbeta^\top\bet_{i}\breve{\bgamma}^\top\bZ_{j} + \breve{\bgamma}^\top\bZ_{i}\bbeta^\top\bet_{j})\bX_{i}^\top\bX_{j}\\
	=:& A_{1}^{\text{c}} + A_{2}^{\text{c}} + A_{3}^{\text{c}} + A_{4}^{\text{c}},
\end{align*}
where the second equality holds by the fact that
\begin{align}
	\label{hatepsilonundermodelmisspecified}
	\hat{\epsilon}_{i} &= \epsilon_{i} + \bgamma^\top\bZ_{i} - \bgamma_{\phi}^\top\bZ_{i} + \breve{\bgamma}^\top\bZ_{i} + \bbeta^\top\bX_{i}\notag\\
	&= \epsilon_{i} + \bigl[\bgamma - \bSigma_{\bZ}^{-1}\mbE\{\bZ(\bZ^\top\bgamma + \bX_{i}^\top\bbeta + \epsilon_{i})\}\bigr]^\top\bZ_{i} + \breve{\bgamma}^\top\bZ_{i} + \bbeta^\top\bX_{i}\notag\\
	&=\epsilon_{i} - \bbeta^\top\bW^\top\bZ_{i} + \breve{\bgamma}^\top\bZ_{i} + \bbeta^\top\bX_{i}\notag\\
	&= \bbeta^\top\bet_{i} + \epsilon_{i} + \breve{\bgamma}^\top\bZ_{i}.
\end{align}
The first term $A_{1}^{c}$ can be rewritten as
\begin{align*}
	A_{1}^{\text{c}} =&\frac{1}{n}\sum_{i\neq j}\epsilon_{i}\epsilon_{j}\bX_{i}^\top\bX_{j} + \frac{1}{n}\sum_{i\neq j}\bbeta^\top\bet_{i}\bX_{i}^\top\bX_{j}\bet_{j}^\top\bbeta\\
	&+ \frac{1}{n}\sum_{i\neq j}(\bbeta^\top\bet_{i}\epsilon_{j} + \epsilon_{i}\bbeta^\top\bet_{j})\bX_{i}^\top\bX_{j}\\
	=:& A_{11}^{\text{c}} + A_{12}^{\text{c}} + A_{13}^{\text{c}}.
\end{align*}
Following Lemma \ref{lemma14} below,  we have
\[
\frac{A_{11}^{\text{c}}}{\sqrt{2\bLambda_{\bX}}}\rightarrow N(0,1)\quad \text{in distribution}
\]
as $(n,p_{\bbeta})\rightarrow\infty$.  Noticing  $\mbE(A_{12}^{\text{c}}) = (n-1)\bbeta^\top\bSigma_{\bet}^{2}\bbeta$ and applying the Hoeffding decomposition we derive
\[
\frac{1}{n-1}A_{12}^{\text{c}} = \bbeta^\top\bSigma_{\bet}^{2}\bbeta + 2A_{121}^{\text{c}} + A_{122}^{\text{c}},
\]
where
\[
A_{121}^{\text{c}} = \frac{1}{n}\sum_{i=1}^{n}\bbeta^\top\bSigma_{\bet}(\bX_{i}\bet_{i}^\top\bbeta - \bSigma_{\bet}\bbeta),
\]
and
\[
A_{122}^{\text{c}} = \frac{1}{n(n-1)}\sum_{i\neq j}(\bX_{i}\bet_{i}^\top\bbeta - \bSigma_{\bet}\bbeta)^\top(\bX_{j}\bet_{j}^\top\bbeta - \bSigma_{\bet}\bbeta).
\]
By the variance formula of U-statistics, we derive the variance of $A_{12}^{c}$
\begin{align*}
	\Var(A_{12}^{\text{c}}) &\lesssim n^2\mbE(A_{121}^{\text{c}})^2 + n^2\mbE(A_{122}^{\text{c}})^2\\
	&\lesssim n\mbE(\bbeta^\top\bSigma_{\bet}\bX_{1}\bet_{1}^\top\bbeta)^2 + \mbE(\bbeta^\top\bet_{1}\bX_{1}^\top\bX_{2}\bet_{2}^\top\bbeta)^2\\
	&= o(\bLambda_{\bX}).
\end{align*}
The last equality holds by Lemma \ref{lemma32} below
and the model structure under the local alternatives.
Similarly, we derive $\mbE(A_{13}^{c}) = 0$ and
\[
\frac{1}{n-1}A_{13}^{\text{c}} = 2A_{131}^{\text{c}} + A_{132}^{\text{c}},
\]
where
\[
A_{131}^{\text{c}} = \frac{1}{n}\sum_{i=1}^{n}\bbeta^\top\bSigma_{\bet}\epsilon_{i}\bX_{i}
\]
and
\[
A_{132}^{\text{c}} = \frac{1}{n(n-1)}\sum_{i\neq j}\bigl\{(\bbeta^\top\bet_{i}\bX_{i}^\top - \bbeta^\top\bSigma_{\bet})\epsilon_{j}\bX_{j} + \epsilon_{i}\bX_{i}^\top(\bX_{j}\bet_{j}^\top\bbeta - \bSigma_{\bet}\bbeta)\bigr\}.
\]
Similar to the derivation of $\Var(A_{12}^{\text{c}})$, we derive
\begin{align*}
	\Var(A_{13}^{\text{c}}) &\lesssim n^2\mbE(A_{131}^{\text{c}})^2 + n^2\mbE(A_{132}^{\text{c}})^2\\
	&\lesssim n\bbeta^\top\bSigma_{\bet}\mbE(\epsilon_{1}^2\bX_{1}\bX_{1}^\top)\bSigma_{\bet}\bbeta + \mbE(\epsilon_{2}\bbeta^\top\bet_{1}\bX_{1}^\top\bX_{2})^2\\
	&\lesssim n\bbeta^\top\bSigma_{\bet}\bSigma_{\bX}\bSigma_{\bet}\bbeta + \tr(\bSigma_{\bX}^{2})\bbeta^\top\bSigma_{\bet}\bbeta\\
	&= o(\bLambda_{\bX}).
\end{align*}
Where the equality holds by the model structure under the local alternatives, and the third inequality holds by using Lemma \ref{lemma32}, Assumption \ref{assumptionb7} and the fact that $\epsilon$ is independent of $\bX$.

In summary, we derive.
\[
\frac{A_{1}^{\text{c}} - n\bbeta^\top\bSigma_{\bet}^{2}\bbeta}{\sqrt{2\bLambda_{\bX}}}\rightarrow N(0,1)\quad \text{in distribution.}
\]
To prove this theorem, it suffices to prove $A_{i}^{\text{c}} = o_p(\sqrt{2\bLambda_{\bX}}),\,i=2,3,4$. By the same argument as getting $I_{2}^{\text{c}}$ and $I_{3}^{\text{c}}$ in Theorem \ref{limitnulldisofguosum}, we can show
\[
A_{2}^{\text{c}} = o_p(\sqrt{2\bLambda_{\bX}})\quad \text{and}\quad A_{3}^{\text{c}} = o_p(\sqrt{2\bLambda_{\bX}}).
\]
Turn to prove $A_{4}^{\text{c}} = o_p(\sqrt{2\bLambda_{\bX}})$. Recall that
\[
A_{4}^{\text{c}} = \frac{1}{n}\sum_{i\neq j}(\bbeta^\top\bet_{i}\breve{\bgamma}^\top\bZ_{j} + \breve{\bgamma}^\top\bZ_{i}\bbeta^\top\bet_{j})\bX_{i}^\top\bX_{j}.
\]
By the H{\" o}lder inequality, we derive
\begin{align}
	\label{ineq5ofproofofguosum}
	\lvert A_{4}^{\text{c}}\rvert \leq \biggl\lVert \frac{1}{n}\sum_{i\neq j}
	(\bbeta^\top\bet_{i}\bZ_{j} + \bZ_{i}\bbeta^\top\bet_{j})\bX_{i}^\top\bX_{j}
	\biggr\rVert_{\infty}\lVert\breve{\bgamma}\rVert_{1}.
\end{align}
Let $\bU_4^{\text{c}}$ be a $p_{\bgamma}$ dimension vector with $k$-th element
\[
U_{4,k}^{\text{c}} =  \frac{1}{n}\sum_{i\neq j}(\bbeta^\top\bet_{i}Z_{j,k} + Z_{i,k}\bbeta^\top\bet_{j})\bX_{i}^\top\bX_{j}.
\]
Note that $\bU_{4}^{\text{c}} = n^{-1}\sum_{i\neq j}(\bbeta^\top\bet_{i}\bZ_{j} + \bZ_{i}\bbeta^\top\bet_{j})\bX_{i}^\top\bX_{j}$ and all of its elements are $U$-statistics. By the Hoeffding decomposition, we derive
\[
\frac{1}{n-1} U_{4,k}^{\text{c}} = 2\bbeta^\top\bSigma_{\bet}\mbE(Z_{k}\bX) + 2S_{4,1,k}^{\text{c}} + S_{4,2,k}^{\text{c}},
\]
where
\[
S_{4,1,k}^{\text{c}} = \frac{1}{n}\sum_{i=1}^{n}g_{4,1,k}^{\text{c}}(\bX_i,\bZ_i)
\]
with $g_{4,1,k}^{\text{c}}(\bX_1,\bZ_1) = 		\mbE(Z_{k}\bX)^\top(\bX_{1}\bet_{1}^\top\bbeta - \bSigma_{\bet}\bbeta) + \bbeta^\top\bSigma_{\bet}\{Z_{1,k}\bX_{1} - \mbE(Z_{k}\bX)\}$,
\[
S_{4,2,k}^{\text{c}} = \frac{1}{n(n-1)}\sum_{i\neq j}^{n}g_{4,2,k}^{\text{c}}(\bX_i,\bZ_i;\bX_j,\bZ_j)
\]
with $g_{4,2,k}^{\text{c}}(\bX_1,\bZ_1;\bX_2,\bZ_2) = (\bX_{1}\bet_{1}^\top\bbeta - \bSigma_{\bet}\bbeta)^\top\{Z_{2,k}\bX_{2} - \mbE(Z_{k}X)\} + \{Z_{1,k}\bX_{1} - \mbE(Z_{k}\bX)\}^\top(\bX_{2}\bet_{2}^\top\bbeta - \bSigma_{\bet}\bbeta)$.

For any $1\leq k\leq p_{\bgamma}$, $Z_{1,k}$ is a sub-Gaussian random variable with bounded sub-Gaussian norm, $\mbE(Z_{k}\bX)^\top\bX_1$ is a sub-Gaussian random variable with norm $\{\mbE(Z_{k}\bX)^\top\bSigma_{\bX}\mbE(Z_{k}\bX)\}^{1/2}$. Also $\bbeta^\top\bet_{1}$ is sub-Gaussian with norm $(\bbeta^\top\bSigma_{\bet}\bbeta)^{1/2}$, $\bbeta^\top\bSigma_{\bet}\bX_{1}$ is sub-Gaussian with norm $(\bbeta^\top\bSigma_{\bet}\bSigma_{\bX}\bSigma_{\bet}\bbeta)^{1/2}$. Thus $g_{4,1,k}^{\text{c}}$ is sub-Exponential with norm
\begin{align}
	\label{ineq1ofproofoflimitpowerofguosum}
	\lVert g_{4,1,k}^{\text{c}}\rVert_{\psi_1}&\leq \lVert\mbE(Z_{k}\bX)^\top(\bX_{1}\bet_{1}^\top\bbeta - \bSigma_{\bet}\bbeta)\rVert_{\psi_1} + \lVert\bbeta^\top\bSigma_{\bet}\{Z_{1,k}\bX_{1} - \mbE(Z_{k}\bX)\}\rVert_{\psi_1}\notag\\
	&\leq \lVert\mbE(Z_{k}\bX)^\top\bX_{1}\rVert_{\psi_2}\lVert\bbeta^\top\bet_{1}\rVert_{\psi_2} + \lVert Z_{1,k}\rVert_{\psi_2}\lVert\bbeta^\top\bSigma_{\bet}\bX_{1}\rVert_{\psi_2}\notag\\
	&\lesssim \bigl\{\bbeta^\top\bSigma_{\bet}\bbeta\mbE(Z_{k}\bX)^\top\bSigma_{\bX}\mbE(Z_{k}\bX) + \bbeta^\top\bSigma_{\bet}\bSigma_{\bX}\bSigma_{\bet}\bbeta\bigr\}^{1/2}.
\end{align}
Therefore,
\begin{align}
	\label{ineq2ofproofoflimitpowerofguosum}
	\lVert\max_{1\leq i\leq n}\max_{1\leq k\leq p_{\bgamma}}g_{4,1,k}^{\text{c}}\rVert_{2}&\lesssim\log (np_{\bgamma})\max_{1\leq i\leq n}\max_{1\leq k\leq p_{\bgamma}}\lVert g_{4,1,k}^{\text{c}}\rVert_{\psi_1}\notag\\
	&\lesssim \log (np_{\bgamma})\max_{1\leq k\leq p_{\bgamma}}\bigl\{\bbeta^\top\bSigma_{\bet}\bbeta\mbE(Z_{k}\bX)^\top\bSigma_{\bX}\mbE(Z_{k}\bX) + \bbeta^\top\bSigma_{\bet}\bSigma_{\bX}\bSigma_{\bet}\bbeta\bigr\}^{1/2}\notag\\
	&\leq \log (np_{\bgamma})\bigl\{\bbeta^\top\bSigma_{\bet}\bbeta\lambda_{\rm{max}}(\bSigma_{\bX})\varrho^2 + \bbeta^\top\bSigma_{\bet}\bSigma_{\bX}\bSigma_{\bet}\bbeta\bigr\}^{1/2}.
\end{align}
The first inequality follows from Lemma \ref{lemma36}, and the second inequality is derived by the inequality \eqref{ineq1ofproofoflimitpowerofguosum}.

Denote $q = \lceil 4/(1-3b) \rceil$. We have
\begin{align}
	\label{ineq3ofproofoflimitpowerofguosum}
	\lVert\max_{1\leq i\neq j\leq n}\max_{1\leq k\leq p_{\bgamma}}g_{4,2,k}^{\text{c}}\rVert_{4}&\leq\lVert\max_{1\leq i\neq j\leq n}\max_{1\leq k\leq p_{\bgamma}}g_{4,2,k}^{\text{c}}\rVert_{q}\notag\\
	&\lesssim \lVert\max_{1\leq i\neq j\leq n}\max_{1\leq k_1,k_2\leq p_{\bgamma}}Z_{i,k}\bbeta^\top\bet_{j}\bX_{i}^\top\bX_{j}\rVert_{q}\notag\\
	&\leq \lVert\max_{1\leq i\neq j\leq n}\max_{1\leq k\leq p_{\bgamma}}Z_{i,k}\bbeta^\top\bet_{j}\rVert_{2q}\lVert\max_{1\leq i\neq j\leq n}\bX_{i}^\top\bX_{j}\rVert_{2q}\notag\\
	&\lesssim \log (np_{\bgamma})\max_{1\leq k\leq p_{\bgamma}}\lVert Z_{1,k}\bbeta^\top\bet_{2}\rVert_{\psi_1}n^{1/q}\tr^{1/2}(\bSigma_{\bX}^{2})\notag\\
	&\lesssim \log (np_{\bgamma})n^{1/q}(\bbeta^\top\bSigma_{\bet}\bbeta)^{1/2}\tr^{1/2}(\bSigma_{\bX}^{2}).
\end{align}
The first inequality is from the Liapounov inequality, the third inequality is due to the Cauchy-Schwartz inequality, and the fourth inequality holds by using the property of Orlicz norm(Page 96 in \cite{vanderVaart.1996}), Lemmas \ref{lemma36} and \ref{lemmaofassumptionb5}. The last inequality is based on the sub-Gaussian property of $Z_{1,k}$ and $\bbeta^\top\bet_{2}$.
Similarly, we derive
\begin{align}
	\label{ineq4ofproofoflimitpowerofguosum}
	\max_{1\leq k\leq p_{\bgamma}}\lVert g_{4,1,k}^{\text{c}}\rVert_{2}\lesssim
	\bigl\{\bbeta^\top\bSigma_{\bet}\bbeta\lambda_{\rm{max}}(\bSigma_{\bX})\varrho^2 + \bbeta^\top\bSigma_{\bet}\bSigma_{\bX}\bSigma_{\bet}\bbeta\bigr\}^{1/2}
\end{align}
and
\begin{align}
	\label{ineq5ofproofoflimitpowerofguosum}
	\max_{1\leq k\leq p_{\bgamma}}\lVert g_{4,2,k}^{\text{c}}\rVert_{2}\leq \max_{1\leq k\leq p_{\bgamma}}\lVert g_{4,2,k}^{\text{c}}\rVert_{4}\lesssim (\bbeta^\top\bSigma_{\bet}\bbeta)^{1/2}\tr^{1/2}(\bSigma_{\bX}^{2}).
\end{align}

Similar to the proof of the inequality \eqref{ineq2ofproofofguosum}, we derive
\begin{align}
	\label{ineq6ofproofofguosum}
	\mbE(\lVert\bU_{4}^{\text{c}}\rVert_{\infty}) \lesssim& n(\bbeta^\top\bSigma_{\bet}^{2}\bbeta)^{1/2}\varrho\notag\\
	&+ \{n^{1/2}(\log p_{\bgamma})^{1/2} + \log p_{\bgamma}\log (np_{\bgamma})\}\bigl\{\bbeta^\top\bSigma_{\bet}\bbeta\lambda_{\rm{max}}(\bSigma_{\bX})\varrho^2 + \bbeta^\top\bSigma_{\bet}\bSigma_{\bX}\bSigma_{\bet}\bbeta\bigr\}^{1/2}\notag\\
	&+ \{\log p_{\bgamma} + n^{-1/2+1/q}(\log p_{\bgamma})^{3/2}\log (np_{\bgamma})\}(\bbeta^\top\bSigma_{\bet}\bbeta)^{1/2}\tr^{1/2}(\bSigma_{\bX}^{2})\notag\\
	\begin{split}
		\lesssim& n(\bbeta^\top\bSigma_{\bet}^{2}\bbeta)^{1/2}\varrho + n^{1/2}(\log p_{\bgamma})^{1/2}\bigl\{\bbeta^\top\bSigma_{\bet}\bbeta\lambda_{\rm{max}}(\bSigma_{\bX})\varrho^2 + \bbeta^\top\bSigma_{\bet}\bSigma_{\bX}\bSigma_{\bet}\bbeta\bigr\}^{1/2}\\
		&+ \log p_{\bgamma}(\bbeta^\top\bSigma_{\bet}\bbeta)^{1/2}\tr^{1/2}(\bSigma_{\bX}^{2}),
	\end{split}
\end{align}
where the first inequality holds by Lemma \ref{lemma26} and inequalities \eqref{ineq2ofproofoflimitpowerofguosum}-\eqref{ineq5ofproofoflimitpowerofguosum}. The last inequality is due to the fact $\log p_{\bgamma} = O(n^b)$, where $0<b<1/3$.
Combining equations \eqref{ineq5ofproofofguosum}, \eqref{ineq6ofproofofguosum} and Assumption \ref{assumptionb6}, we have
\begin{align*}
	A_{4}^{\text{c}} = O_p\biggl[n^{1/2}s(\log p_{\bgamma}\bbeta^\top\bSigma_{\bet}^{2}\bbeta)^{1/2}\varrho + s&\log p_{\bgamma}\bigl\{\bbeta^\top\bSigma_{\bet}\bbeta\lambda_{\rm{max}}(\bSigma_{\bX})\varrho^2 + \bbeta^\top\bSigma_{\bet}\bSigma_{\bX}\bSigma_{\bet}\bbeta\bigr\}^{1/2}\notag\\
	&+ n^{-1/2}s(\log p_{\bgamma})^{3/2}(\bbeta^\top\bSigma_{\bet}\bbeta)^{1/2}\tr^{1/2}(\bSigma_{\bX}^{2})\biggr].
\end{align*}
Thus $A_{4}^{\text{c}} = o_p(\sqrt{2\Lambda_{\bX}})$ by the conditions in Theorem \ref{limitnulldisofguosum} and the model structure under the local alternatives.
In summary, this theorem is proved.


\subsection{The proof of Theorem \ref{limitnulldisoforaclesum}}
As $\bgamma_{\phi} = \bgamma$ under $\bH_{0}$, we can decompose $M_{n}^{o}$ as
\begin{align*}
	M_{n}^{o} &= \frac{1}{n}\sum_{i\neq j}\hat{\epsilon}_{i}\hat{\epsilon}_{j}\bet_{i}^\top\bet_{j}\\
	&= \frac{1}{n}\sum_{i\neq j}(\epsilon_i + \breve{\bgamma}^\top\bZ_i)(\epsilon_{j} + \breve{\bgamma}^\top\bZ_j)\bet_{i}^\top\bet_{j}\\
	&= \frac{1}{n}\sum_{i\neq j}(
	\epsilon_{i}\epsilon_{j} + \breve{\bgamma}^\top\bZ_i\breve{\bgamma}^\top\bZ_j + \epsilon_{i}\breve{\bgamma}^\top\bZ_j + \breve{\bgamma}^\top\bZ_i\epsilon_{j})\bet_{i}^\top\bet_{j}\\
	&= \frac{1}{n}\sum_{i\neq j}\epsilon_{i}\epsilon_{j}\bet_{i}^\top\bet_{j} + \frac{1}{n}\sum_{i\neq j}\breve{\bgamma}^\top\bZ_i\breve{\bgamma}^\top\bZ_j\bet_{i}^\top\bet_{j}+ \frac{2}{n}\sum_{i\neq j}\epsilon_{i}\breve{\bgamma}^\top\bZ_j\bet_{i}^\top\bet_{j}\\
	&=:I_{1}^{o} + I_{2}^{o} + I_{3}^{o}.
\end{align*}

Similar to the arguments of proving Theorem~3 in \cite{guojrssb2016}, we can see
\[
\frac{I_1^{o}}{\sqrt{2\bLambda_{\bet}}}\rightarrow N(0,1)\quad \text{in distribution}
\]
as $(n,p_{\bbeta})\rightarrow\infty$. See Lemma \ref{lemma14} below for more details.
Thus to prove this theorem,  it suffices to  $I_{2}^{o}$ and $I_{3}^{o}$ are $o_p(\sqrt{2\bLambda_{\bet}})$.

Following Lemma \ref{lemma37} below, we derive that
%
%
\[
|I_{2}^{o}|\leq \biggl\lVert\frac{1}{n}\sum_{i\neq j}\bZ_i\bZ_j^\top\bet_i^\top\bet_{j}\biggr\rVert_{\infty} \lVert\bgamma - \hat\bgamma\rVert_{1}^2,
\]
where $\lVert n^{-1}\sum_{i\neq j}\bZ_i\bZ_j^\top\bet_i^\top\bet_{j}\rVert_{\infty} = \max_{1\leq k_1,k_2\leq p_{\bgamma}}\lvert n^{-1}\sum_{i\neq j} Z_{i,k_1}Z_{j,k_2}\bet_i^\top\bet_j\rvert$.
Denote $q = \lceil 4/(1-3b) \rceil$. Similar to the proof of inequality \eqref{ineq3ofproofoflimitnulldisofguosum} in the main text, it can be shown that
\begin{align*}
	\lVert\max_{1\leq i\neq j\leq n}\max_{1\leq k_1,k_2\leq p_{\bgamma}} Z_{i,k_1}Z_{j,k_2}\bet_{i}^\top\bet_{j}\rVert_{4} &\leq \lVert\max_{1\leq i\neq j\leq n}\max_{1\leq k_1,k_2\leq p_{\bgamma}} Z_{i,k_1}Z_{j,k_2}\bet_{i}^\top\bet_{j}\rVert_{q}\notag\\
	&\leq \lVert\max_{1\leq i\neq j\leq n}\max_{1\leq k_1,k_2\leq p_{\bgamma}}Z_{i,k_1}Z_{j,k_2}\rVert_{2q}\lVert\max_{1\leq i\neq j\leq n}\bet_{i}^\top\bet_{j}\rVert_{2q}\notag\\
	&\lesssim \log (np_{\bgamma})\max_{1\leq k_1,k_2\leq p_{\bgamma}}\lVert Z_{1,k_1}Z_{2,k_2}\rVert_{\psi_1}n^{1/q}\lVert\bet_{1}^\top\bet_{2}\rVert_{2q}\notag\\
	&\lesssim \log (np_{\bgamma})n^{1/q}\tr^{1/2}(\bSigma_{\bet}^2).
\end{align*}
Similar to proving the inequality \eqref{ineq5ofproofoflimitnulldisofguosum} in the main text and Theorem 5.1 in \cite{chen2018gaussian}, we derive
\begin{align*}
	\max_{1\leq k_1,k_2\leq p_{\bgamma}}\lVert Z_{1,k_1}Z_{2,k_2}\bet_1^\top\bet_2\rVert_{2}\leq \max_{1\leq k_1,k_2\leq p_{\bgamma}}\lVert Z_{1,k_1}Z_{2,k_2}\bet_1^\top\bet_2\rVert_{4} \lesssim \tr^{1/2}(\bSigma_{\bet}^2),
\end{align*}
and
\[
\biggl\lVert\frac{1}{n}\sum_{i\neq j}\bZ_i\bZ_j^\top\bet_i^\top\bet_{j}\biggr\rVert_{\infty}  = O_p\biggl[\bigl\{\log p_{\bgamma} + n^{1/q-1/2}(\log p)^{3/2}\log (np_{\bgamma})\bigr\}\tr^{1/2}(\bSigma_{\bet}^2)\biggr].
\]
It follows that $\lVert n^{-1}\sum_{i\neq j}\bZ_i\bZ_j^\top\bet_i^\top\bet_{j}\rVert_{\infty} = O_p\bigl(\log p_{\bgamma}\sqrt{\bLambda_{\bet}}\bigr)$ as $\tr^{1/2}(\bSigma_{\bet}^2)\lesssim \sqrt{\bLambda_{\bet}}$ and $\log p_{\bgamma} = O(n^b)$ for $0< b< 1/3$. Thus $\lvert I_{2}^{o}\rvert = o_p(\sqrt{2\bLambda_{\bet}})$ is proved when $s\log p_{\bgamma}/\sqrt{n} = o(1)$.

Similarly, the argument of proving $I_{3}^{c}$ yields
\[
\lvert I_3^{\text{o}}\rvert\leq \biggl\lVert\frac{1}{n}\sum_{i\neq j}(\epsilon_{i}\bZ_j + \bZ_{i}\epsilon_{j})\bet_{i}^\top\bet_{j}\biggr\rVert_{\infty}\lVert\breve{\bgamma}\rVert_{1}.
\]
Note that
\begin{align*}
	\lVert\max_{1\leq i\neq j\leq n}\max_{1\leq k\leq p_{\bgamma}}(\epsilon_{i}Z_{j,k} + Z_{i,k}\epsilon_{j})\bet_{i}^\top\bet_{j}\rVert_{4}	&\leq \lVert\max_{1\leq i\neq j\leq n}\max_{1\leq k\leq p_{\bgamma}}(\epsilon_{i}Z_{j,k} + Z_{i,k}\epsilon_{j})\bet_{i}^\top\bet_{j}\rVert_{q}\\
	&\leq \lVert\max_{1\leq i\neq j\leq n}\max_{1\leq k\leq p_{\bgamma}}\epsilon_{i}Z_{j,k}\rVert_{2q}\lVert\max_{1\leq i\neq j\leq n}\bet_{i}^\top\bet_{j}\rVert_{2q}\\
	&\lesssim \log (np_{\bgamma})n^{1/q}\tr^{1/2}(\bSigma_{\bet}^2),
\end{align*}
and
\[
\max_{1\leq k\leq p_{\bgamma}}\lVert (\epsilon_{i}Z_{j,k} + Z_{i,k}\epsilon_{j})\bet_{i}^\top\bet_{j}\rVert_{2}\leq\max_{1\leq k\leq p_{\bgamma}}\lVert (\epsilon_{i}Z_{j,k} + Z_{i,k}\epsilon_{j})\bet_{i}^\top\bet_{j}\rVert_{4}\lesssim\tr^{1/2}(\bSigma_{\bet}^2).
\]
Theorem 5.1 in \cite{chen2018gaussian} again shows
\[
\biggl\lVert\frac{1}{n}\sum_{i\neq j}(\epsilon_i\bZ_j + \bZ_{i}\epsilon_{j})\bet_i^\top\bet_{j}\biggr\rVert_{\infty}  = O_p\biggl[\bigl\{\log p_{\bgamma} + n^{1/q-1/2}(\log p_{\bgamma})^{3/2}\log (np_{\bgamma})\bigr\}\tr^{1/2}(\bSigma_{\bet}^2)\biggr].
\]
It follows that $\lVert n^{-1}\sum_{i\neq j}(\epsilon_i\bZ_j + \bZ_{i}\epsilon_{j})\bet_i^\top\bet_{j}\rVert_{\infty} = O_p(\log p_{\bgamma}\sqrt{\bLambda_{\bet}})$.
Thus $\lvert I_{3}^{o}\rvert = o_p(\sqrt{2\bLambda_{\bet}})$ is proved when Assumption \ref{assumptionb6} and $s(\log p_{\bgamma})^{3/2}/\sqrt{n} = o(1)$ hold.
In summary, the proof is finished.

%


\subsection{The proof of Theorem \ref{limitpoweroforaclesum}}

The proof of this theorem is similar to the proof of Theorem \ref{limitpowerofguosum}. Under the local alternatives, we decompose $M_{n}^{\text{o}}$ as:
\begin{align*}
	M_{n}^{\text{o}} =& \frac{1}{n}\sum_{i\neq j}\hat{\epsilon}_{i}\hat{\epsilon}_{j}\bet_{i}^\top\bet_{j}\\
	=& \frac{1}{n}\sum_{i\neq j}(\bbeta^\top\bet_{i} + \epsilon_{i} + \breve{\bgamma}^\top\bZ_{i})(\bbeta^\top\bet_{j} + \epsilon_{j} + \breve{\bgamma}^\top\bZ_{j})\bet_{i}^\top\bet_{j}\\
	=& \frac{1}{n}\sum_{i\neq j}(\bbeta^\top\bet_{i} + \epsilon_{i})(\bbeta^\top\bet_{j} + \epsilon_{j})\bet_{i}^\top\bet_{j}\\
	&+ \frac{1}{n}\sum_{i\neq j}\breve{\bgamma}^\top\bZ_{i}\breve{\bgamma}^\top\bZ_{j}\bet_{i}^\top\bet_{j}\\
	&+ \frac{1}{n}\sum_{i\neq j}(\epsilon_{i}\breve{\bgamma}^\top\bZ_{j} + \breve{\bgamma}^\top\bZ_{i}\epsilon_{j})\bet_{i}^\top\bet_{j}\\
	&+ \frac{1}{n}\sum_{i\neq j}(\bbeta^\top\bet_{i}\breve{\bgamma}^\top\bZ_{j} + \breve{\bgamma}^\top\bZ_{i}\bbeta^\top\bet_{j})\bet_{i}^\top\bet_{j}\\
	=:& A_{1}^{\text{o}} + A_{2}^{\text{o}} + A_{3}^{\text{o}} + A_{4}^{\text{o}}.
\end{align*}
Rewrite the first term $A_{1}^{\text{o}}$ as
\begin{align*}
	A_{1}^{\text{o}} =&\frac{1}{n}\sum_{i\neq j}\epsilon_{i}\epsilon_{j}\bet_{i}^\top\bet_{j} + \frac{1}{n}\sum_{i\neq j}\bbeta^\top\bet_{i}\bet_{i}^\top\bet_{j}\bet_{j}^\top\bbeta\\
	&+ \frac{1}{n}\sum_{i\neq j}(\bbeta^\top\bet_{i}\epsilon_{j} + \epsilon_{i}\bbeta^\top\bet_{j})\bet_{i}^\top\bet_{j}\\
	=:& A_{11}^{\text{o}} + A_{12}^{\text{o}} + A_{13}^{\text{o}}.
\end{align*}
Following Lemma \ref{lemma14} below, it can be seen
\[
\frac{A_{11}^{\text{o}}}{\sqrt{2\bLambda_{\bet}}}\rightarrow N(0,1)\quad \text{in distribution}
\]
as $(n,p_{\bbeta})\rightarrow\infty$. 
Note that $\mbE(A_{12}^{\text{o}}) = (n-1)\bbeta^\top\bSigma_{\bet}^{2}\bbeta$. The Hoeffding decomposition yields
\[
\frac{1}{n-1}A_{12}^{\text{o}} = \bbeta^\top\bSigma_{\bet}^{2}\bbeta + 2A_{121}^{\text{o}} + A_{122}^{\text{o}},
\]
where
\[
A_{121}^{\text{o}} = \frac{1}{n}\sum_{i=1}^{n}\bbeta^\top\bSigma_{\bet}(\bet_{i}\bet_{i}^\top\bbeta - \bSigma_{\bet}\bbeta),
\]
and
\[
A_{122}^{\text{o}} = \frac{1}{n(n-1)}\sum_{i\neq j}(\bet_{i}\bet_{i}^\top\bbeta - \bSigma_{\bet}\bbeta)^\top(\bet_{j}\bet_{j}^\top\bbeta - \bSigma_{\bet}\bbeta).
\]
The variance of $A_{12}^{\text{o}}$ can be bounded by
\begin{align*}
	\Var(A_{12}^{\text{o}}) &\lesssim n^2\mbE(A_{121}^{\text{o}})^2 + n^2\mbE(A_{122}^{\text{o}})^2\\
	&\lesssim n\mbE(\bbeta^\top\bSigma_{\bet}\bet_{1}\bet_{1}^\top\bbeta)^2 + \mbE(\bbeta^\top\bet_{1}\bet_{1}^\top\bet_{2}\bet_{2}^\top\bbeta)^2\\
	&= o(\bLambda_{\bet}).
\end{align*}
The last equality holds by equations \eqref{ineq1oflemma33} and \eqref{ineq2oflemma33} in Lemma \ref{lemma33}, the fact that $(\bbeta^\top\bSigma_{\bet}^{2}\bbeta)^2\leq \bbeta^\top\bSigma_{\bet}\bbeta\bbeta^\top\bSigma_{\bet}^{3}\bbeta$.
Similarly, we derive $\mbE(A_{13}^{\text{o}}) = 0$ and
\[
\frac{1}{n-1}A_{13}^{\text{o}} = 2A_{131}^{\text{o}} + A_{132}^{\text{o}},
\]
where
\[
A_{131}^{\text{o}} = \frac{1}{n}\sum_{i=1}^{n}\bbeta^\top\bSigma_{\bet}\epsilon_{i}\bet_{i}
\]
and
\[
A_{132}^{\text{o}} = \frac{1}{n(n-1)}\sum_{i\neq j}\bigl\{(\bbeta^\top\bet_{i}\bet_{i}^\top - \bbeta^\top\bSigma_{\bet})\epsilon_{j}\bet_{j} + \epsilon_{i}\bet_{i}^\top(\bet_{j}\bet_{j}^\top\bbeta - \bSigma_{\bet}\bbeta)\bigr\}.
\]
Similar to the derivation of $\Var(A_{12}^{\text{o}})$, we derive
\begin{align*}
	\Var(A_{13}^{\text{o}}) &\lesssim n^2\mbE(A_{131}^{\text{o}})^2 + n^2\mbE(A_{132}^{\text{o}})^2\\
	&\lesssim n\bbeta^\top\bSigma_{\bet}\mbE(\epsilon_{1}^2\bet_{1}\bet_{1}^\top)\bSigma_{\bet}\bbeta + \mbE(\epsilon_{2}\bbeta^\top\bet_{1}\bet_{1}^\top\bet_{2})^2\\
	&\lesssim n\bbeta^\top\bSigma_{\bet}^{3}\bbeta + \tr(\bSigma_{\bet}^{2})\bbeta^\top\bSigma_{\bet}\bbeta\\
	&= o(\bLambda_{\bet}),
\end{align*}
where the equality holds by the model structure under the local alternatives, and the third inequality holds by \eqref{ineq3oflemma33} in Lemma \ref{lemma33}, Assumption \ref{assumptionb7} and the fact that $\epsilon$ is independent of $\bnu$.

Altogether, we have
\[
\frac{A_{1}^{\text{o}} - n\bbeta^\top\bSigma_{\bet}^{2}\bbeta}{\sqrt{2\bLambda_{\bet}}}\rightarrow N(0,1)\quad \text{in distribution.}
\]
To conclude the proof, we now prove $A_{i}^{\text{o}} = o_p(\sqrt{2\bLambda_{\bet}}),\,i=2,3,4$. By the same arguments for handling  $I_{2}^{\text{o}}$ and $I_{3}^{\text{o}}$ in Theorem \ref{limitnulldisoforaclesum}, we can show
\[
A_{2}^{\text{o}} = o_p(\sqrt{2\bLambda_{\bet}})\quad \text{and}\quad A_{3}^{\text{o}} = o_p(\sqrt{2\bLambda_{\bet}}).
\]
For $A_{4}^{\text{o}} = o_p(\sqrt{2\bLambda_{\bet}})$, we recall that
\[
A_{4}^{\text{o}} = \frac{1}{n}\sum_{i\neq j}(\bbeta^\top\bet_{i}\breve{\bgamma}^\top\bZ_{j} + \breve{\bgamma}^\top\bZ_{i}\bbeta^\top\bet_{j})\bet_{i}^\top\bet_{j}.
\]
By the H{\" o}lder inequality, we derive
\begin{align}
	\label{ineq5ofproofoforaclesum}
	\lvert A_{4}^{\text{o}}\rvert \leq \biggl\lVert \frac{1}{n}\sum_{i\neq j}
	(\bbeta^\top\bet_{i}\bZ_{j} + \bZ_{i}\bbeta^\top\bet_{j})\bet_{i}^\top\bet_{j}
	\biggr\rVert_{\infty}\lVert\breve{\bgamma}\rVert_{1}.
\end{align}
Let $\bU_4^{\text{o}}$ be a $p_{\bgamma}$ dimension vector with $k$-th element
\[
U_{4,k}^{\text{o}} =  \frac{1}{n}\sum_{i\neq j}(\bbeta^\top\bet_{i}Z_{j,k} + Z_{i,k}\bbeta^\top\bet_{j})\bet_{i}^\top\bet_{j}.
\]
Note that $\bU_{4}^{\text{o}} = n^{-1}\sum_{i\neq j}(\bbeta^\top\bet_{i}\bZ_{j} + \bZ_{i}\bbeta^\top\bet_{j})\bet_{i}^\top\bet_{j}$ and all of its elements are $U$-statistics. The Hoeffding decomposition yields
\[
\frac{1}{n-1} U_{4,k}^{\text{o}} = 2S_{4,1,k}^{\text{o}} + S_{4,2,k}^{\text{o}},
\]
where
\[
S_{4,1,k}^{\text{o}} = \frac{1}{n}\sum_{i=1}^{n}g_{4,1,k}^{\text{o}}(\bnu_{i})
\]
with $g_{4,1,k}^{\text{o}}(\bnu_{1}) = \bbeta^\top\bSigma_{\bet}Z_{1,k}\bet_{1}$,
\[
S_{4,2,k}^{\text{o}} = \frac{1}{n(n-1)}\sum_{i\neq j}^{n}g_{4,2,k}^{\text{o}}(\bnu_i;\bnu_j)
\]
with $g_{4,2,k}^{\text{o}}(\bnu_1;\bnu_2) = (\bet_{1}\bet_{1}^\top\bbeta - \bSigma_{\bet}\bbeta)^\top Z_{2,k}\bet_{2} + Z_{1,k}\bet_{1}^\top(\bet_{2}\bet_{2}^\top\bbeta - \bSigma_{\bet}\bbeta)$.

For any $1\leq k\leq p_{\bgamma}$, according to Assumption \ref{assumptionb5}, $Z_{1,k}$ is sub-Gaussian with bounded norm, and $\bbeta^\top\bSigma_{\bet}\bet_{1}$ is sub-Gaussian with norm $\lVert\bbeta^\top\bSigma_{\bet}\bet_{1}\rVert_{\psi_2}\leq (\bbeta^\top\bSigma_{\bet}^{3}\bbeta)^{1/2}$. Then  $g_{4,1,k}^{\text{o}}(\bnu_{1})$ is sub-Exponential with norm
\[
\lVert g_{4,1,k}^{\text{o}}\rVert_{\psi_1}\leq \lVert Z_{1,k}\rVert_{\psi_2}\lVert\bbeta^\top\bSigma_{\bet}\bet_{1}\rVert_{\psi_2}\lesssim (\bbeta^\top\bSigma_{\bet}^{3}\bbeta)^{1/2}.
\]
Therefore
\begin{align*}
	\lVert\max_{1\leq i\leq n}\max_{1\leq k\leq p_{\bgamma}}g_{4,1,k}^{\text{o}}\rVert_{2} &\lesssim \log (np_{\bgamma})\max_{1\leq i\leq n}\max_{1\leq k\leq p_{\bgamma}}\lVert g_{4,1,k}^{\text{o}}\rVert_{\psi_1}\\
	&\lesssim \log (np_{\bgamma})(\bbeta^\top\bSigma_{\bet}^{3}\bbeta)^{1/2}.
\end{align*}
Similarly, we derive $\bbeta^\top\bet_{1}Z_{2,k}$ is sub-Exponential with norm $\lVert\bbeta^\top\bX_{1}Z_{2,k}\rVert_{\psi_1}\lesssim \bbeta^\top\bSigma_{\bet}\bbeta$. Write $q = \lceil 4/(1-3b)\rceil$. We have
\begin{align*}
	\lVert \max_{1\leq i\neq j\leq n}\max_{1\leq k\leq p_{\bgamma}}g_{4,2,k}^{\text{o}}\rVert_{4}&\leq \lVert \max_{1\leq i\neq j\leq n}\max_{1\leq k\leq p_{\bgamma}}g_{4,2,k}^{\text{o}}\rVert_{q}\\
	&\leq \lVert\max_{1\leq i\neq j\leq n}\max_{1\leq k\leq p_{\bgamma}}\bbeta^\top\bet_{i}Z_{j,k}\bet_{i}^\top\bet_{j}\rVert_{q}\\
	&\leq \lVert\max_{1\leq i\neq j\leq n}\max_{1\leq k\leq p_{\bgamma}}\bbeta^\top\bet_{i}Z_{j,k}\rVert_{2q}\lVert\max_{1\leq i\neq j\leq n}\bet_{i}^\top\bet_{j}\rVert_{2q}\\
	&\lesssim \log (np_{\bgamma})(\bbeta^\top\bSigma_{\bet}\bbeta)^{1/2} n^{1/q}\tr^{1/2}(\bSigma_{\bet}^2).
\end{align*}
Similarly,
\[
\max_{1\leq k\leq p_{\bgamma}}\lVert g_{4,1,k}^{\text{o}}\rVert_{2}\lesssim (\bbeta^\top\bSigma_{\bet}^{3}\bbeta)^{1/2}
\]
and
\[
\max_{1\leq k\leq p_{\bgamma}}\lVert g_{4,2,k}^{\text{o}}\rVert_{2}\leq \max_{1\leq k\leq p_{\bgamma}}\lVert g_{4,2,k}^{\text{o}}\rVert_{4}\lesssim (\bbeta^\top\bSigma_{\bet}\bbeta)^{1/2}\tr^{1/2}(\bSigma_{\bet}^2).
\]
Lemma \ref{lemma26} and Assumption \ref{assumptionb8} imply
\begin{align}
	\label{ineq6ofproofoforaclesum}
	\mbE(\lVert\bU_{4}^{\text{o}}\rVert_{\infty}) \lesssim n^{1/2}(\log p_{\bgamma})^{1/2}(\bbeta^\top\bSigma_{\bet}^{3}\bbeta)^{1/2} + \log p_{\bgamma}(\bbeta^\top\bSigma_{\bet}\bbeta)^{1/2}\tr^{1/2}(\bSigma_{\bet}^{2}).
\end{align}
Combining equations \eqref{ineq5ofproofoforaclesum} and \eqref{ineq6ofproofoforaclesum}, Assumption \ref{assumptionb6}, we have
\begin{align*}
	A_{4}^{\text{o}} = O_p\biggl(s\log p_{\bgamma}(\bbeta^\top\bSigma_{\bet}^{3}\bbeta)^{1/2} + n^{-1/2}s(\log p_{\bgamma})^{3/2}(\bbeta^\top\bSigma_{\bet}\bbeta)^{1/2}\tr^{1/2}(\bSigma_{\bet}^{2})\biggr).
\end{align*}
Thus $A_4^{\text{o}} = o_p(\sqrt{2\bLambda_{\bet}})$ holds by the conditions in Theorem \ref{limitnulldisoforaclesum} and the model structure under the local alternatives.
Altogether, the proof is done.


\subsection{The proof of Theorem \ref{limitnulldisofsum}}

Denote $\mA = \bigl\{\lVert\breve{\bW}\rVert_{F}^2 \lesssim s^\prime\log p_{\bgamma}/n^\prime\bigr\}$. According to Assumption \ref{assumption8},
$\rmPr\bigl(\mA^{c}\bigr) = o(1)$.
Thus it suffices to prove
\[
\frac{M_{n}}{\sqrt{2\bLambda_{\bet}}}\rightarrow N(0,1)\quad \text{in distribution}
\]
as $(n,p_{\bbeta},p_{\bgamma})\rightarrow\infty$ conditional on $\mA$. For brevity, in the rest of proof we abbreviate $\mbE(\cdot\vert\mA),\,\rmPr(\cdot\vert\mA)$ as $\mbE(\cdot),\,\rmPr(\cdot)$ respectively.
As $\bgamma_{\phi} = \bgamma$ under $\bH_{0}$, we can decompose $M_{n}$ as
\begin{align*}
	M_{n} &= \frac{1}{n}\sum_{i\neq j}\hat{\epsilon}_{i}\hat{\epsilon}_j\hat\bet_i^\top\hat\bet_j\\
	&= \frac{1}{n}\sum_{i\neq j}(\epsilon_{i} + \breve{\bgamma}^\top\bZ_i)(\epsilon_{j} + \breve{\bgamma}^\top\bZ_j)\hat\bet_i^\top\hat\bet_j\\
	&= \frac{1}{n}\sum_{i\neq j}\epsilon_{i}\epsilon_{j}\hat\bet_i^\top\hat\bet_j + \frac{1}{n}\sum_{i\neq j}\breve{\bgamma}^\top\bZ_i\breve{\bgamma}^\top\bZ_j\hat\bet_i^\top\hat\bet_j\\
	&\quad\,\,+\frac{1}{n}\sum_{i\neq j}(\epsilon_{i}\breve{\bgamma}^\top\bZ_j + \breve{\bgamma}^\top\bZ_i\epsilon_{j})\hat\bet_i^\top\hat\bet_j\\
	&=: I_{1} + I_{2} + I_{3}.
\end{align*}
Similar to the proof of Theorem 3 in \cite{guojrssb2016}, it can be seen
\[
\frac{I_{1}}{\sqrt{2\bLambda_{\hat\bet}}}\rightarrow N(0,1)\quad \text{in distribution}
\]
as $(n,p_{\bbeta})\rightarrow\infty$, where $\bLambda_{\hat\bet}=\sigma^4 \mathrm{tr}(\bSigma_{\hat\bet}^2)$. See Lemma \ref{lemma14} for more details. As the inequality \eqref{ineq1oflemma34} in Lemma \ref{lemma34} implies $\bLambda_{\hat\bet}/\bLambda_{\bet}\rightarrow 1$ in probability, we  can then use Sluskty theorem to derive
\[
\frac{I_{1}}{\sqrt{2\bLambda_{\bet}}}\rightarrow N(0,1)\quad \text{in distribution}
\]
as $(n,p_{\bbeta},p_{\bgamma})\rightarrow\infty$.
Further, similarly, we prove $I_{2}$ and $I_{3}$ are $o_p(\sqrt{2\bLambda_{\hat\bet}})$ to finish the proof. As the proof is very similar, we give a sketch below.

Lemma \ref{lemma37} below yields
\begin{align}
	\label{ineq1ofproofoflimitnulldisofsum}
	I_{2}\leq \biggl\lVert\frac{1}{n}\sum_{i\neq j}\bZ_i\bZ_j^\top\hat\bet_i^\top\hat\bet_j\biggr\rVert_{\infty}\lVert\breve{\bgamma}\rVert_{1}^{2}.
\end{align}
Let $\bU_{1} = n^{-1}\sum_{i\neq j}\bZ_i\bZ_j^\top\hat\bet_i^\top\hat\bet_j$ and all of its elements are $U$-statistics: for $(k_1,k_2)$-th element,
\[
U_{1,(k_1,k_2)} =  \frac{1}{n}\sum_{i\neq j}\frac{1}{2}(Z_{i,k_1}Z_{j,k_2} + Z_{i,k_2}Z_{j,k_1})\hat\bet_i^\top\hat\bet_j.
\]
The Hoeffding decomposition implies
\[
\frac{1}{n-1} U_{1,(k_1,k_2)} = \mbE(Z_{k_1}\hat\bet)^\top\mbE(Z_{k_2}\hat\bet) + 2S_{1,1,(k_1,k_2)} + S_{1,2,(k_1,k_2)},
\]
where
\[
S_{1,1,(k_1,k_2)} = \frac{1}{n}\sum_{i=1}^{n}g_{1,1,(k_1,k_2)}(\bnu_{i})
\]
with $g_{1,1,(k_1,k_2)}(\bnu_{1}) = \{Z_{1,k_1}\hat\bet_1 - \mbE(Z_{k_1}\hat\bet)\}^\top\mbE(Z_{k_2}\hat\bet)/2 + \{Z_{1,k_2}\hat\bet_1 - \mbE(Z_{k_2}\hat\bet)\}^\top\mbE(Z_{k_1}\hat\bet)/2$,
\[
\bS_{1,2,(k_1,k_2)} = \frac{1}{n(n-1)}\sum_{i\neq j}^{n}g_{1,2,(k_1,k_2)}(\bnu_i;\bnu_j)
\]
with $g_{1,2,(k_1,k_2)}(\bnu_1;\bnu_2) = \{Z_{1,k_1}\hat\bet_1 - \mbE(Z_{k_1}\hat\bet)\}^\top\{Z_{2,k_2}\hat\bet_2 - \mbE(Z_{k_2}\hat\bet)\}/2 + \{Z_{1,k_2}\hat\bet_1 - \mbE(Z_{k_2}\hat\bet)\}^\top\{Z_{2,k_1}\hat\bet_2 - \mbE(Z_{k_1}\hat\bet)\}/2$.

Similar to the derivation of $I_{2}^{\text{c}}$ in the proof of Theorem \ref{limitnulldisofguosum}, we derive
\begin{align*}
	\lVert\max_{1\leq i\leq n}\max_{1\leq k_1,k_2\leq p_{\bgamma}}g_{1,1,(k_1,k_2)}\rVert_{2}&\lesssim \log (np_{\bgamma})\lambda_{\rm{max}}^{1/2}(\bSigma_{\hat\bet})\bigl\{\max_{1\leq k\leq p_{\bgamma}}\lVert\mbE(Z_{k}\hat\bet)\rVert_{2}^{2}\bigr\}^{1/2}\\
	\lVert\max_{1\leq i\neq j\leq n}\max_{1\leq k_1,k_2\leq p_{\bgamma}}g_{1,2,(k_1,k_2)}\rVert_{4}&\lesssim \log (np_{\bgamma})n^{1/q}\tr^{1/2}(\bSigma_{\hat\bet}^2)\\
	\max_{1\leq k_1,k_2\leq p_{\bgamma}}\lVert g_{1,1,(k_1,k_2)}\rVert_{2}&\lesssim \lambda_{\rm{max}}^{1/2}(\bSigma_{\hat\bet})\bigl\{\max_{1\leq k\leq p_{\bgamma}}\lVert\mbE(Z_{k}\hat\bet)\rVert_{2}^{2}\bigr\}^{1/2}\\
	\max_{1\leq k_1,k_2\leq p_{\bgamma}}\lVert g_{1,2,(k_1,k_2)}\rVert_{2}&\leq \max_{1\leq k_1,k_2\leq p_{\bgamma}}\lVert g_{1,2,(k_1,k_2)}\rVert_{4}\lesssim \tr^{1/2}(\bSigma_{\hat\bet}^2).
\end{align*}
Lemma \ref{lemma26}, \ref{lemma34} and Assumption \ref{assumptionb8} yield
\begin{align}
	\label{ineq2ofproofoflimitnulldisofsum}
	\mbE(\lVert\bU_{1}\rVert_{\infty}) \lesssim n\frac{s^\prime\log p_{\bgamma}}{n^\prime}\varphi^2 + n^{1/2}(\log p_{\bgamma})^{1/2}&\biggl\{\lambda_{\max}^{1/2}(\bSigma_{\bet})\biggl(\frac{s^\prime\log p_{\bgamma}}{n^\prime}\biggr)^{1/2} \vee \lambda_{\max}^{1/2}(\bSigma_{\bZ})\frac{s^\prime\log p_{\bgamma}}{n^\prime}\biggr\}\varphi\notag\\
	&+ \log p_{\bgamma}\tr^{1/2}(\bSigma_{\hat\bet}^2).
\end{align}
Together equations \eqref{ineq1ofproofoflimitnulldisofsum} with \eqref{ineq2ofproofoflimitnulldisofsum} and Assumption \ref{assumptionb6}, we have
\begin{align}
	\label{eq1ofproofoflimitnulldisofsum}
	I_2 = O_p\biggl[s^2\log p_{\bgamma}\frac{s^\prime\log p_{\bgamma}}{n^\prime}\varphi^2 + n^{-1/2}s^2(\log p_{\bgamma})^{3/2}&\biggl\{\lambda_{\max}^{1/2}(\bSigma_{\bet})\biggl(\frac{s^\prime\log p_{\bgamma}}{n^\prime}\biggr)^{1/2} \vee \lambda_{\max}^{1/2}(\bSigma_{\bZ})\frac{s^\prime\log p_{\bgamma}}{n^\prime}\biggr\}\varphi\notag\\
	&+ n^{-1}s^2(\log p_{\bgamma})^{2}\tr^{1/2}(\bSigma_{\hat\bet}^2)\biggr].
\end{align}
Thus $I_2 = o_p(\sqrt{\bLambda_{\hat\bet}})$ when $s^2\log p_{\bgamma}\frac{s^\prime\log p_{\bgamma}}{n^\prime}\varphi^2 = o(\sqrt{\bLambda_{\hat\bet}})$,
\[
n^{-1/2}s^{2}(\log p_{\bgamma})^{3/2}\biggl\{\lambda_{\max}^{1/2}(\bSigma_{\bet})\biggl(\frac{s^\prime\log p_{\bgamma}}{n^\prime}\biggr)^{1/2} \vee \lambda_{\max}^{1/2}(\bSigma_{\bZ})\frac{s^\prime\log p_{\bgamma}}{n^\prime}\biggr\}\varphi = o(\sqrt{\bLambda_{\hat\bet}})
\]
and $n^{-1}s^2(\log p_{\bgamma})^{2} = o(1)$ hold simultaneously.

Similar to the proof of $I_{2}$,
\begin{align}
	\label{ineq3ofproofoflimitnulldisofsum}
	I_3\leq \biggl\lVert\frac{1}{n}\sum_{i\neq j}(\epsilon_{i}\bZ_j + \bZ_i\epsilon_{j})\hat\bet_i^\top\hat\bet_j\biggr\rVert_{\infty}\lVert\breve{\bgamma}\rVert_{1}=:\lVert \bU_2 \rVert_{\infty}\lVert\breve{\bgamma}\rVert_{1}.
\end{align}
Applying the Hoeffding decomposition to every element of $\bU_2$, we have
\[
\frac{1}{n-1} U_{2,k} = 2S_{2,1,k} + S_{2,2,k},
\]
where
\[
S_{2,1,k} = \frac{1}{n}\sum_{i=1}^{n}g_{2,1,k}(\bnu_i)
\]
with $g_{2,1,k}(\bnu_1) = \epsilon_{1}\hat\bet_1^\top\mbE(Z_{k}\hat\bet)$,
\[
S_{2,2,k} = \frac{1}{n(n-1)}\sum_{i\neq j}^{n}g_{2,2,k}(\bnu_i,\bnu_j)
\]
with $g_{2,2,k}(\bnu_1;\bnu_2) = \epsilon_{1}\hat\bet_1^\top\{Z_{2,k}\hat\bet_2 - \mbE(Z_{k}\hat\bet)\} + \epsilon_{2}\hat\bet_2^\top\{Z_{1,k}\hat\bet_1 - \mbE(Z_{k}\hat\bet)\}$.

Similar to the proof for $I_{3}^{\text{c}}$ in the proof of Theorem \ref{limitnulldisofguosum}, and for the inequality \eqref{ineq2ofproofoflimitnulldisofsum}, we can then have
\begin{align}
	\label{ineq4ofproofoflimitnulldisofsum}
	\mbE(\lVert\bU_{2}\rVert_{\infty}) \lesssim n^{1/2}(\log p_{\bgamma})^{1/2}\biggl\{\lambda_{\max}^{1/2}(\bSigma_{\bet})\biggl(\frac{s^\prime\log p_{\bgamma}}{n^\prime}\biggr)^{1/2} \vee \lambda_{\max}^{1/2}(\bSigma_{\bZ})\frac{s^\prime\log p_{\bgamma}}{n^\prime}\biggr\}\varphi + \log p_{\bgamma}\tr^{1/2}(\bSigma_{\hat\bet}^2).
\end{align}
Combining equations \eqref{ineq3ofproofoflimitnulldisofsum} and \eqref{ineq4ofproofoflimitnulldisofsum}, Assumption \ref{assumptionb6}, we have
\begin{align}
	\label{eq2ofproofoflimitnulldisofsum}
	I_3 = O_p\biggl[s\log p_{\bgamma}\biggl\{\lambda_{\max}^{1/2}(\bSigma_{\bet})\biggl(\frac{s^\prime\log p_{\bgamma}}{n^\prime}\biggr)^{1/2} \vee \lambda_{\max}^{1/2}(\bSigma_{\bZ})\frac{s^\prime\log p_{\bgamma}}{n^\prime}\biggr\}\varphi + n^{-1/2}s(\log p_{\bgamma})^{3/2}\tr^{1/2}(\bSigma_{\hat\bet}^2)\biggr].
\end{align}
Thus $I_3 = o_p(\sqrt{\bLambda_{\hat\bet}})$ when
\[
s\log p_{\bgamma}\biggl\{\lambda_{\max}^{1/2}(\bSigma_{\bet})\biggl(\frac{s^\prime\log p_{\bgamma}}{n^\prime}\biggr)^{1/2} \vee \lambda_{\max}^{1/2}(\bSigma_{\bZ})\frac{s^\prime\log p_{\bgamma}}{n^\prime}\biggr\}\varphi = o(\sqrt{\bLambda_{\hat\bet}})
\]
and $n^{-1/2}s(\log p_{\bgamma})^{3/2} = o(1)$ hold simultaneously.
The proof is concluded.


\subsection{The proof of Theorem \ref{limitpoweroftruesum}}

As we can follow almost the same lines of proving Theorem \ref{limitnulldisofsum} to have,
for $\bbeta\in\mathscr{L}(\bbeta)$,
\[
\frac{M_{n} - n\bbeta^\top\bSigma_{\bet}^{2}\bbeta}{\sqrt{2\bLambda_{\bet}}}\rightarrow N(0,1)\quad \text{in distribution}
\]
as $(n,p_{\bbeta},p_{\bgamma})\rightarrow\infty$ conditional on $\mA$, we then omit the details here. 

\vspace{3mm}

\section{Some useful lemmas}
\label{someusefullemmas}

\begin{lemma}
	\label{lemma14}
	Let $\bxi = \bA\bnu$, where $\bA$ is a $p_{\bxi}\times m$ matrix with $p_{\bxi}\leq m$ and $\bSigma_{\bxi} = \bA\bA^\top$. Let $\bnu$ be a $m-$dimensional sub-Gaussian random vector with mean zero and identity covariance matrix. $\epsilon$ is a zero mean sub-Gaussian random variable with variance $\sigma^2$ and $\epsilon$ is independent of $\bxi$. Assume that $\tr(\bSigma^4_{\bxi})=o(\tr^2(\bSigma_{\bxi}^2))$ and $\tr(\bSigma_{\bxi}^2)\rightarrow\infty$ as $(n,p_{\bxi})\rightarrow\infty$. Then
	\[
	\frac{\sum_{i\neq j}\epsilon_{i}\epsilon_{j}\bxi_{i}^\top\bxi_{j}}{n\sqrt{2\Lambda_{\bxi}}}\rightarrow N(0,1)\quad \text{in distribution}
	\]
	as $(n,p_{\bxi})\rightarrow\infty$, where $\Lambda_{\bxi} = \sigma^4\tr(\bSigma_{\bxi}^{2})$. 	
\end{lemma}
\begin{proof}
	Denote
	\[
	R_{ni} = 2n^{-1}\sum_{j=1}^{i-1}\epsilon_{i}\epsilon_{j}\bxi_{i}^\top\bxi_{j},
	\]
	$S_{nk} = \sum_{i=2}^{k}R_{ni}$ and $\mathscr{F}_{k} = \sigma\{(\bxi_{i},\epsilon_{i}),\,i=1,2,\ldots,k\}$. Obviously,  $\mbE(R_{nk}\vert \mathscr{F}_{k-1}) = 0$ as $(S_{nk},\mathscr{F}_{k})$ is a zero-mean martingale sequence. Denote $v_{ni} = \Var(R_{ni}\vert\mathscr{F}_{i-1})$ and $V_{n} = \sum_{i=2}^{n}v_{ni}$. To apply the martingale central limit theorem, it suffices to show  the following two conditions:
	\begin{align}
		\label{eq1oflemma14}
		\frac{V_{n}}{\Var(S_{nn})}\rightarrow 1\quad \text{in probability}
	\end{align}
	and
	\begin{align}
		\label{eq2oflemma14}
		\frac{\sum_{i=2}^{n}\mbE\bigl[R_{ni}^{2}I\{\lvert R_{ni}\rvert/\sqrt{\tr(\bSigma_{\bxi}^{2})}>\eta\}\vert \mathscr{F}_{i-1}\bigr]}{\tr(\bSigma_{\bxi}^{2})}\rightarrow0\quad\text{in probability}
	\end{align}
	for any $\eta>0$.
	To prove equation \eqref{eq1oflemma14} first. Observe that
	\begin{align*}
		v_{ni} = 4\sigma^2n^{-2}\sum_{j=1}^{i-1}\epsilon_{j}^{2}\bxi_{j}^\top\bSigma_{\bxi}\bxi_{j} + 4\sigma^2n^{-2}\sum_{1\leq j_1\neq j_2\leq i-1}\epsilon_{j_1}\epsilon_{j_2}\bxi_{j_1}^\top\bSigma_{\bxi}\bxi_{j_2},
	\end{align*}
	and
	\begin{align*}
		V_n = 4\sigma^2n^{-2}\sum_{i=2}^{n}\biggl(\sum_{j=1}^{i-1}\epsilon_{j}^{2}\bxi_{j}^\top\bSigma_{\bxi}\bxi_{j} + \sum_{1\leq j_1\neq j_2\leq i-1}\epsilon_{j_1}\epsilon_{j_2}\bxi_{j_1}^\top\bSigma_{\bxi}\bxi_{j_2}\biggr).
	\end{align*}
	Since $\Var(S_{nn}) = 2n^{-1}(n-1)\sigma^4\tr(\bSigma_{\bxi}^{2})$, we have
	\begin{align*}
		\frac{V_{n}}{\Var(S_{nn})} &= \frac{2}{n(n-1)\sigma^2\tr(\bSigma_{\bxi}^{2})}\biggl(\sum_{i=2}^{n}\sum_{j=1}^{i-1}\epsilon_{j}^{2}\bxi_{j}^\top\bSigma_{\bxi}\bxi_{j} + \sum_{i=2}^{n}\sum_{1\leq j_1\neq j_2\leq i-1}\epsilon_{j_1}\epsilon_{j_2}\bxi_{j_1}^\top\bSigma_{\bxi}\bxi_{j_2}\biggr)\\
		&=: G_{1} + G_{2},
	\end{align*}
	where $\mbE(G_{1}) = 1$ and $\mbE(G_{2}) = 0$. Observe that
	\begin{align*}
		\Var(G_{1}) &\lesssim \frac{1}{n^{4}\tr^2(\bSigma_{\bxi}^{2})}\sum_{j=1}^{n-1}(n-j)^2\bigl\{\mbE(\bxi_{j}^\top\bSigma_{\bxi}\bxi_{j})^2 - \tr^2(\bSigma_{\bxi}^{2})\bigr\}\\
		&\lesssim \frac{1}{n^{4}}\sum_{j=1}^{n-1}(n-j)^2
		\lesssim n^{-1},
	\end{align*}
	where the second inequality holds by using Lemma \ref{lemma31} to derive
	\[
	\mbE(\bxi_{1}^\top\bSigma_{\bxi}\bxi_{1})^2\lesssim \tr^2(\bSigma_{\bxi}^2).
	\]
	Similarly, we derive that
	\begin{align*}
		\Var(G_2) &\lesssim \frac{1}{n^4\tr^2(\bSigma_{\bxi}^2)}\sum_{j_1<k_1}\sum_{j_2<k_2}(n-k_1)(n-k_2)\mbE\bigl(\epsilon_{j_1}\epsilon_{j_2}\epsilon_{k_1}\epsilon_{k_2}\bxi_{j_1}^\top\bSigma_{\bxi}\bxi_{k_1}\bxi_{j_2}^\top\bSigma_{\bxi}\bxi_{k_2}\bigr)\\
		&= \frac{\sum_{k=1}^{n}(n-k)^2(k-1)}{n^4}\cdot\frac{\tr(\bSigma_{\bxi}^{4})}{\tr^2(\bSigma_{\bxi}^{2})} = o(1)
	\end{align*}
	where the last equality holds by condition $\tr(\bSigma_{\bxi}^{4}) = o(\tr^2(\bSigma_{\bxi}^{2}))$. The Markov inequality yields $G_{1}\rightarrow 1$ in probability and $G_{2}\rightarrow 0$ in probability. Thus equation \eqref{eq1oflemma14} is proved.
	
	To handle equation \eqref{eq2oflemma14}. Notice that for any $\eta>0$,
	\begin{align}
		\label{ineq1oflemma14}
		\frac{\sum_{i=2}^{n}\mbE\bigl[R_{ni}^{2}I\{\lvert R_{ni}\rvert/\sqrt{\tr(\bSigma_{\bxi}^{2})}>\eta\}\vert \mathscr{F}_{i-1}\bigr]}{\tr(\bSigma_{\bxi}^{2})}\leq \frac{1}{\eta^2}\cdot\frac{\sum_{i=2}^{n}\mbE(R_{ni}^{4}\vert\mathscr{F}_{i-1})}{\tr^2(\bSigma_{\bxi}^{2})},
	\end{align}
	and
	\begin{align}
		\label{ineq2oflemma14}
		\mbE\biggl\{\frac{\sum_{i=2}^{n}\mbE(R_{ni}^{4}\vert\mathscr{F}_{i-1})}{\tr^2(\bSigma_{\bxi}^{2})}\biggr\} &= \frac{\sum_{i=2}^{n}\mbE(R_{ni}^{4})}{\tr^2(\bSigma_{\bxi}^{2})}\notag\\
		&\lesssim \frac{1}{n^4\tr^2(\bSigma_{\bxi}^{2})}\sum_{i=2}^{n}\mbE\biggl(\sum_{j=1}^{i-1}\bxi_{i}^\top\bxi_{j}\biggr)^4\notag\\
		&\lesssim \frac{1}{n^4\tr^2(\bSigma_{\bxi}^{2})}\biggl[\sum_{i=2}^{n}\sum_{s\neq t}\mbE\bigl\{(\bxi_{i}^\top\bxi_{s})^{2}(\bxi_{i}^\top\bxi_{t})^{2}\bigr\} + \sum_{i=2}^{n}\sum_{j=1}^{i-1}\mbE(\bxi_{i}^\top\bxi_{j})^{4}\biggr]\notag\\
		&=o(1),
	\end{align}
	where the last equality holds by the fact that
	\[
	\sum_{i=2}^{n}\sum_{s\neq t}\mbE\bigl\{(\bxi_{i}^\top\bxi_{s})^{2}(\bxi_{i}^\top\bxi_{t})^{2}\bigr\} \lesssim n^3\mbE\bigl\{(\bxi_{3}^\top\bxi_{1})^{2}(\bxi_{3}^\top\bxi_{2})^{2}\bigr\}\lesssim n^3\tr^2(\bSigma_{\bxi}^{2}),
	\]
	Lemma \ref{lemma31} ensures
	\begin{align*}
		\mbE\bigl\{(\bxi_{3}^\top\bxi_{1})^{2}(\bxi_{3}^\top\bxi_{2})^{2}\bigr\} &= \mbE\bigl[\mbE\bigl\{(\bxi_{3}^\top\bxi_{1})^{2}(\bxi_{3}^\top\bxi_{2})^{2}\vert\bxi_{3}\bigr\}\bigr]\\
		&\lesssim\mbE(\bxi_{3}^\top\bSigma_{\bxi}\bxi_{3})^2\\
		&\lesssim\tr^2(\bSigma_{\bxi}^{2}),
	\end{align*}
	and Lemma \ref{lemmaofassumptionb5} yields
	\begin{align*}
		\sum_{i=2}^{n}\sum_{j=1}^{i-1}\mbE(\bxi_{i}^\top\bxi_{j})^{4} \lesssim n^2 \mbE(\bxi_{1}^\top\bxi_{2})^{4}\lesssim n^2\tr^2(\bSigma_{\bxi}^{2}).
	\end{align*}
	Combining inequalities \eqref{ineq1oflemma14} and \eqref{ineq2oflemma14}, equation \eqref{eq2oflemma14} is verified.  The proof is concluded.
\end{proof}

\vspace{3mm}

\begin{lemma}
	\label{lemma26}
	(A maximal inequality for U-statistics of order two). Let $\{X_i\}_{i=1}^{n}$ be a sample of i.i.d. random variables in a separable and measurable space $(S,\mathcal{S})$. Let $\bm{f}:S\times S\rightarrow\mathbb{R}^{p}$ be an $\mathcal{S}\bigotimes\mathcal{S}$-measurable, symmetric kernel such that $\mbE\{\lvert f_k(X_1,X_2)\rvert\}<\infty$ for all $k=1,\ldots,p$. Let $\bU_n = \{n(n-1)\}^{-1}\sum_{1\leq i\neq j\leq n}\bm{f}(X_i,X_j)$ and $\btheta = \mbE\{\bm{f}(X_1,X_2)\}$. Suppose $2\leq p\leq \exp(bn)$ for some constant $b>0$, then
	\[
	\mbE(\lVert\bU_n\rVert_{\infty})\lesssim \lVert\btheta\rVert_{\infty} + \biggl(\frac{\log p}{n}\biggr)^{1/2}D_2' + \frac{\log p}{n}M_2' + \frac{\log p}{n}D_2'' + \biggl(\frac{\log p}{n}\biggr)^{5/4}D_4'' + \biggl(\frac{\log p}{n}\biggr)^{3/2}M_4'',
	\]
	where $D_2' = \max_{1\leq k\leq p}[E\{f_{1,k}^{2}(X_1)\}]^{1/2}$, $M_2' = (\mbE[\{\max_{1\leq i\leq n}\max_{1\leq k\leq p}\lvert f_{1,k}(X_i)\rvert\}^2])^{1/2}$, $D_q'' = \max_{1\leq k\leq p}[E\{f_{2,k}^{q}(X_1,X_2)\}]^{1/q}$ for $q=2,4$, $M_4'' = (\mbE[\{\max_{1\leq i\neq j\leq n}\max_{1\leq k\leq p}\lvert f_{2,k}(X_i,X_j)\rvert\}^4])^{1/4}$ with kernel function
	\[
	\bm{f}_1(x_1) = \mbE\{\bm{f}(X_1,X_2)\vert X_1 = x_1\} - \btheta
	\]	
	and
	\[
	\bm{f}_2(x_1,x_2) = \bm{f}(x_1,x_2) - \bm{f}_1(x_1) - \bm{f}_1(x_2) - \btheta.
	\]
	
\end{lemma}
\begin{proof}
	By the Hoeffding decomposition of $U$-statistic, we derive
	\[
	\bU_n = \btheta + 2\bS_{1n} + \bS_{2n},
	\]
	where
	\[
	\bS_{1n} = \frac{1}{n}\sum_{i=1}^{n}\bm{f}_1(X_i)
	\]
	and
	\[
	\bS_{2n} = \frac{1}{n(n-1)}\sum_{i\neq j}\bm{f}_2(X_i,X_j).
	\]
	By the triangle inequality of norm, we derive
	\begin{align}
		\label{ineq1oflemma26}
		\mbE(\lVert\bU_n\rVert_{\infty})\lesssim \lVert\btheta\rVert_{\infty} + \mbE(\lVert\bS_{1n}\rVert_{\infty}) + \mbE(\lVert\bS_{2n}\rVert_{\infty}).
	\end{align}
	Note that $\bS_{1n}$ is the sample mean of an i.i.d. random vector sequence. Thus
	\begin{align}
		\label{ineq2oflemma26}
		\mbE(\lVert\bS_{1n}\rVert_{\infty})\lesssim \biggl(\frac{\log p}{n}\biggr)^{1/2}D_2' + \frac{\log p}{n}M_2'
	\end{align}
	by Lemma 8 in \cite{chernozhukov2015comparison}. Observe that $\bS_{2n}$ is a degenerate U-statistic, we derive
	\begin{align}
		\label{ineq3oflemma26}
		\mbE(\lVert\bS_{2n}\rVert_{\infty})\lesssim \frac{\log p}{n}D_2'' + \biggl(\frac{\log p}{n}\biggr)^{5/4}D_4'' + \biggl(\frac{\log p}{n}\biggr)^{3/2}M_4''
	\end{align}
	by Theorem 5 in \cite{chen2018gaussian}.
	Altogether, the proof is finished.
\end{proof}

\begin{lemma}
	\label{lemma31}
	Let $\{\bA_{i}\}_{i=1}^{k}$ be a sequence of $p\times p$ semi-positive matrices and $k$ is a fixed positive integer.
	Suppose $\bnu$ is a $p-$dimensional sub-Gaussian random vector with sub-Gaussian norm $\lVert\bnu\rVert_{\psi_2}$. Then
	\[
	\mbE\biggl(\prod_{i=1}^{k}\bnu^\top\bA_{i}\bnu\biggr) \lesssim \lVert\bnu\rVert_{\psi_2}^{2k}\prod_{i=1}^{k}\tr(\bA_{i}).
	\]
\end{lemma}
\begin{proof}
	Matrix eigen-decomposition yields, for $i=1,\ldots,k$,
	\[
	\bA_{i} = \bC_{i}^\top\bD_{i}\bC_{i},
	\]
	where $\bD_{i} = \diag\{\lambda_{1}(\bA_{i}),\lambda_{2}(\bA_{i}),\cdots,\lambda_{p}(\bA_{i})\}$, $\lambda_{j}(\bA_{i})$ is the $j$-th eigenvalue of $\bA_{i}$ and $\bC_{i}$ is a unit orthogonal matrix. Denote $\bS_{i} = \bC_{i}\bnu_{1}$. Then
	\begin{align*}
		\mbE\biggl(\prod_{i=1}^{k}\bnu^\top\bA_{i}\bnu\biggr) &= \mbE\biggl(\prod_{i=1}^{k}\sum_{j_i=1}^{p}\lambda_{j_i}(\bA_{i})S_{i,j_i}^2\biggr)\\
		&=\sum_{j_1 = 1}^{p}\sum_{j_2=1}^{p}\cdots\sum_{j_k=1}^{p}\lambda_{j_1}(\bA_1)\lambda_{j_2}(\bA_2)\cdots\lambda_{j_k}(\bA_k)\mbE\bigl(S_{1,j_1}S_{2,j_2}\cdots S_{k,j_k}\bigr)^2\\
		&\leq \max_{1\leq j_1,j_2,\ldots,j_k\leq p}\mbE\bigl(S_{1,j_1}S_{2,j_2}\cdots S_{k,j_k}\bigr)^2\prod_{i=1}^{k}\tr(\bA_{i}),
	\end{align*}
	where $S_{i,j_i}$ is the $j_i$-th element of $\bS_{i}$. We then verify the following to conclude the proof:
	\begin{align}
		\label{ineq1oflemma31}
		\max_{1\leq j_1,j_2,\ldots,j_k\leq p}\mbE\bigl(S_{1,j_1}S_{2,j_2}\cdots S_{k,j_k}\bigr)^2\lesssim \lVert\bnu\rVert_{\psi_2}^{2k}.
	\end{align}
	Observe that
	\begin{align*}
		&\max_{1\leq j_1,j_2,\ldots,j_k\leq p}\mbE\bigl(S_{1,j_1}S_{2,j_2}\cdots S_{k,j_k}\bigr)^2\\
		\leq& \max_{1\leq j_1,j_2,\ldots,j_k\leq p}\bigl\{\mbE\bigl(S_{1,j_1}\bigr)^{2^k}\bigr\}^{1/2^{k-1}}\bigl\{\mbE\bigl(S_{2,j_2}\bigr)^{2^k}\bigr\}^{1/2^{k-1}}\cdots \bigl\{\mbE\bigl(S_{k,j_k}\bigr)^{2^k}\bigr\}^{1/2^{k-1}}\\
		\leq&\bigl\{\max_{1\leq i\leq k}\max_{1\leq j\leq p}\mbE\bigl(S_{i,j}\bigr)^{2^k}\bigr\}^{k/2^{k-1}},
	\end{align*}
	where the first inequality holds by Cauchy-Schwartz inequality and Liapounov inequality.
	By the sub-Gaussian assumption of $\bnu$, $\max_{1\leq i\leq k}\max_{1\leq j\leq p}\mbE\bigl(S_{i,j}\bigr)^{2^k}\lesssim \lVert\bnu\rVert_{\psi_2}^{2^k}$. Thus the inequality \eqref{ineq1oflemma31} is proved.
\end{proof}

\begin{lemma}
	\label{lemmaofassumptionb5}
	(A corollary of Lemma \ref{lemma31}) Let $\bxi = \bA\bnu$, where $\bA$ is a $p_{\bxi}\times m$ matrix with $p_{\bxi}\leq m$ and $\bSigma_{\bxi} = \bA\bA^\top$. Under Assumption \ref{assumptionb5}, for a fixed positive integer $q$,
	\[
	\mbE(\bm{\xi}_1^\top\bm{\xi}_2)^{2q} \lesssim \tr^{q}(\bSigma_{\bm{\xi}}^2).
	\]
\end{lemma}
\begin{proof}
	Denote $\mbE_{\bxi_{1}}(\cdot) = \mbE(\cdot\vert\bxi_{1})$, it can be shown that
	\begin{align*}
		\mbE(\bm{\xi}_1^\top\bm{\xi}_2)^{2q} &= \mbE(\bxi_{2}^\top\bxi_{1}\bxi_{1}^\top\bxi_{2})^{q}
		= \mbE\bigl\{\mbE_{\bxi_{1}}(\bnu_{2}^\top\bA^\top\bxi_{1}\bxi_{1}^\top\bA\bnu_{2})^{q}\bigr\}\\
		&\lesssim \mbE(\bxi_{1}^\top\bSigma_{\bxi}\bxi_{1})^{q}
		= \mbE(\bnu_{1}\bA^\top\bSigma_{\bxi}\bA\bnu_{1})^{q}
		\lesssim \tr^{q}(\bSigma_{\bxi}^2)
	\end{align*}
	where the first inequality holds by Lemma \ref{lemma31}, $\bxi_{1}$ is independent of $\bxi_{2}$, and the second inequality holds also by Lemma \ref{lemma31}.
\end{proof}

\begin{definition}
	(Some notations used in Lemmas \ref{lemma32}, \ref{lemma33} and \ref{lemma34}.)
	Recall that $\bW$ is defined in \eqref{bW} in subsection \ref{poweranalysisofsec2} and $\hat{\bW}$ is defined in \eqref{teststatistic1} in section \ref{sec5}.
	We give some notations here to avoid repeating the notations in the following proofs. Let $\bet_i = \bX_i - \bW^\top\bZ_i$, $\hat{\bet}_i = \bX_i - \hat\bW^\top\bZ_i$, $\bGamma_{\bet} = \bGamma_{\bX} - \bW^\top\bGamma_{\bZ}$, $\bGamma_{\hat\bet} = \bGamma_{\bX} - \hat\bW^\top\bGamma_{\bZ}$ and $\breve{\bW} = \bW - \hat{\bW}$, where $\bGamma = (\bGamma_{\bX}^\top,\bGamma_{\bZ}^\top)^\top$ and $\bGamma$ is defined in Assumption \ref{assumptionb5}. Write $\bX_{i} = \bGamma_{\bX}\bnu_{i},\,\bZ_{i} = \bGamma_{\bZ}\bnu_{i},\,\bet_{i} = \bGamma_{\bet}\bnu_{i}$ and $\hat{\bet}_{i} = \bGamma_{\hat\bet}\bnu_{i}$. Further, write $\bSigma_{\bX} = \bGamma_{\bX}\bGamma_{\bX}^\top$, $\bSigma_{\bZ} = \bGamma_{\bZ}\bGamma_{\bZ}^\top$, $\bSigma_{\bet} = \bGamma_{\bet}\bGamma_{\bet}^\top$ and $\bSigma_{\hat\bet} = \bGamma_{\hat\bet}\bGamma_{\hat\bet}^\top$.
\end{definition}

\begin{lemma}
	\label{lemma32}
	(Some technical results for the proof of Theorem \ref{limitpowerofguosum}.)
	Under the conditions in Theorem \ref{limitpowerofguosum},
	\begin{align}
		\label{ineq1oflemma32}
		\mbE(\bbeta^\top\bSigma_{\bet}\bX_{1}\bet_{1}^\top\bbeta)^2&\lesssim\bbeta^\top\bSigma_{\bet}\bbeta\bbeta^\top\bSigma_{\bet}\bSigma_{\bX}\bSigma_{\bet}\bbeta,\\
		\label{ineq2oflemma32}
		\mbE(\bbeta^\top\bet_{1}\bX_{1}^\top\bX_{2}\bet_{2}^\top\bbeta)^2&\lesssim\bbeta^\top\bSigma_{\bet}\bbeta\bbeta^\top\bSigma_{\bet}\bSigma_{\bX}\bSigma_{\bet}\bbeta,\\
		\label{ineq3oflemma32}
		\mbE(\bbeta^\top\bet_{1}\bX_{1}^\top\bX_{2})^2&\lesssim \bbeta^\top\bSigma_{\bet}\bbeta\,\tr(\bSigma_{\bX}^{2})
	\end{align}
\end{lemma}
\begin{proof}
	Denote $\mbE_{\bnu_{1}}(\cdot) = \mbE(\cdot\vert\bnu_{1})$ where $\nu$ is defined in Assumption~\ref{assumptionb5}.  To prove the inequality \eqref{ineq1oflemma32}, we have
	\begin{align*}
		\mbE(\bbeta^\top\bSigma_{\bet}\bX_{1}\bet_{1}^\top\bbeta)^2 &= \mbE(\bnu_{1}^\top\bGamma_{\bet}^\top\bbeta\bbeta^\top\bGamma_{\bet}\bnu_{1}\bnu_{1}^\top\bGamma_{\bX}^\top\bSigma_{\bet}\bbeta\bbeta^\top\bSigma_{\bet}\bGamma_{\bX}\bnu_{1})\\
		&\lesssim \bbeta^\top\bSigma_{\bet}\bbeta\bbeta^\top\bSigma_{\bet}\bSigma_{\bX}\bSigma_{\bet}\bbeta.
	\end{align*}
	The inequality then follows from Lemma \ref{lemma31}. To prove the inequality \eqref{ineq2oflemma32}, we have
	\begin{align*}
		\mbE(\bbeta^\top\bet_{1}\bX_{1}^\top\bX_{2}\bet_{2}^\top\bbeta)^2 &= \mbE\bigl\{\mbE_{\bnu_{1}}(\bnu_{2}^\top\bGamma_{\bet}^\top\bbeta\bbeta^\top\bet_{1}\bX_{1}^\top\bGamma_{\bX}\bnu_{2})^2\bigr\}\\
		&\lesssim\mbE(\bX_{1}^\top\bSigma_{\bet}\bbeta\bbeta^\top\bet_{1})^2=\mbE(\bbeta^\top\bSigma_{\bet}\bX_{1}\bet_{1}^\top\bbeta)^2.
	\end{align*}
	Combining the inequality \eqref{ineq1oflemma32}, the inequality \eqref{ineq2oflemma32} is proved. Consider the inequality \eqref{ineq3oflemma32}. Observe that
	\begin{align*}
		\mbE(\bbeta^\top\bet_{1}\bX_{1}^\top\bX_{2})^2&= \mbE\bigl\{\mbE_{\bnu_{1}}(\bnu_{2}^\top\bGamma_{\bX}^\top\bX_{1}\bet_{1}^\top\bbeta\bbeta^\top\bet_{1}\bX_{1}^\top\bGamma_{\bX}\bnu_{2})\bigr\}\\
		&\lesssim \mbE(\bnu_{1}^\top\bGamma_{\bet}^\top\bbeta\bbeta^\top\bGamma_{\bet}\bnu_{1}\bnu_{1}^\top\bGamma_{\bX}^\top\bGamma_{\bX}\bGamma_{\bX}^\top\bGamma_{\bX}\bnu_{1})\\
		&\lesssim\bbeta^\top\bSigma_{\bet}\bbeta\,\tr(\bSigma_{\bX}^{2}).
	\end{align*}
	Thus the inequality \eqref{ineq3oflemma32} is proved.
\end{proof}

\begin{lemma}
	\label{lemma33}
	(Some technical results for the proof of Theorem \ref{limitpoweroforaclesum}.) Under conditions in Theorem \ref{limitpoweroforaclesum},
	\begin{align}
		\label{ineq1oflemma33}
		\mbE(\bbeta^\top\bSigma_{\bet}\bet_{1}\bet_{1}^\top\bbeta)^2 &\lesssim (\bbeta^\top\bSigma_{\bet}^{2}\bbeta)^{2},\\
		\label{ineq2oflemma33}
		\mbE(\bbeta^\top\bet_{1}\bet_{1}^\top\bet_{2}\bet_{2}^\top\bbeta)^2 &\lesssim (\bbeta^\top\bSigma_{\bet}^{2}\bbeta)^{2},\\
		\label{ineq3oflemma33}
		\mbE(\bbeta^\top\bet_{1}\bet_{1}^\top\bet_{2})^2 &\lesssim \bbeta^\top\bSigma_{\bet}\bbeta\,\tr(\bSigma_{\bet}^2).
	\end{align}
\end{lemma}
\begin{proof}
	Similar to Lemma \ref{lemma32}, we can prove this lemma by Lemma \ref{lemma31}. At first, we prove inequality \eqref{ineq1oflemma33}. Note that
	\begin{align*}
		\mbE(\bbeta^\top\bSigma_{\bet}\bet_{1}\bet_{1}^\top\bbeta)^2 &= \mbE(\bnu_{1}^\top\bGamma_{\bet}^\top\bbeta\bbeta^\top\bSigma_{\bet}\bGamma_{\bet}\bnu_{1})^2\\
		&\lesssim (\bbeta^\top\bSigma_{\bet}^{2}\bbeta)^2.
	\end{align*}
	Where the equality holds by Assumption \ref{assumptionb5} and the inequality follows from Lemma \ref{lemma31}. Now we prove inequality \eqref{ineq2oflemma33}. Denote $\mbE_{\bnu_{1}}(\cdot) = \mbE(\cdot\vert\bnu_{1})$. Observe that
	\begin{align*}
		\mbE(\bbeta^\top\bet_{1}\bet_{1}^\top\bet_{2}\bet_{2}^\top\bbeta)^2 &= \mbE\bigl\{\mbE_{\bnu_{1}}(\bnu_{2}^\top\bGamma_{\bet}^\top\bbeta\bbeta^\top\bet_{1}\bet_{1}^\top\bGamma_{\bet}\bnu_{2})^2\bigr\}\\
		&\lesssim \mbE(\bet_{1}^\top\bGamma_{\bet}\bGamma_{\bet}^\top\bbeta\bbeta^\top\bet_{1})^2 = \mbE(\bbeta^\top\bSigma_{\bet}\bet_{1}\bet_{1}^\top\bbeta)^2.
	\end{align*}
	Combing inequality \eqref{ineq1oflemma33}, expression \eqref{ineq2oflemma33} is proved.
	Now we prove inequality \eqref{ineq3oflemma33}. It can be shown that
	\begin{align*}
		\mbE(\bbeta^\top\bet_{1}\bet_{1}^\top\bet_{2})^2 &= \mbE\bigl\{\mbE_{\bnu_{1}}(\bnu_{2}^\top\bGamma_{\bet}^\top\bet_{1}\bet_{1}^\top\bbeta\bbeta^\top\bet_{1}\bet_{1}^\top\bGamma_{\bet}\bnu_{2})\bigr\}\\
		&\lesssim\mbE(\bet_{1}^\top\bbeta\bbeta^\top\bet_{1}\bet_{1}^\top\bGamma_{\bet}\bGamma_{\bet}^\top\bet_{1})\\
		&=\mbE(\bnu_{1}^\top\bGamma_{\bet}^\top\bbeta\bbeta^\top\bGamma_{\bet}\bnu_{1}\bnu_{1}^\top\bGamma_{\bet}^\top\bGamma_{\bet}\bGamma_{\bet}^\top\bGamma_{\bet}\bnu_{1})\\
		&\lesssim\bbeta^\top\bSigma_{\bet}\bbeta\,\tr(\bSigma_{\bet}^2).
	\end{align*}
	The inequalities follow from Lemma \ref{lemma31} and Assumption \ref{assumptionb5}.
\end{proof}

\begin{lemma}
	\label{lemma34}
	(Some technical results for the proof of Theorem \ref{limitnulldisofsum}.) Under conditions in Theorem \ref{limitnulldisofsum},
	\begin{align}
		\label{ineq1oflemma34}
		\tr(\bSigma_{\hat\bet}^{2}) &\leq \bigl\{\tr^{1/2}(\bSigma_{\bet}^2) + \lambda_{\max}(\bSigma_{\bZ})\lVert\brebW\rVert_{F}^{2}\bigr\}^{2}\\
		\label{ineq2oflemma34}
		\lambda_{\max}(\bSigma_{\hat\bet}) &\leq \lambda_{\max}(\bSigma_{\bet}) + \lambda_{\max}(\bSigma_{\bZ})\lVert\brebW\rVert_{F}^{2}\\
		\label{ineq3oflemma34}
		\max_{1\leq k\leq p_{\bgamma}}\lVert\mbE(Z_{k}\hat\bet)\rVert_{2}^{2}&\leq \lVert\brebW\rVert_{F}^{2}\varphi^2.
	\end{align}
\end{lemma}
\begin{proof}
	Note that
	\begin{align*}
		\bSigma_{\hat\bet} &= \bGamma_{\hat\bet}\bGamma_{\hat\bet}^\top
		= (\bGamma_{\bet} + \brebW^\top\bGamma_{\bZ})(\bGamma_{\bet} + \brebW^\top\bGamma_{\bZ})^\top\\
		&= \bGamma_{\bet}\bGamma_{\bet}^\top + \bGamma_{\bet}\bGamma_{\bZ}^\top\brebW + \brebW^\top\bGamma_{\bZ}\bGamma_{\bet}^\top + \brebW^\top\bGamma_{\bZ}\bGamma_{\bZ}^\top\brebW\\
		&= \bSigma_{\bet} + \brebW^\top\bSigma_{\bZ}\brebW
	\end{align*}
	where the last equality holds because of
	\begin{align*}
		\bGamma_{\bZ}\bGamma_{\bet}^\top &= \bGamma_{\bZ}(\bGamma_{\bX} - \bGamma_{\bX}\bGamma_{\bZ}^\top(\bGamma_{\bZ}\bGamma_{\bZ}^\top)^{-1}\bGamma_{\bZ})^\top\\
		&=\bGamma_{\bZ}\bGamma_{\bX}^\top - \bGamma_{\bZ}\bGamma_{\bZ}^\top(\bGamma_{\bZ}\bGamma_{\bZ}^\top)^{-1}\bGamma_{\bZ}\bGamma_{\bX}^\top = \bzero.
	\end{align*}
	Now we prove the inequality \eqref{ineq1oflemma34}. It can be shown that
	\begin{align*}
		\tr(\bSigma_{\hat\bet}^{2}) &= \tr\bigl\{(\bSigma_{\bet} + \brebW^\top\bSigma_{\bZ}\brebW)^2\bigr\}\\
		&= \tr(\bSigma_{\bet}^2) + \tr(\bSigma_{\bet}\brebW^\top\bSigma_{\bZ}\brebW) + \tr(\brebW^\top\bSigma_{\bZ}\brebW\bSigma_{\bet}) + \tr\bigl\{(\brebW^\top\bSigma_{\bZ}\brebW)^2\bigr\}\\
		&\leq \tr(\bSigma_{\bet}^{2}) + 2\tr^{1/2}(\bSigma_{\bet}^2)\tr^{1/2}\bigl\{(\brebW^\top\bSigma_{\bZ}\brebW)^2\bigr\} + \tr\bigl\{(\brebW^\top\bSigma_{\bZ}\brebW)^2\bigr\}\\
		&\leq \bigl\{\tr^{1/2}(\bSigma_{\bet}^2) + \lambda_{\max}(\bSigma_{\bZ})\lVert\brebW\rVert_{F}^{2}\bigr\}^{2},
	\end{align*}
	where the first inequality holds by the matrix Cauchy-Schwartz inequality, the second inequality holds by the fact that
	\[
	\tr^{1/2}\bigl\{(\brebW^\top\bSigma_{\bZ}\brebW)^2\bigr\}\leq \tr(\brebW^\top\bSigma_{\bZ}\brebW)\leq \lambda_{\max}(\bSigma_{\bZ})\lVert\brebW\rVert_{F}^{2}.
	\]
	Similarly, 	(\ref{ineq2oflemma34}) can be derive as
	\begin{align*}
		\lambda_{\max}(\bSigma_{\hat\bet})  = \lambda_{\max}(\bSigma_{\bet}) + \tr(\brebW^\top\bSigma_{\bZ}\brebW) \leq \lambda_{\max}(\bSigma_{\bet}) + \lambda_{\max}(\bSigma_{\bZ})\lVert\brebW\rVert_{F}^{2}.
	\end{align*}
	For \eqref{ineq3oflemma34}, we derive
	\begin{align*}
		\max_{1\leq k\leq p_{\bgamma}}\lVert\mbE(Z_{k}\hat\bet)\rVert_{2}^{2} = \max_{1\leq k\leq p_{\bgamma}}\lVert\mbE(Z_{k}\brebW^\top\bZ)\rVert_{2}^{2} &\leq \lVert\brebW\rVert_{F}^{2}\varphi^2,
	\end{align*}
	where the inequality holds by the fact that
	\[
	\max_{1\leq k\leq p_{\bgamma}}\lVert\mbE(Z_{k}\brebW^\top\bZ)\rVert_{2}^{2} \leq \lambda_{\max}(\brebW\brebW^\top)\varphi^2
	\]
	and $\lambda_{\max}(\brebW\brebW^\top)\leq \lVert\brebW\rVert_{F}^{2}$.
\end{proof}

\begin{lemma}
	\label{lemma36}
	For any random variables $X_1,\ldots,X_m$(without independence assumption) and any fixed integer $q$,
	\[
	\lVert\max_{1\leq i\leq m}X_i\rVert_{q}\lesssim \log m\max_{1\leq i\leq m}\lVert X_i\rVert_{\psi_1}, \quad \text{and}\quad \lVert\max_{1\leq i\leq m}X_i\rVert_{q}\lesssim \sqrt{\log m}\max_{1\leq i\leq m}\lVert X_i\rVert_{\psi_2}.
	\]
\end{lemma}
\begin{proof}
	By the property of Orlicz norm and Lemma 2.2.2 in \cite{vanderVaart.1996} (the first is on Page 95), we have
	\[
	\lVert\max_{1\leq i\leq m}X_i\rVert_{q}\lesssim \lVert\max_{1\leq i\leq m}X_i\rVert_{\psi_1}\lesssim \log m\max_{1\leq i\leq m}\lVert X_i\rVert_{\psi_1}.
	\]
	Similarly,
	\[
	\lVert\max_{1\leq i\leq m}X_i\rVert_{q}\lesssim \lVert\max_{1\leq i\leq m}X_i\rVert_{\psi_2}\lesssim \sqrt{\log m}\max_{1\leq i\leq m}\lVert X_i\rVert_{\psi_2}.
	\]
\end{proof}

\begin{lemma}
	\label{lemma37}
	Suppose $\bA$ is a $p\times p$ dimensional matrix. For any $\bx\in\mbR^{p}$, it can be shown that
	\begin{align*}
		\lvert\bx^\top\bA\bx\rvert\leq \lVert\bA\rVert_{\infty}\lVert\bx\rVert_{1}^{2}.
	\end{align*}
\end{lemma}
\begin{proof}
	Observe that
	\begin{align*}
		\lvert\bx^\top\bA\bx\rvert &= \biggl\lvert\sum_{i,j=1}^{p}a_{ij}x_{i}x_{j}\biggr\rvert
		\leq \sum_{i,j=1}^{p}\lvert a_{ij}\rvert\cdot\lvert x_{i}\rvert\cdot\lvert x_{j}\rvert\\
		&\leq \lVert\bA\rVert_{\infty}\sum_{i,j=1}^{p}\lvert x_{i}\rvert\cdot\lvert x_{j}\rvert
		=  \lVert\bA\rVert_{\infty}\lVert\bx\rVert_{1}^{2}.
	\end{align*}
	Where the first inequality holds by the triangle inequality of norm.
\end{proof}

\section{Additional simulation results}
\label{additionalsimulationresults}

Figure \ref{plotsecnarioQF} displays the empirical size-power curves of $WALD$ in scenario 1 in section \ref{sec6} of the main text. We find that with different values $\tau= 0, 0.5, 1.0, 1.5, 2.0, 2.5, 3.0, 3.5$ the test can be either very liberal or very conservative. Thus selecting a proper tuning parameter is difficult in general.

\begin{figure}[ht]
	\centering
	\includegraphics[width=\textwidth]{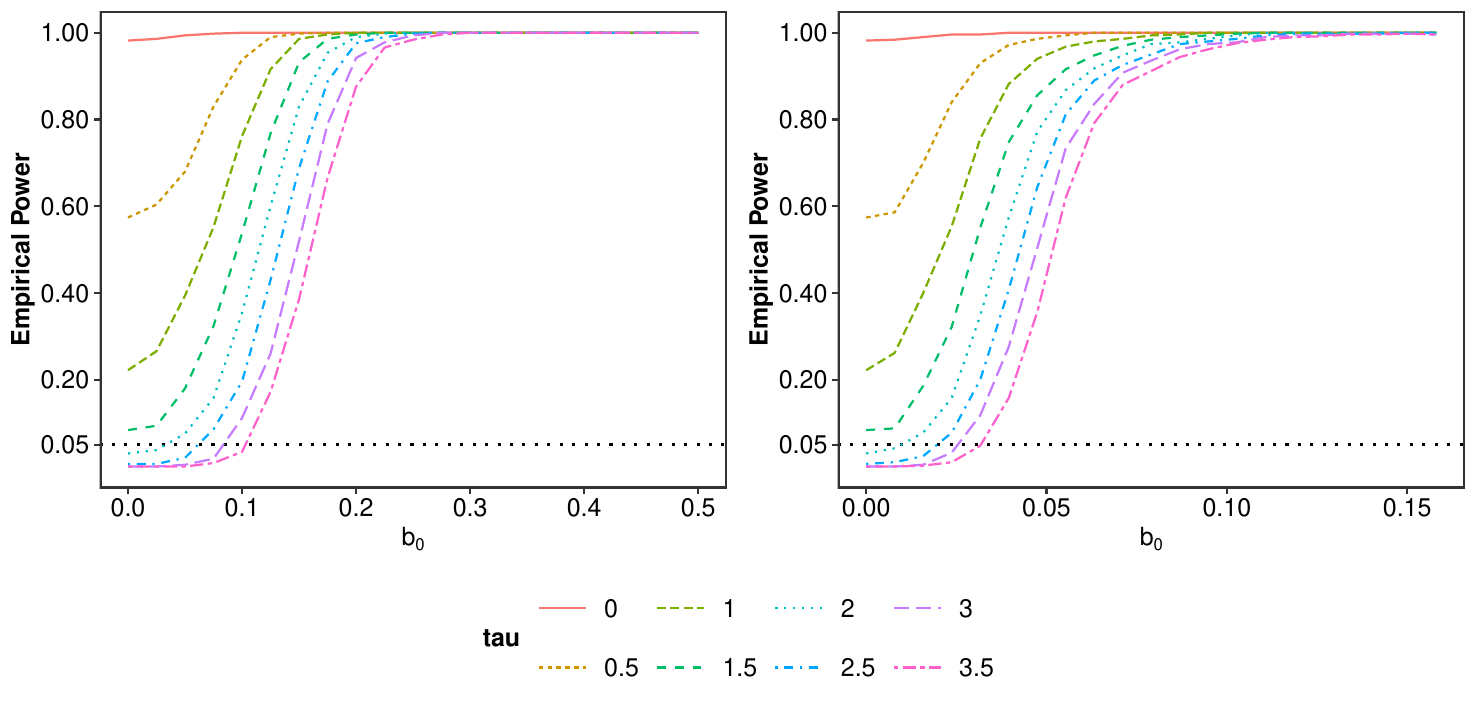}
	\caption{This figure presents the finite sample performance of $WALD$ in scenario 1. Left panel presents the empirical sizes and powers of $WALD$ with different tuning parameter $\tau(\tau=0,0.5,1.0,1.5,2.0,2.5,3.0,3.5)$ in the sparse alternative (setting 2).
		Right panel corresponds to dense alternative (setting 3).}
	\label{plotsecnarioQF}
\end{figure}

Based on the settings in section \ref{sec6} in the main text, we consider the following five scenarios.

\noindent {\bf{Scenario 3.}} This scenario investigates the performance of our tests  thoroughly when the correlation between covariates of interest and nuisance covariates is weak. Generate the covariates from the multivariate normal distribution $N_p(\bm{0}_p, \bSigma)$, where $\bSigma = (\sigma_{ij})_{p\times p}$ with $\sigma_{ij} = \rho^{|i-j|}, i,j = 1, \ldots, p$ and $\rho = 0.3,0.5,0.7$. The sample size $n = 100,200$, the covariate dimension $p = 600,1000,2000,4000$ and $p_{\bbeta} = p_{\bgamma}$. In the sparse alternative (setting 2) and the dense alternative (setting 3), we set $b_0 = \sqrt{\lVert\bgamma\rVert_{2}^{2}/s_{\bbeta}}$.

\noindent {\bf{Scenario 4.}} This scenario is same to Example \ref{example2} in the main text.

\noindent {\bf{Scenario 5.}} This scenario investigates the finite performance of our methods thoroughly when $\bX$ and $\bZ$ are not weakly correlated.
Covariates are generated according to the model defined in Scenario 3.
Throughout the simulation study, we set $d_{\bZ} = 3$ and vary $d_{\bX} = 0, 5, 10, 15$, where $d_{\bX} = 0$ represents there exists no correlation between $\bX$ and $\bZ$.
The sample size $n = 100,200$, the predictor dimension $p = 600,1000,2000,4000$ and $p_{\bbeta} = p_{\bgamma}$. In the sparse alternative (setting 2) and the dense alternative (setting 3), we set $b_0 = \sqrt{\lVert\bgamma\rVert_{2}^{2}/s_{\bbeta}}$.

\noindent \textbf{Scenario 6.} We aim to compare our tests with other testing methods in this scenario. The covariates $\bV = (\bX^\top,\bZ^\top)^\top$ are generated according to $\bV = \bSigma^{1/2}\bnu$, where $\bnu = (\nu_{1},\ldots,\nu_{p})^\top$ and $\{\nu_{k}\}_{k=1}^{p}$ are generated independently from the $U(-2,2)$ distribution. Here $\bSigma = (\sigma_{ij})_{p\times p}$ follows the Toeplitz design, that is, $\sigma_{ij} = 0.5^{|i-j|}, i,j = 1, \ldots, p$. The regression error $\epsilon\sim U(-2,2)$ is independent of $\bV$. The sample size $n = 100$, the covariate dimension $p = 600$ and $p_{\bbeta} = p_{\bgamma}=300$. In the sparse alternative (setting 2) and the dense alternative (setting 3), we vary $b_0$ from $0$ to $\sqrt{\lVert\bgamma\rVert_{2}^{2}/s_{\bbeta}}$.

\noindent \textbf{Scenario 7.} This scenario investigates the performance of our tests when $\bX$ and $\bZ$ are highly correlated.
The covariates are generated according to the following model:
\begin{align*}
    \bX &= \bW^\top\bZ + \bet,
\end{align*}
where $\bet$ is a $p_{\bbeta}$-dimensional random vector and $\bet$ is independent of $\bZ$. $\bet=\bSigma_{\bet}^{1/2}\bnu_{\bet}$ and the elements of $\bnu_{\bet}$ are generated independently from the $U(-2,2)$ distribution, $\bZ=\bSigma_{\bZ}^{1/2}\bnu_{\bZ}$ and the elements of $\bnu_{\bZ}$ are generated independently from the $U(-2,2)$ distribution. $\bSigma_{\bet}$ and $\bSigma_{\bZ}$ follow the Toeplitz design with $\rho=0.5$ respectively. $\bW$ is defined in \eqref{bWexample} in subsection \ref{sec2.3}.
Throughout the scenario, $d_{\bZ} = 3$ and $d_{\bX} = 10$. The regression error $\epsilon\sim U(-2,2)$ is independent of covariates.
The sample size $n = 100$, the predictor dimension $p = 600$ and $p_{\bbeta} = p_{\bgamma} = 300$. In the sparse alternative (setting 2) and the dense alternative (setting 3), we vary $b_0$ from $0$ to $\sqrt{\lVert\bgamma\rVert_{2}^{2}/s_{\bbeta}}$.

\noindent \textbf{Scenario 8.} This scenario investigates the power performance of $T_{n}$ when $\lVert\bbeta\rVert_{2}$ is fixed but the orientation of $\bbeta$ is changed. The settings of this scenario are the same as those in Scenario 2 of the main text, except for the following modifications. We set $\bW = \bzero$ and $\bbeta = b_{0}\cdot\be_{k}$, where $e_k$ represents a $p_{\bbeta}$-dimensional vector where the $k$-th element is 1, and all other elements are 0. We set $b_{0} = 3$. As $k$ varies from 1 to 10, the power performances of $T_{n}$ are $0.808$, $0.914$, $0.910$, $0.906$, $0.932$, $0.902$, $0.910$, $0.890$, $0.928$, and $0.926$, respectively. And the difference in power is appreciable with the orientation of $\bbeta$.

Table \ref{tabletoeplitz} reports the simulation results of scenario 3. We have the following observations. First, $T_n$ and $M_n$ control the type I error well, even when the dimension is  $4000$. Second, the empirical powers increase when the dimension decreases and the sample size increases. Third, there is no significant difference in power between sparse and dense alternatives as long as $\lVert\bbeta\rVert_{2}$ stays the same.

{\small
	\begin{table}[htbp]
		\renewcommand\arraystretch{0.8}
		\centering
		\caption{The empirical type-I errors and powers for $T_n$  and $M_n$.}
		\label{tabletoeplitz}
		
		\begin{threeparttable}
			\begin{tabular}{cccccccccccc}
				\toprule
				& & & \multicolumn{3}{c}{Type-I Error} & \multicolumn{3}{c}{Power (Sparse)} & \multicolumn{3}{c}{Power (Dense)}  \\
				\cmidrule(r){4-6} \cmidrule(lr){7-9} \cmidrule(l){10-12}
				$n$ & $p$ & $\rho$  & $0.3$ & $0.5$ & $0.7$  & $0.3$ & $0.5$ & $0.7$  & $0.3$ & $0.5$ & $0.7$   \\
				\midrule
				$100$ & $600$   & $T_n$ & $0.046$ & $0.070$ & $0.060$ & $0.950$ & $0.998$ & $1.000$ & $0.956$ & $1.000$ & $1.000$ \\
				&         & $M_n$ & $0.050$ & $0.058$ & $0.054$ & $0.964$ & $0.998$ & $1.000$ & $0.950$ & $1.000$ & $1.000$ \\
				$100$ & $1000$  & $T_n$ & $0.038$ & $0.050$ & $0.076$ & $0.820$ & $0.972$ & $1.000$ & $0.826$ & $0.980$ & $1.000$ \\
				&         & $M_n$ & $0.042$ & $0.050$ & $0.076$ & $0.808$ & $0.974$ & $1.000$ & $0.824$ & $0.980$ & $1.000$ \\
				$100$ & $2000$  & $T_n$ & $0.052$ & $0.058$ & $0.048$ & $0.552$ & $0.800$ & $0.982$ & $0.522$ & $0.760$ & $0.996$ \\
				&         & $M_n$ & $0.042$ & $0.062$ & $0.042$ & $0.532$ & $0.796$ & $0.984$ & $0.504$ & $0.780$ & $0.990$ \\
				$100$ & $4000$  & $T_n$ & $0.052$ & $0.066$ & $0.068$ & $0.382$ & $0.534$ & $0.816$ & $0.356$ & $0.546$ & $0.848$ \\
				&         & $M_n$ & $0.046$ & $0.066$ & $0.064$ & $0.376$ & $0.516$ & $0.826$ & $0.348$ & $0.528$ & $0.852$ \\
				$200$ & $600$   & $T_n$ & $0.048$ & $0.056$ & $0.060$ & $1.000$ & $1.000$ & $1.000$ & $1.000$ & $1.000$ & $1.000$ \\
				&         & $M_n$ & $0.048$ & $0.052$ & $0.052$ & $1.000$ & $1.000$ & $1.000$ & $1.000$ & $1.000$ & $1.000$ \\
				$200$ & $1000$  & $T_n$ & $0.062$ & $0.062$ & $0.054$ & $1.000$ & $1.000$ & $1.000$ & $1.000$ & $1.000$ & $1.000$ \\
				&         & $M_n$ & $0.070$ & $0.064$ & $0.060$ & $1.000$ & $1.000$ & $1.000$ & $1.000$ & $1.000$ & $1.000$ \\
				$200$ & $2000$  & $T_n$ & $0.070$ & $0.048$ & $0.068$ & $0.982$ & $1.000$ & $1.000$ & $0.978$ & $0.998$ & $1.000$ \\
				&         & $M_n$ & $0.070$ & $0.042$ & $0.060$ & $0.986$ & $1.000$ & $1.000$ & $0.976$ & $0.998$ & $1.000$ \\
				$200$ & $4000$  & $T_n$ & $0.040$ & $0.050$ & $0.052$ & $0.818$ & $0.956$ & $1.000$ & $0.796$ & $0.986$ & $1.000$ \\
				&         & $M_n$ & $0.040$ & $0.054$ & $0.052$ & $0.818$ & $0.960$ & $1.000$ & $0.794$ & $0.980$ & $1.000$ \\
				
				\bottomrule
			\end{tabular}
			
			\begin{tablenotes}
				\footnotesize
				\item ``Type-I error", ``Power (Sparse)," and ``Power (Dense)" correspond to Setting 1, Setting 2 and Setting 3.
			\end{tablenotes}
		\end{threeparttable}
	\end{table}
}

The left panel of figure \ref{figureofscenario4} displays
\begin{figure}[ht]
	\centering
	\includegraphics[width=\textwidth]{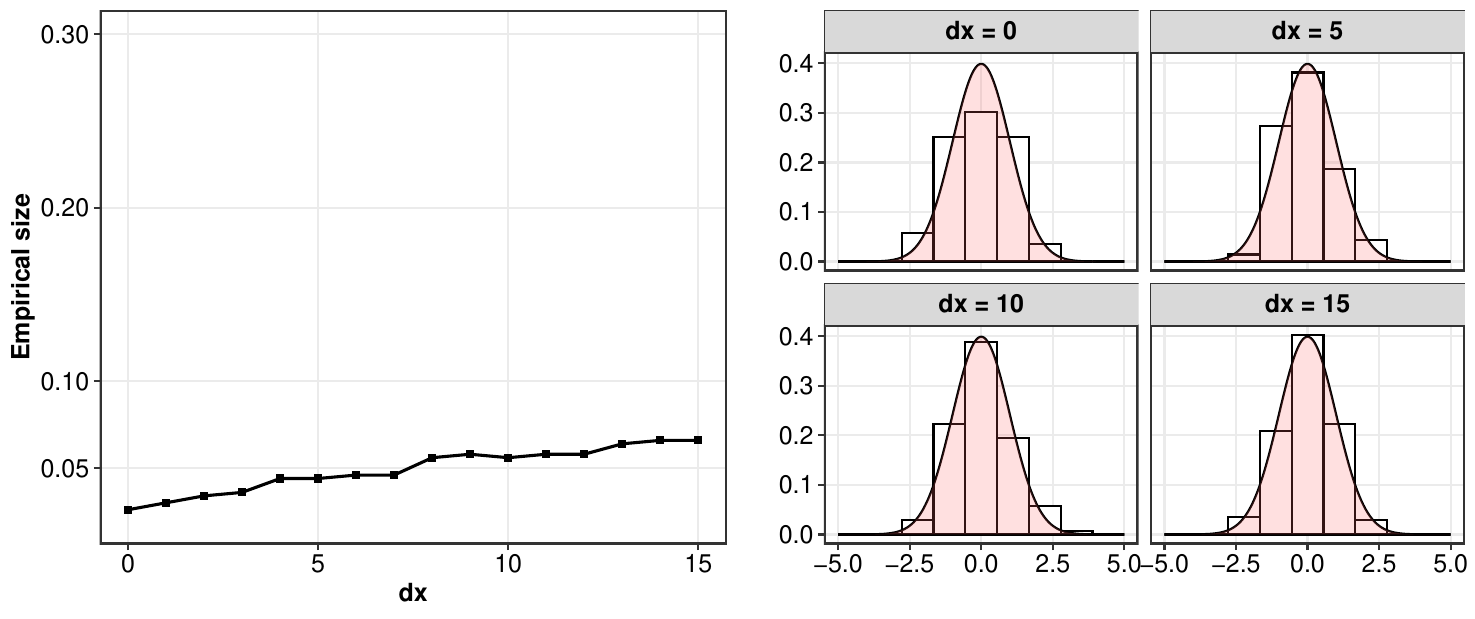}
	\caption{Left panel presents the empirical sizes of $M_n$ with different $d_{\bX}$.
		Right panel presents the empirical probability density function of $M_n$ with different $d_{\bX}$ ($d_{\bX}$ = 0,5,10,15). The pink shade represents the probability density function of the standard normal distribution. Set $n=100$, $p = 600$ and $p_{\bbeta} = p_{\bgamma}$. We generate 500 replications and reject the null hypothesis at the significance level $\alpha = 0.05$. More details can be seen in Scenario 3 in Section \ref{sec6} in the main text.}
	\label{figureofscenario4}
\end{figure}
the empirical sizes of $M_n$ in scenario 4. The empirical sizes are close to the nominal level $\alpha=0.05$ regardless of the increase of $d_{\bX}$. The right panel of figure \ref{figureofscenario4} presents the empirical probability density function of $M_n$, which can be well approximated by the standard normal distribution no matter what $d_{\bX}$ is. These results illustrate that $M_n$ still maintains type I error well when $\bX$ and $\bZ$ are not weakly dependent.

Table \ref{tablesxneq0} summarizes the empirical sizes and powers of the proposed tests in scenario 5.
Both perform satisfactorily when $d_{\bX} = 0$.
But when $d_{\bX}\neq 0$, $M_n$ performs better than $T_n$ in terms of both the empirical type I error rate and empirical power. As $d_{\bX}$ increases, the empirical size of the $T_n$ test deviates significantly from the significant level while the $M_n$ test maintains the empirical size well. Besides, the empirical power of $M_{n}$ is significantly larger than $T_{n}$ when $d_{\bX}\neq 0$.
The results show that the $M_n$ test is more applicable when covariates of interest and nuisance covariates have a strong relationship.
{\small
	\begin{table}[htbp]
		\renewcommand\arraystretch{0.8}
		\centering
		\caption{The empirical type-I errors and powers for $T_n$ test and $M_n$ test.}
		\label{tablesxneq0}
		
		\begin{threeparttable}
			\begin{tabular}{ccccccccccccccc}
				\toprule
				& & & \multicolumn{4}{c}{Type-I Error} & \multicolumn{4}{c}{Power (Sparse)} & \multicolumn{4}{c}{Power (Dense)}  \\
				\cmidrule(r){4-7} \cmidrule(lr){8-11} \cmidrule(l){12-15}
				$n$ & $p$ & $d_{\bX}$ & $0$ & $5$ & $10$ & $15$ & $0$ & $5$ & $10$ & $15$ & $0$ & $5$ & $10$ & $15$   \\
				\midrule
				$100$ & $600$ & $T_{n}$ & $0.038$ & $0.226$ & $0.282$ & $0.296$ & $0.996$ & $0.976$ & $0.816$ & $0.660$ & $1.000$ & $0.982$ & $0.850$ & $0.668$ \\
				&       & $M_{n}$ & $0.038$ & $0.052$ & $0.062$ & $0.062$ & $0.994$ & $0.994$ & $0.996$ & $0.998$ & $0.998$ & $1.000$ & $0.998$ & $0.998$ \\
				$100$ & $1000$ & $T_{n}$ & $0.042$ & $0.210$ & $0.270$ & $0.296$ & $0.980$ & $0.914$ & $0.738$ & $0.584$ & $0.990$ & $0.948$ & $0.748$ & $0.576$ \\
				&        & $M_{n}$ & $0.044$ & $0.046$ & $0.054$ & $0.074$ & $0.980$ & $0.986$ & $0.984$ & $0.988$ & $0.992$ & $0.988$ & $0.990$ & $0.990$ \\
				$100$ & $2000$ & $T_{n}$ & $0.054$ & $0.168$ & $0.242$ & $0.274$ & $0.770$ & $0.706$ & $0.536$ & $0.412$ & $0.786$ & $0.720$ & $0.532$ & $0.426$ \\
				&        & $M_{n}$ & $0.052$ & $0.050$ & $0.048$ & $0.062$ & $0.758$ & $0.764$ & $0.772$ & $0.780$ & $0.770$ & $0.776$ & $0.780$ & $0.782$ \\
				$100$ & $4000$ & $T_{n}$ & $0.038$ & $0.098$ & $0.144$ & $0.192$ & $0.514$ & $0.508$ & $0.380$ & $0.316$ & $0.536$ & $0.476$ & $0.382$ & $0.300$ \\
				&        & $M_{n}$ & $0.032$ & $0.038$ & $0.054$ & $0.050$ & $0.520$ & $0.528$ & $0.528$ & $0.520$ & $0.524$ & $0.520$ & $0.530$ & $0.532$ \\
				$200$ & $600$ & $T_{n}$ & $0.062$ & $0.246$ & $0.294$ & $0.308$ & $1.000$ & $1.000$ & $1.000$ & $0.994$ & $1.000$ & $1.000$ & $1.000$ & $1.000$ \\
				&       & $M_{n}$ & $0.062$ & $0.070$ & $0.072$ & $0.076$ & $1.000$ & $1.000$ & $1.000$ & $1.000$ & $1.000$ & $1.000$ & $1.000$ & $1.000$ \\
				$200$ & $1000$ & $T_{n}$ & $0.072$ & $0.230$ & $0.282$ & $0.298$ & $1.000$ & $1.000$ & $1.000$ & $0.994$ & $1.000$ & $1.000$ & $1.000$ & $0.994$ \\
				&        & $M_{n}$ & $0.080$ & $0.084$ & $0.086$ & $0.096$ & $1.000$ & $1.000$ & $1.000$ & $1.000$ & $1.000$ & $1.000$ & $1.000$ & $1.000$ \\
				$200$ & $2000$ & $T_{n}$ & $0.060$ & $0.184$ & $0.256$ & $0.278$ & $1.000$ & $1.000$ & $0.996$ & $0.954$ & $1.000$ & $1.000$ & $0.992$ & $0.960$ \\
				&        & $M_{n}$ & $0.060$ & $0.068$ & $0.072$ & $0.080$ & $1.000$ & $1.000$ & $1.000$ & $1.000$ & $1.000$ & $1.000$ & $1.000$ & $1.000$ \\
				$200$ & $4000$ & $T_{n}$ & $0.066$ & $0.152$ & $0.248$ & $0.288$ & $0.980$ & $0.962$ & $0.910$ & $0.772$ & $0.974$ & $0.958$ & $0.894$ & $0.772$ \\
				&        & $M_{n}$ & $0.064$ & $0.068$ & $0.074$ & $0.080$ & $0.974$ & $0.978$ & $0.980$ & $0.978$ & $0.974$ & $0.976$ & $0.974$ & $0.974$ \\
				
				\bottomrule
			\end{tabular}
			
			\begin{tablenotes}
				\footnotesize
				\item ``Type-I error", ``Power (Sparse)," and ``Power (Dense)" correspond to Setting 1, Setting 2, and Setting 3.
			\end{tablenotes}
		\end{threeparttable}
	\end{table}
}

Figure \ref{figure2unif} displays the empirical size-power curves of the four tests in Scenario 6. The empirical results in Scenario 6 are similar to those in Scenario 1 of subsection \ref{sec6.1}.
It can be observed that $T_{n}$, $M_{n}$ and $ST$ tests control the size well.  $T_n$ and $M_n$ are generally more powerful than $ST$ under  the sparse and dense alternative hypotheses. Under the dense alternative, the empirical powers of $ST$ can be as low as the significance level. The empirical powers of $T_n$ and $M_n$ increase quickly as the signal strength $b_0$ becomes stronger.
On the other hand, the $WALD$ test is very liberal to have very large empirical size when we use the tuning parameter $\tau =1$ recommended by \cite{guo2021group}. 
It is worth noticing that $T_n$ and $M_n$ have similar performances in this scenario. $T_n$ performs well enough when the correlation between $\bX$ and $\bZ$ is relatively weak.
\begin{figure}[ht]
\centering
\includegraphics[width=\textwidth]{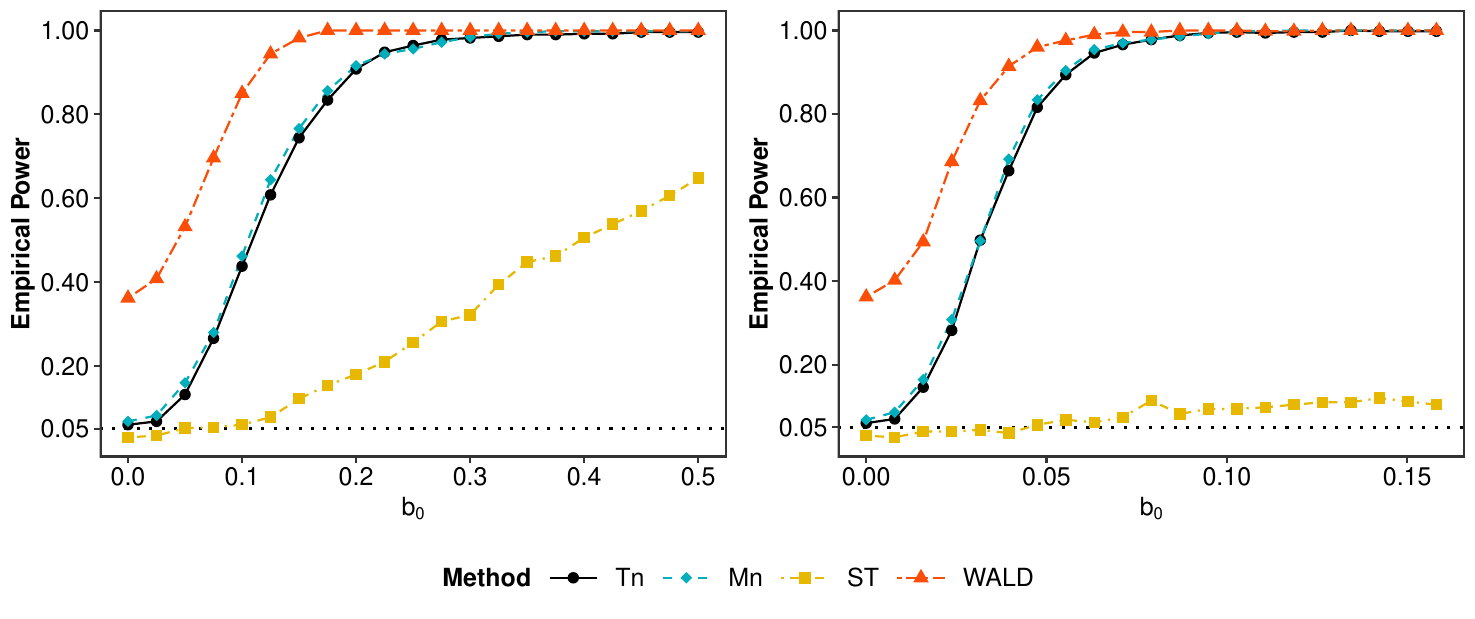}
\caption{The left panel represents empirical sizes and powers of the $T_n, M_n$, $ST$ and $WALD$ in the    sparse alternative (setting 2). The right panel corresponds to dense alternative (setting 3). The solid     line with circle points, dash line with diamond points, dot-dash line with square points and two-dash       line with triangle points represent the empirical sizes and powers of $T_n$, $M_n$, $ST$ and $WALD$,        respectively.}
\label{figure2unif}
\end{figure}

Figure \ref{figure3unif} displays the empirical size-power curves of $T_n$ and $M_n$ in Scenario 7. It can be verified that the empirical results in Scenario 7 are similar to those in Scenario 2 of subsection \ref{sec6.1}. We find that  $T_n$ is too liberal to maintain the significance level. On the contrary, $M_n$ maintains the level well. As $b_0$ increases, although the empirical powers of  $T_n$ and $M_n$ increase rapidly, the empirical power of $T_n$ does not go to $1$ as the increase of $b_0$. In contrast, the empirical power of $M_n$ can increase to 1 quickly. The results show that  $M_n$ can also improve the power compared with $T_n$. This confirms the theory.
\begin{figure}[ht]
\centering
\includegraphics[width=\textwidth]{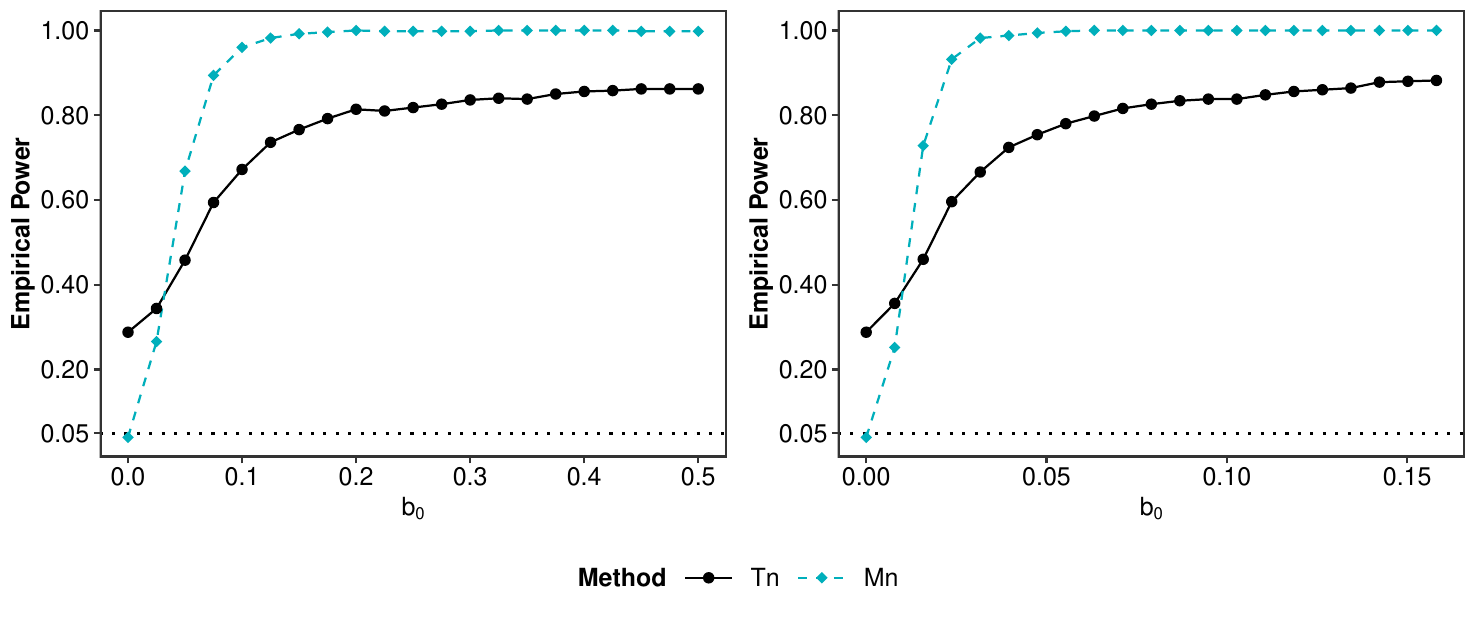}
    \caption{The left panel represents empirical sizes and powers of the $T_n$ and $M_n$ in the sparse alternatives (setting 2). The right panel corresponds to dense alternatives (setting 3). The solid line with circle points and dash line with diamond points represent the empirical sizes and powers of $T_n$ and $M_n$, respectively.}
    \label{figure3unif}
\end{figure}

\section{Significant DNA methylation sets in real data analysis in the main text}
\label{selectedgenesinrealdataanalysis}

Table \ref{selecteddnamethylationsetfortnofrealdata} represents the significant DNA methylation sets for $T_{n}$ in real data analysis in section \ref{sec6} in the main text. Table \ref{selecteddnamethylationsetformnofrealdata} correspond to $M_{n}$.

{\small
	\begin{table}[ht]
		\renewcommand\arraystretch{0.8}
		\centering
		\caption{Numbers of 126 significant DNA methylation sets for $T_{n}$.}
		\label{selecteddnamethylationsetfortnofrealdata}
		\begin{threeparttable}
			\begin{tabular}{p{16cm}}
				\toprule
				6   
				28   
				33   
				46   
				62   
				83   
				86   
				95  
				103  
				117  
				118  
				144
				155  
				166  
				168  
				175  
				196  
				197  
				323  
				346  
				365  
				367  
				374  
				386
				391  
				392  
				407  
				413  
				417  
				424  
				440  
				448  
				457  
				471  
				489  
				495
				496  
				518  
				539  
				557  
				564  
				587  
				594  
				678  
				706  
				713  
				717  
				718
				719  
				739  
				757  
				779  
				802  
				807  
				808  
				839  
				843  
				858  
				883  
				924
				977 
				1008 
				1036 
				1043 
				1080 
				1106 
				1114 
				1120 
				1165 
				1192 
				1203 
				1209
				1218 
				1224 
				1234 
				1237 
				1250 
				1255 
				1290 
				1309 
				1320 
				1327 
				1328 
				1333
				1357 
				1379 
				1385 
				1409 
				1410 
				1414 
				1415 
				1467 
				1503 
				1534 
				1543 
				1550
				1585 
				1614 
				1632 
				1638 
				1648 
				1652 
				1654 
				1672 
				1692 
				1712 
				1731 
				1733
				1743 
				1757 
				1761 
				1781 
				1782 
				1786 
				1787 
				1788 
				1797 
				1807 
				1825 
				1862
				1863 
				1865 
				1887 
				1909 
				1924 
				1927\\
				\bottomrule
			\end{tabular}
		\end{threeparttable}
	\end{table}
}

{\small
	\begin{table}[ht]
		\renewcommand\arraystretch{0.8}
		\centering
		\caption{Numbers of 149 significant DNA methylation sets for $M_{n}$.}
		\label{selecteddnamethylationsetformnofrealdata}
		\begin{threeparttable}
			\begin{tabular}{p{16cm}}
				\toprule
				6    
				9   
				26   
				28   
				33   
				35   
				62   
				83   
				86   
				95  
				103  
				117
				118  
				144  
				155  
				159  
				166  
				168  
				175  
				178  
				196  
				197  
				201  
				323
				329  
				346  
				359  
				365  
				367  
				374  
				386  
				391  
				392  
				394  
				407  
				413
				417  
				424  
				427  
				432  
				440  
				448  
				457  
				468  
				474  
				489  
				495  
				496
				510  
				518  
				539  
				557  
				564  
				587  
				594  
				662  
				678  
				706  
				713  
				717
				719  
				739  
				757  
				779  
				802  
				807  
				808  
				843  
				858  
				871  
				880  
				883
				891  
				924  
				977 
				1008 
				1024 
				1036 
				1043 
				1080 
				1106 
				1114 
				1120 
				1191
				1192 
				1203 
				1209 
				1218 
				1224 
				1234 
				1237 
				1250 
				1255 
				1290 
				1309 
				1320
				1327 
				1328 
				1333 
				1357 
				1379 
				1385 
				1393 
				1409 
				1410 
				1414 
				1415 
				1426
				1441 
				1461 
				1467 
				1475 
				1503 
				1522 
				1534 
				1543 
				1550 
				1585 
				1614 
				1632
				1638 
				1648 
				1652 
				1654 
				1672 
				1692 
				1693 
				1712 
				1731 
				1733 
				1737 
				1743
				1757 
				1761 
				1781 
				1782 
				1786 
				1787 
				1788 
				1797 
				1802 
				1807 
				1825 
				1863
				1865 
				1887 
				1909 
				1924 
				1927
				\\
				\bottomrule
			\end{tabular}
		\end{threeparttable}
	\end{table}
}

\section{Additional real data analysis}
\label{additionalrealdataanalysis}
We apply our tests to a data set about riboflavin (vitamin B2) production rate with Bacillus Subtilis. This data set was made publicly by \cite{buhlmann2014high} and  analyzed by several authors, for instance \cite{meinshausen2009p},  \cite{van2014asymptotically}, \cite{javanmard2014confidence}, \cite{dezeure2017high}, and \cite{fei2019drawing}. It consists of $71$ observations of strains of Bacillus Subtilis and $4088$ covariates, measuring the log-expression levels of 4088 genes. The response variable is the logarithm of the riboflavin production rate.

Several genes have been previously identified as having associations with the response variable. \cite{meinshausen2009p} identified YXLD\_at using a multisample-splitting method, while \cite{javanmard2014confidence} detected YXLD\_at and YXLE\_at. \cite{van2014asymptotically} claimed none. Additionally, \cite{fei2019drawing} reported the presence of YCKE\_at, XHLA\_at, YXLD\_at, YDAR\_at, and YCGN\_at. In summary, prior literature has identified six genes: YXLD\_at, YXLE\_at, YCKE\_at, XHLA\_at, YDAR\_at, and YCGN\_at, as being associated with the response variable. Denote $\calG$ as the set of six selected genes and $\calG^c$ as its complement.
A natural question is whether the selected genes contribute to the response given the other genes? Consider the following regression modeling:
\begin{align*}
	Y = \bbeta^\top\bV_{\calG} + \bgamma^\top\bV_{\calG^c} + \epsilon,
\end{align*}
where $Y$ is the response variable, $\bV_{\calG}$ is the vector of selected genes, $\bV_{\calG^c}$ denotes the genes in set $\calG^c$, and $\epsilon$ is the regression error. Further $\lVert\bbeta\rVert_{0} = \lvert\calG\rvert = 6$, $\lVert\bgamma\rVert_{0} = \lvert\calG^c\rvert = 4082$. The null hypothesis of interest is $\bH_{01}:\bbeta = \bzero$. Thus the testing parameter is $\bbeta$, and the nuisance parameter is $\bgamma$.  
To verify whether some genes in $\calG^{c}$ contribute to the response given the other genes, we also consider the null hypothesis $\bH_{02}:\bgamma = \bzero$. In this testing problem $\bH_{02}$, the testing parameter is changed to be $\bgamma$, and $\bbeta$ is the nuisance parameter correspondingly.

Apply  $T_n$, $M_n$, $ST$, and $WALD$. We standardize the data and report the $p$-values in Table \ref{pvalueofrealdata}. For the testing problem $\bH_{01}$,  only $T_n$  and $M_n$ reject the null hypothesis at the significance level $\alpha = 0.05$. For the testing problem $\bH_{02}$, all tests do not reject the null hypothesis at the significance level $\alpha = 0.05$. The results suggest that the selected gene set $\calG$ contributes to the response, and there is no significant gene in $\calG^c$.	
{\small
	\begin{table}[ht]
		\renewcommand\arraystretch{0.8}
		\centering
		\caption{The $p$-values for a riboflavin data. }
		\label{pvalueofrealdata}
		
		\begin{threeparttable}
			\begin{tabular}{ccccc}
				\toprule
				Method & $T_n$ & $M_n$ & $ST$ & $WALD$ \\
				\midrule
				& \multicolumn{4}{c}{$p$-value} \\
				\cmidrule(l){2-5}
				$\bH_{01}$ & 0.021 & 0.045  & 0.160 & 0.279\\
				$\bH_{02}$ & 0.644 & 0.632  & 0.430 & 0.082\\
				\bottomrule
			\end{tabular}
		\end{threeparttable}
	\end{table}
}

\end{appendix}


\begin{funding}
Xu Guo was supported by the National Key R \& D Program of China (grant No. 2023YFA1011100), National Natural Science Foundation of China (grant No. 12322112, 12071038), and the Fundamental Research Funds for the Central Universities.

\noindent Lixing Zhu was supported by the National Natural Science Foundation of China (grant No. 12131006).
\end{funding}

\bibliographystyle{imsart-number} 
\bibliography{bibliography}       

\begin{thebibliography}{41}

\bibitem{bai1996effect}
\begin{barticle}[author]
\bauthor{\bsnm{Bai},~\bfnm{Zhidong}\binits{Z.}} \AND
  \bauthor{\bsnm{Saranadasa},~\bfnm{Hewa}\binits{H.}}
(\byear{1996}).
\btitle{Effect of high dimension: by an example of a two sample problem}.
\bjournal{Statistica Sinica}
\bvolume{6}
\bpages{311--329}.
\end{barticle}
\endbibitem

\bibitem{belloni2012sparse}
\begin{barticle}[author]
\bauthor{\bsnm{Belloni},~\bfnm{Alexandre}\binits{A.}},
  \bauthor{\bsnm{Chen},~\bfnm{Daniel}\binits{D.}},
  \bauthor{\bsnm{Chernozhukov},~\bfnm{Victor}\binits{V.}} \AND
  \bauthor{\bsnm{Hansen},~\bfnm{Christian}\binits{C.}}
(\byear{2012}).
\btitle{Sparse models and methods for optimal instruments with an application
  to eminent domain}.
\bjournal{Econometrica}
\bvolume{80}
\bpages{2369--2429}.
\end{barticle}
\endbibitem

\bibitem{belloni2018uniformly}
\begin{barticle}[author]
\bauthor{\bsnm{Belloni},~\bfnm{Alexandre}\binits{A.}},
  \bauthor{\bsnm{Chernozhukov},~\bfnm{Victor}\binits{V.}},
  \bauthor{\bsnm{Chetverikov},~\bfnm{Denis}\binits{D.}} \AND
  \bauthor{\bsnm{Wei},~\bfnm{Ying}\binits{Y.}}
(\byear{2018}).
\btitle{Uniformly valid post-regularization confidence regions for many
  functional parameters in z-estimation framework}.
\bjournal{Annals of statistics}
\bvolume{46}
\bpages{3643--3675}.
\end{barticle}
\endbibitem

\bibitem{belloni2015uniform}
\begin{barticle}[author]
\bauthor{\bsnm{Belloni},~\bfnm{Alexandre}\binits{A.}},
  \bauthor{\bsnm{Chernozhukov},~\bfnm{Victor}\binits{V.}} \AND
  \bauthor{\bsnm{Kato},~\bfnm{Kengo}\binits{K.}}
(\byear{2015}).
\btitle{Uniform post-selection inference for least absolute deviation
  regression and other Z-estimation problems}.
\bjournal{Biometrika}
\bvolume{102}
\bpages{77--94}.
\end{barticle}
\endbibitem

\bibitem{buhlmann2014high}
\begin{barticle}[author]
\bauthor{\bsnm{B{\"u}hlmann},~\bfnm{Peter}\binits{P.}},
  \bauthor{\bsnm{Kalisch},~\bfnm{Markus}\binits{M.}} \AND
  \bauthor{\bsnm{Meier},~\bfnm{Lukas}\binits{L.}}
(\byear{2014}).
\btitle{High-dimensional statistics with a view toward applications in
  biology}.
\bjournal{Annual Review of Statistics and Its Application}
\bvolume{1}
\bpages{255--278}.
\end{barticle}
\endbibitem

\bibitem{bulik2015ld}
\begin{barticle}[author]
\bauthor{\bsnm{Bulik-Sullivan},~\bfnm{Brendan~K}\binits{B.~K.}},
  \bauthor{\bsnm{Loh},~\bfnm{Po-Ru}\binits{P.-R.}},
  \bauthor{\bsnm{Finucane},~\bfnm{Hilary~K}\binits{H.~K.}},
  \bauthor{\bsnm{Ripke},~\bfnm{Stephan}\binits{S.}},
  \bauthor{\bsnm{Yang},~\bfnm{Jian}\binits{J.}}, \bauthor{\bparticle{of~the}
  \bsnm{Psychiatric Genomics~Consortium},~\bfnm{Schizophrenia
  Working~Group}\binits{S.~W.~G.}},
  \bauthor{\bsnm{Patterson},~\bfnm{Nick}\binits{N.}},
  \bauthor{\bsnm{Daly},~\bfnm{Mark~J}\binits{M.~J.}},
  \bauthor{\bsnm{Price},~\bfnm{Alkes~L}\binits{A.~L.}} \AND
  \bauthor{\bsnm{Neale},~\bfnm{Benjamin~M}\binits{B.~M.}}
(\byear{2015}).
\btitle{LD Score regression distinguishes confounding from polygenicity in
  genome-wide association studies}.
\bjournal{Nature genetics}
\bvolume{47}
\bpages{291--295}.
\end{barticle}
\endbibitem

\bibitem{candes2018panning}
\begin{barticle}[author]
\bauthor{\bsnm{Candes},~\bfnm{Emmanuel}\binits{E.}},
  \bauthor{\bsnm{Fan},~\bfnm{Yingying}\binits{Y.}},
  \bauthor{\bsnm{Janson},~\bfnm{Lucas}\binits{L.}} \AND
  \bauthor{\bsnm{Lv},~\bfnm{Jinchi}\binits{J.}}
(\byear{2018}).
\btitle{Panning for gold:‘model-X’knockoffs for high dimensional controlled
  variable selection}.
\bjournal{Journal of the Royal Statistical Society: Series B (Statistical
  Methodology)}
\bvolume{80}
\bpages{551--577}.
\end{barticle}
\endbibitem

\bibitem{chen2022testing}
\begin{barticle}[author]
\bauthor{\bsnm{Chen},~\bfnm{Jinsong}\binits{J.}},
  \bauthor{\bsnm{Li},~\bfnm{Quefeng}\binits{Q.}} \AND
  \bauthor{\bsnm{Chen},~\bfnm{Hua~Yun}\binits{H.~Y.}}
(\byear{2023}).
\btitle{Testing generalized linear models with high-dimensional nuisance
  parameters}.
\bjournal{Biometrika}
\bvolume{110}
\bpages{83--99}.
\end{barticle}
\endbibitem

\bibitem{chen2009effects}
\begin{barticle}[author]
\bauthor{\bsnm{Chen},~\bfnm{Song~Xi}\binits{S.~X.}},
  \bauthor{\bsnm{Peng},~\bfnm{Liang}\binits{L.}} \AND
  \bauthor{\bsnm{Qin},~\bfnm{Ying-Li}\binits{Y.-L.}}
(\byear{2009}).
\btitle{Effects of data dimension on empirical likelihood}.
\bjournal{Biometrika}
\bvolume{96}
\bpages{711--722}.
\end{barticle}
\endbibitem

\bibitem{chen2010two}
\begin{barticle}[author]
\bauthor{\bsnm{Chen},~\bfnm{Song~Xi}\binits{S.~X.}} \AND
  \bauthor{\bsnm{Qin},~\bfnm{Ying-Li}\binits{Y.-L.}}
(\byear{2010}).
\btitle{A two-sample test for high-dimensional data with applications to
  gene-set testing}.
\bjournal{Annals of Statistics}
\bvolume{38}
\bpages{808--835}.
\end{barticle}
\endbibitem

\bibitem{chen2018gaussian}
\begin{barticle}[author]
\bauthor{\bsnm{Chen},~\bfnm{Xiaohui}\binits{X.}}
(\byear{2018}).
\btitle{Gaussian and bootstrap approximations for high-dimensional U-statistics
  and their applications}.
\bjournal{Annals of Statistics}
\bvolume{46}
\bpages{642--678}.
\end{barticle}
\endbibitem

\bibitem{chernozhukov2018double}
\begin{barticle}[author]
\bauthor{\bsnm{Chernozhukov},~\bfnm{Victor}\binits{V.}},
  \bauthor{\bsnm{Chetverikov},~\bfnm{Denis}\binits{D.}},
  \bauthor{\bsnm{Demirer},~\bfnm{Mert}\binits{M.}},
  \bauthor{\bsnm{Duflo},~\bfnm{Esther}\binits{E.}},
  \bauthor{\bsnm{Hansen},~\bfnm{Christian}\binits{C.}},
  \bauthor{\bsnm{Newey},~\bfnm{Whitney}\binits{W.}} \AND
  \bauthor{\bsnm{Robins},~\bfnm{James}\binits{J.}}
(\byear{2018}).
\btitle{Double/debiased machine learning for treatment and structural
  parameters}.
\bjournal{The Econometrics Journal}
\bvolume{21}
\bpages{C1--C68}.
\end{barticle}
\endbibitem

\bibitem{chernozhukov2015comparison}
\begin{barticle}[author]
\bauthor{\bsnm{Chernozhukov},~\bfnm{Victor}\binits{V.}},
  \bauthor{\bsnm{Chetverikov},~\bfnm{Denis}\binits{D.}} \AND
  \bauthor{\bsnm{Kato},~\bfnm{Kengo}\binits{K.}}
(\byear{2015}).
\btitle{Comparison and anti-concentration bounds for maxima of Gaussian random
  vectors}.
\bjournal{Probability Theory and Related Fields}
\bvolume{162}
\bpages{47--70}.
\end{barticle}
\endbibitem

\bibitem{cui2018test}
\begin{barticle}[author]
\bauthor{\bsnm{Cui},~\bfnm{Hengjian}\binits{H.}},
  \bauthor{\bsnm{Guo},~\bfnm{Wenwen}\binits{W.}} \AND
  \bauthor{\bsnm{Zhong},~\bfnm{Wei}\binits{W.}}
(\byear{2018}).
\btitle{Test for high-dimensional regression coefficients using refitted
  cross-validation variance estimation}.
\bjournal{Annals of Statistics}
\bvolume{46}
\bpages{958--988}.
\end{barticle}
\endbibitem

\bibitem{cui2024estimation}
\begin{barticle}[author]
\bauthor{\bsnm{Cui},~\bfnm{Shijie}\binits{S.}},
  \bauthor{\bsnm{Guo},~\bfnm{Xu}\binits{X.}} \AND
  \bauthor{\bsnm{Zhang},~\bfnm{Zhe}\binits{Z.}}
(\byear{2024}).
\btitle{Estimation and Inference in Ultrahigh Dimensional Partially Linear
  Single-Index Models}.
\bjournal{arXiv preprint arXiv:2404.04471}.
\end{barticle}
\endbibitem

\bibitem{dezeure2017high}
\begin{barticle}[author]
\bauthor{\bsnm{Dezeure},~\bfnm{Ruben}\binits{R.}},
  \bauthor{\bsnm{B{\"u}hlmann},~\bfnm{Peter}\binits{P.}} \AND
  \bauthor{\bsnm{Zhang},~\bfnm{Cun-Hui}\binits{C.-H.}}
(\byear{2017}).
\btitle{High-dimensional simultaneous inference with the bootstrap}.
\bjournal{Test}
\bvolume{26}
\bpages{685--719}.
\end{barticle}
\endbibitem

\bibitem{fei2019drawing}
\begin{barticle}[author]
\bauthor{\bsnm{Fei},~\bfnm{Zhe}\binits{Z.}},
  \bauthor{\bsnm{Zhu},~\bfnm{Ji}\binits{J.}},
  \bauthor{\bsnm{Banerjee},~\bfnm{Moulinath}\binits{M.}} \AND
  \bauthor{\bsnm{Li},~\bfnm{Yi}\binits{Y.}}
(\byear{2019}).
\btitle{Drawing inferences for high-dimensional linear models: A
  selection-assisted partial regression and smoothing approach}.
\bjournal{Biometrics}
\bvolume{75}
\bpages{551--561}.
\end{barticle}
\endbibitem

\bibitem{glmnet2010}
\begin{barticle}[author]
\bauthor{\bsnm{Friedman},~\bfnm{Jerome}\binits{J.}},
  \bauthor{\bsnm{Tibshirani},~\bfnm{Robert}\binits{R.}} \AND
  \bauthor{\bsnm{Hastie},~\bfnm{Trevor}\binits{T.}}
(\byear{2010}).
\btitle{Regularization Paths for Generalized Linear Models via Coordinate
  Descent}.
\bjournal{Journal of Statistical Software}
\bvolume{33}
\bpages{1--22}.
\end{barticle}
\endbibitem

\bibitem{goeman2006testing}
\begin{barticle}[author]
\bauthor{\bsnm{Goeman},~\bfnm{Jelle~J}\binits{J.~J.}}, \bauthor{\bsnm{Van
  De~Geer},~\bfnm{Sara~A}\binits{S.~A.}} \AND
  \bauthor{\bsnm{Van~Houwelingen},~\bfnm{Hans~C}\binits{H.~C.}}
(\byear{2006}).
\btitle{Testing against a high dimensional alternative}.
\bjournal{Journal of the Royal Statistical Society: Series B (Statistical
  Methodology)}
\bvolume{68}
\bpages{477--493}.
\end{barticle}
\endbibitem

\bibitem{guojrssb2016}
\begin{barticle}[author]
\bauthor{\bsnm{Guo},~\bfnm{Bin}\binits{B.}} \AND
  \bauthor{\bsnm{Chen},~\bfnm{Song~Xi}\binits{S.~X.}}
(\byear{2016}).
\btitle{Tests for high dimensional generalized linear models}.
\bjournal{Journal of the Royal Statistical Society: Series B (Statistical
  Methodology)}
\bvolume{78}
\bpages{1079-1102}.
\end{barticle}
\endbibitem

\bibitem{guo2022conditional}
\begin{barticle}[author]
\bauthor{\bsnm{Guo},~\bfnm{Wenwen}\binits{W.}},
  \bauthor{\bsnm{Zhong},~\bfnm{Wei}\binits{W.}},
  \bauthor{\bsnm{Duan},~\bfnm{Sunpeng}\binits{S.}} \AND
  \bauthor{\bsnm{Cui},~\bfnm{Hengjian}\binits{H.}}
(\byear{2022}).
\btitle{Conditional Test for Ultrahigh Dimensional Linear Regression
  Coefficients}.
\bjournal{Statistica Sinica}
\bvolume{32}
\bpages{1381--1409}.
\end{barticle}
\endbibitem

\bibitem{guo2022high}
\begin{barticle}[author]
\bauthor{\bsnm{Guo},~\bfnm{Xu}\binits{X.}},
  \bauthor{\bsnm{Li},~\bfnm{Runze}\binits{R.}},
  \bauthor{\bsnm{Liu},~\bfnm{Jingyuan}\binits{J.}} \AND
  \bauthor{\bsnm{Zeng},~\bfnm{Mudong}\binits{M.}}
(\byear{2022}).
\btitle{High-dimensional mediation analysis for selecting DNA methylation Loci
  mediating childhood trauma and cortisol stress reactivity}.
\bjournal{Journal of the American Statistical Association}
\bvolume{117}
\bpages{1110--1121}.
\end{barticle}
\endbibitem

\bibitem{guo2021group}
\begin{barticle}[author]
\bauthor{\bsnm{Guo},~\bfnm{Zijian}\binits{Z.}},
  \bauthor{\bsnm{Renaux},~\bfnm{Claude}\binits{C.}},
  \bauthor{\bsnm{B{\"u}hlmann},~\bfnm{Peter}\binits{P.}} \AND
  \bauthor{\bsnm{Cai},~\bfnm{Tony}\binits{T.}}
(\byear{2021}).
\btitle{Group inference in high dimensions with applications to hierarchical
  testing}.
\bjournal{Electronic Journal of Statistics}
\bvolume{15}
\bpages{6633--6676}.
\end{barticle}
\endbibitem

\bibitem{houtepen2016genome}
\begin{barticle}[author]
\bauthor{\bsnm{Houtepen},~\bfnm{Lotte~C}\binits{L.~C.}},
  \bauthor{\bsnm{Vinkers},~\bfnm{Christiaan~H}\binits{C.~H.}},
  \bauthor{\bsnm{Carrillo-Roa},~\bfnm{Tania}\binits{T.}},
  \bauthor{\bsnm{Hiemstra},~\bfnm{Marieke}\binits{M.}},
  \bauthor{\bsnm{Van~Lier},~\bfnm{Pol~A}\binits{P.~A.}},
  \bauthor{\bsnm{Meeus},~\bfnm{Wim}\binits{W.}},
  \bauthor{\bsnm{Branje},~\bfnm{Susan}\binits{S.}},
  \bauthor{\bsnm{Heim},~\bfnm{Christine~M}\binits{C.~M.}},
  \bauthor{\bsnm{Nemeroff},~\bfnm{Charles~B}\binits{C.~B.}},
  \bauthor{\bsnm{Mill},~\bfnm{Jonathan}\binits{J.}} \betal{et~al.}
(\byear{2016}).
\btitle{Genome-wide DNA methylation levels and altered cortisol stress
  reactivity following childhood trauma in humans}.
\bjournal{Nature communications}
\bvolume{7}
\bpages{10967}.
\end{barticle}
\endbibitem

\bibitem{javanmard2014confidence}
\begin{barticle}[author]
\bauthor{\bsnm{Javanmard},~\bfnm{Adel}\binits{A.}} \AND
  \bauthor{\bsnm{Montanari},~\bfnm{Andrea}\binits{A.}}
(\byear{2014}).
\btitle{Confidence intervals and hypothesis testing for high-dimensional
  regression}.
\bjournal{Journal of Machine Learning Research}
\bvolume{15}
\bpages{2869--2909}.
\end{barticle}
\endbibitem

\bibitem{listgarten2010correction}
\begin{barticle}[author]
\bauthor{\bsnm{Listgarten},~\bfnm{Jennifer}\binits{J.}},
  \bauthor{\bsnm{Kadie},~\bfnm{Carl}\binits{C.}},
  \bauthor{\bsnm{Schadt},~\bfnm{Eric~E}\binits{E.~E.}} \AND
  \bauthor{\bsnm{Heckerman},~\bfnm{David}\binits{D.}}
(\byear{2010}).
\btitle{Correction for hidden confounders in the genetic analysis of gene
  expression}.
\bjournal{Proceedings of the National Academy of Sciences}
\bvolume{107}
\bpages{16465--16470}.
\end{barticle}
\endbibitem

\bibitem{loh2015regularized}
\begin{barticle}[author]
\bauthor{\bsnm{Loh},~\bfnm{Po-Ling}\binits{P.-L.}} \AND
  \bauthor{\bsnm{Wainwright},~\bfnm{Martin~J}\binits{M.~J.}}
(\byear{2015}).
\btitle{Regularized M-estimators with nonconvexity: Statistical and algorithmic
  theory for local optima}.
\bjournal{Journal of Machine Learning Research}
\bvolume{16}
\bpages{559--616}.
\end{barticle}
\endbibitem

\bibitem{ma2021global}
\begin{barticle}[author]
\bauthor{\bsnm{Ma},~\bfnm{Rong}\binits{R.}},
  \bauthor{\bsnm{Tony~Cai},~\bfnm{T}\binits{T.}} \AND
  \bauthor{\bsnm{Li},~\bfnm{Hongzhe}\binits{H.}}
(\byear{2021}).
\btitle{Global and simultaneous hypothesis testing for high-dimensional
  logistic regression models}.
\bjournal{Journal of the American Statistical Association}
\bvolume{116}
\bpages{984--998}.
\end{barticle}
\endbibitem

\bibitem{meinshausen2009p}
\begin{barticle}[author]
\bauthor{\bsnm{Meinshausen},~\bfnm{Nicolai}\binits{N.}},
  \bauthor{\bsnm{Meier},~\bfnm{Lukas}\binits{L.}} \AND
  \bauthor{\bsnm{B{\"u}hlmann},~\bfnm{Peter}\binits{P.}}
(\byear{2009}).
\btitle{P-values for high-dimensional regression}.
\bjournal{Journal of the American Statistical Association}
\bvolume{104}
\bpages{1671--1681}.
\end{barticle}
\endbibitem

\bibitem{ning2017general}
\begin{barticle}[author]
\bauthor{\bsnm{Ning},~\bfnm{Yang}\binits{Y.}} \AND
  \bauthor{\bsnm{Liu},~\bfnm{Han}\binits{H.}}
(\byear{2017}).
\btitle{A general theory of hypothesis tests and confidence regions for sparse
  high dimensional models}.
\bjournal{Annals of Statistics}
\bvolume{45}
\bpages{158--195}.
\end{barticle}
\endbibitem

\bibitem{serfling1980approximation}
\begin{bbook}[author]
\bauthor{\bsnm{Serfling},~\bfnm{Robert~J}\binits{R.~J.}}
(\byear{1980}).
\btitle{Approximation theorems of mathematical statistics}.
\bpublisher{John Wiley \& Sons}.
\end{bbook}
\endbibitem

\bibitem{sun2012scaled}
\begin{barticle}[author]
\bauthor{\bsnm{Sun},~\bfnm{Tingni}\binits{T.}} \AND
  \bauthor{\bsnm{Zhang},~\bfnm{Cun-Hui}\binits{C.-H.}}
(\byear{2012}).
\btitle{Scaled sparse linear regression}.
\bjournal{Biometrika}
\bvolume{99}
\bpages{879--898}.
\end{barticle}
\endbibitem

\bibitem{van2014asymptotically}
\begin{barticle}[author]
\bauthor{\bparticle{Van~de} \bsnm{Geer},~\bfnm{Sara}\binits{S.}},
  \bauthor{\bsnm{B{\"u}hlmann},~\bfnm{Peter}\binits{P.}},
  \bauthor{\bsnm{Ritov},~\bfnm{Ya’acov}\binits{Y.}} \AND
  \bauthor{\bsnm{Dezeure},~\bfnm{Ruben}\binits{R.}}
(\byear{2014}).
\btitle{On asymptotically optimal confidence regions and tests for
  high-dimensional models}.
\bjournal{Annals of Statistics}
\bvolume{42}
\bpages{1166--1202}.
\end{barticle}
\endbibitem

\bibitem{vanderVaart.1996}
\begin{bbook}[author]
\bauthor{\bsnm{{Van der Vaart}},~\bfnm{Aad~W.}\binits{A.~W.}} \AND
  \bauthor{\bsnm{Wellner},~\bfnm{Jon~A.}\binits{J.~A.}}
(\byear{1996}).
\btitle{Weak Convergence and Empirical Processes: With Applications to
  Statistics}.
\bpublisher{{Springer, New York}}.
\end{bbook}
\endbibitem

\bibitem{van2019exploratory}
\begin{barticle}[author]
\bauthor{\bparticle{van} \bsnm{Kesteren},~\bfnm{Erik-Jan}\binits{E.-J.}} \AND
  \bauthor{\bsnm{Oberski},~\bfnm{Daniel~L}\binits{D.~L.}}
(\byear{2019}).
\btitle{Exploratory mediation analysis with many potential mediators}.
\bjournal{Structural Equation Modeling: A Multidisciplinary Journal}
\bvolume{26}
\bpages{710--723}.
\end{barticle}
\endbibitem

\bibitem{highprobability2019}
\begin{bbook}[author]
\bauthor{\bsnm{Wainwright},~\bfnm{Martin~J}\binits{M.~J.}}
(\byear{2019}).
\btitle{High-Dimensional Statistics: A Non-Asymptotic Viewpoint}.
\bpublisher{Cambridge University Press}.
\end{bbook}
\endbibitem

\bibitem{wu2021model}
\begin{barticle}[author]
\bauthor{\bsnm{Wu},~\bfnm{Yunan}\binits{Y.}},
  \bauthor{\bsnm{Wang},~\bfnm{Lan}\binits{L.}} \AND
  \bauthor{\bsnm{Fu},~\bfnm{Haoda}\binits{H.}}
(\byear{2023}).
\btitle{Model-Assisted Uniformly Honest Inference for Optimal Treatment Regimes
  in High Dimension}.
\bjournal{Journal of the American Statistical Association}
\bvolume{118}
\bpages{305--314}.
\end{barticle}
\endbibitem

\bibitem{zhang2014confidence}
\begin{barticle}[author]
\bauthor{\bsnm{Zhang},~\bfnm{Cun-Hui}\binits{C.-H.}} \AND
  \bauthor{\bsnm{Zhang},~\bfnm{Stephanie~S}\binits{S.~S.}}
(\byear{2014}).
\btitle{Confidence intervals for low dimensional parameters in high dimensional
  linear models}.
\bjournal{Journal of the Royal Statistical Society: Series B (Statistical
  Methodology)}
\bvolume{76}
\bpages{217--242}.
\end{barticle}
\endbibitem

\bibitem{zhang2017simultaneous}
\begin{barticle}[author]
\bauthor{\bsnm{Zhang},~\bfnm{Xianyang}\binits{X.}} \AND
  \bauthor{\bsnm{Cheng},~\bfnm{Guang}\binits{G.}}
(\byear{2017}).
\btitle{Simultaneous inference for high-dimensional linear models}.
\bjournal{Journal of the American Statistical Association}
\bvolume{112}
\bpages{757--768}.
\end{barticle}
\endbibitem

\bibitem{zhong2011tests}
\begin{barticle}[author]
\bauthor{\bsnm{Zhong},~\bfnm{Ping-Shou}\binits{P.-S.}} \AND
  \bauthor{\bsnm{Chen},~\bfnm{Song~Xi}\binits{S.~X.}}
(\byear{2011}).
\btitle{Tests for high-dimensional regression coefficients with factorial
  designs}.
\bjournal{Journal of the American Statistical Association}
\bvolume{106}
\bpages{260--274}.
\end{barticle}
\endbibitem

\bibitem{zhu2006empirical}
\begin{barticle}[author]
\bauthor{\bsnm{Zhu},~\bfnm{Lixing}\binits{L.}} \AND
  \bauthor{\bsnm{Xue},~\bfnm{Liugen}\binits{L.}}
(\byear{2006}).
\btitle{Empirical likelihood confidence regions in a partially linear
  single-index model}.
\bjournal{Journal of the Royal Statistical Society: Series B (Statistical
  Methodology)}
\bvolume{68}
\bpages{549--570}.
\end{barticle}
\endbibitem

\end{thebibliography}

%
%
%
%

\end{document}